\documentclass{amsart}
\usepackage{amssymb}

\usepackage{amsmath}

\newtheorem{theorem}{Theorem}

\newtheorem{claim}[theorem]{Claim}

\newtheorem{definition}[theorem]{Definition}

\newtheorem{lemma}[theorem]{Lemma}
\newtheorem{notation}[theorem]{Notation}

\newtheorem{proposition}[theorem]{Proposition}

\theoremstyle{remark}
\newtheorem{remark}[theorem]{Remark}
\newtheorem{example}[theorem]{Example}
\numberwithin{equation}{section}   
\numberwithin{theorem}{section}

\begin{document}

\title[Yang-Mills and Segal-Bargmann]{Yang-Mills Theory and the
Segal-Bargmann Transform}
\author[Driver]{Bruce K. Driver\footnotemark {$^\dagger$}}
\thanks{\footnotemark {$^\dagger$}This research was partially supported by
NSF Grant
DMS 96-12651.}
\address{Department of Mathematics, 0112\\
University of California, San Diego \\
La Jolla, CA 92093-0112 }
\email{driver{@}math.ucsd.edu}
\author[Hall]{Brian C. Hall\footnotemark {$^*$}}
\thanks{\footnotemark {$^*$}Supported by an NSF Postdoctoral Fellowship.}
\email{bhall{@}math.ucsd.edu}


\maketitle

\begin{abstract}
We use a variant of the Segal-Bargmann transform to study canonically
quantized Yang-Mills theory on a space-time cylinder with a compact
structure group $K.$ The non-existent Lebesgue measure on the space of
connections is ``approximated'' by a Gaussian measure with large variance.
The Segal-Bargmann transform is then a unitary map from the $L^{2}$ space
over the space of connections to a \textit{holomorphic} $L^{2}$ space over
the space of complexified connections with a certain Gaussian measure. This
transform is given roughly by $e^{t\Delta _{\mathcal{A}}/2}$ followed by
analytic continuation. Here $\Delta _{\mathcal{A}}$ is the Laplacian on the
space of connections and is the Hamiltonian for the quantized theory.

On the gauge-trivial subspace, consisting of functions of the holonomy
around the spatial circle, the Segal-Bargmann transform becomes $e^{t\Delta
_{K}/2}$ followed by analytic continuation, where $\Delta _{K}$ is the
Laplacian for the structure group $K.$ This result gives a rigorous meaning
to the idea that $\Delta _{\mathcal{A}}$ reduces to $\Delta _{K}$ on
functions of the holonomy. By letting the variance of the Gaussian measure
tend to infinity we recover the standard realization of the quantized
Yang-Mills theory on a space-time cylinder, namely, $-\frac{1}{2}\Delta _{K}$
is the Hamiltonian and $L^{2}(K)$ is the Hilbert space. As a byproduct of
these considerations, we find a new one-parameter family of unitary
transforms from $L^{2}(K)$ to certain holmorphic $L^{2}$-spaces over the
complexification of $K.$ This family of transformations interpolates between
the two previously known unitary transformations.

Our work is motivated by results of Landsman and Wren and uses probabilistic
techniques similar to those of Gross and Malliavin.
\end{abstract}

\tableofcontents

\section{Introduction\label{s.1}}

This paper uses techniques of stochastic analysis to address the problem of
canonically quantizing Yang-Mills theory on a space-time cylinder. We
outline our results briefly here, leaving a detailed description of the
Yang-Mills interpretation to Section \ref{s.2}. Let $K$ be a connected
compact Lie group and $\frak{k}$ be its Lie algebra endowed with a fixed Ad-$%
K$-invariant inner product. Let $\mathcal{\bar{A}}$ be a certain subspace of
the $\frak{k}$-valued distributions on $\left[ 0,1\right] $ and $\tilde{P}%
_{s}$ be a scaled white noise measure on $\mathcal{\bar{A}}.$ See (\ref
{e.4.2}) and Definition \ref{d.4.1} below. By taking the indefinite
``integrals'' of elements of $\mathcal{\bar{A}},$ the measure space $(%
\mathcal{\bar{A}},\tilde{P}_{s})$ may be identified with the space of $\frak{%
k}$-valued paths on $[0,1]$ starting at $0$ equipped with a Wiener measure
of variance $s.$ Elements of $\mathcal{\bar{A}}$ are to be interpreted as
(generalized) connections on the spatial circle.

Our objective is to understand infinite-dimensional Laplacian operator $%
\Delta _{\mathcal{A}},$ where $\mathcal{A}$ is the Cameron-Martin subspace
of $\mathcal{\bar{A}},$ namely, the Hilbert space of square-integrable $%
\frak{k}$-valued functions. Since $\Delta _{\mathcal{A}}$ is poorly behaved
(e.g., non-closable) as a operator on $L^{2}(\mathcal{\bar{A}},\tilde{P}%
_{s}),$ we work with a variant $\tilde{S}_{s,t}$ of the \textit{%
Segal-Bargmann transform}. This is defined to be $e^{t\Delta _{\mathcal{A}%
}/2}$ followed by analytic continuation. The transform is defined at first
on cylinder functions but extends to a unitary map of $L^{2}(\mathcal{\bar{A}%
},\tilde{P}_{s})$ onto the \textit{holomorphic} subspace of $L^{2}(\mathcal{%
\bar{A}}_{\mathbb{C}},\tilde{M}_{s,t}).$ Here $\mathcal{\bar{A}}_{\mathbb{C}%
} $ is a space of $\frak{k}_{\mathbb{C}}$-valued distributions (where $\frak{%
k}_{\mathbb{C}}=\frak{k}+i\frak{k}$) and $\tilde{M}_{s,t}$ is a certain
Gaussian measure on $\mathcal{\bar{A}}_{\mathbb{C}}.$

We are particularly interested in the It\^{o} map $\theta ,$ which
associates to almost every $A\in \mathcal{\bar{A}}$ a continuous $K$-valued
path $\theta _{\cdot }\left( A\right) .$ Geometrically, $\theta _{\tau
}\left( A\right) $ represents the parallel transport of the connection $A$
from $0$ to $\tau ,$ and $h\left( A\right) :=\theta _{1}\left( A\right) $
represents the \textit{holonomy} of $A$ around the spatial circle. We
similarly consider the It\^{o} map $\theta ^{\mathbb{C}}$ and the holonomy $%
h_{\mathbb{C}}$ for complex connections $C\in \mathcal{\bar{A}}_{\mathbb{C}%
}, $ where $\theta _{\tau }^{\mathbb{C}}\left( C\right) $ and $h_{\mathbb{C}%
}\left( C\right) $ take values in the complexification $K_{\mathbb{C}}$ of
the compact group $K.$

The main result is Theorem \ref{t.5.2} of Section \ref{s.5}, which states:

\begin{quotation}
Suppose $f\in L^{2}(\mathcal{\bar{A}},\tilde{P}_{s})$ is of the form 
\begin{equation*}
f\left( A\right) =\phi \left( h\left( A\right) \right)
\end{equation*}
where $\phi $ is a function on $K.$ Then there exists a unique holomorphic
function $\Phi $ on $K_{\mathbb{C}}$ such that 
\begin{equation*}
\tilde{S}_{s,t}f\left( C\right) =\Phi \left( h_{\mathbb{C}}\left( C\right)
\right) .
\end{equation*}
The function $\Phi $ is determined by the condition that 
\begin{equation*}
\left. \Phi \right| _{K}=e^{t\Delta _{K}/2}\phi .
\end{equation*}
\end{quotation}

Here $\Delta _{K}$ refers to the Laplacian for the compact group $K.$ Recall
that $\tilde{S}_{s,t}$ is defined to be $e^{t\Delta _{\mathcal{A}}/2}$
followed by analytic continuation. Theorem \ref{t.5.2} says that on
functions of the holonomy $\tilde{S}_{s,t}$ reduces to $e^{t\Delta _{K}/2}$
followed by analytic continuation. This is formally equivalent to the
following imprecise principle.

\begin{quotation}
On functions of the holonomy $\Delta _{\mathcal{A}}$ reduces to $\Delta
_{K}. $
\end{quotation}

As a consequence of Theorem \ref{t.5.2} and the ``averaging lemma'' in \cite
{H1} we obtain a Segal-Bargmann type transform for the compact group $K$
(Theorem \ref{t.5.3}). To describe this theorem let $\rho _{s}$ denote the
distribution of $h\left( A\right) $ with respect to $\tilde{P}_{s}$ and $\mu
_{s,t}$ denote the distribution of $h_{\mathbb{C}}\left( C\right) $ with
respect to $\tilde{M}_{s,t}.$ The measures $\rho _{s}$ and $\mu _{s,t}$ are
certain heat kernel measures on $K$ and $K_{\mathbb{C}},$ respectively. Then
Theorem \ref{t.5.3} asserts:

\begin{quotation}
The map 
\begin{equation*}
\phi \rightarrow \text{ analytic continuation of }e^{t\Delta _{K}/2}\phi
\end{equation*}
is an isometric isomorphism of $L^{2}\left( K,\rho _{s}\right) $ onto the
holomorphic subspace of $L^{2}\left( K_{\mathbb{C}},\mu _{s,t}\right) .$
\end{quotation}

This generalized Segal-Bargmann transform for $K$ was known previously \cite
{H1} in the case $s=t$ and also in the limiting case $s\rightarrow \infty .$
(See also \cite{H2,H3,H4,D2,DG}.) For general $s$ and $t$ this transform is
new, and interpolates continuously between the two previously known cases.
An analysis of this new transform, from purely finite-dimensional point of
view, is given in \cite{H5}.

There is a simple explanation (not a proof) for Theorem \ref{t.5.2}. The
Hilbert space $\mathcal{A}$ (the Cameron-Martin subspace) may be thought of
as an infinite-dimensional flat Riemannian manifold. Let $H\left( K\right) $
be the infinite-dimensional group of finite-energy paths with values in $K,$
starting at the identity. This has a natural right-invariant Riemannian
metric. Theorem \ref{t.8.1} in Section \ref{s.8} asserts that the It\^{o}
map $\theta ,$ restricted to the Cameron-Martin subspace $\mathcal{A},$ is
an isometry of $\mathcal{A}$ onto $H\left( K\right) .$ From this
observation, one formally concludes that $\Delta _{\mathcal{A}}\left( f\circ
\theta \right) =\left( \Delta _{H\left( K\right) }f\right) \circ \theta $
where $\Delta _{H\left( K\right) }$ is the ``Laplace-Beltrami'' operator
associated to the right-invariant Riemannian structure on $H(K).$
Furthermore, if $f$ depends only on the endpoint of the path (i.e., $f\circ
\theta $ is a function of the holonomy) then an easy calculation shows that $%
\Delta _{H\left( K\right) }$ reduces to $\Delta _{K}$ (Theorem \ref{t.8.11}%
). However, even working on $\mathcal{A}$ there are serious domain issues to
deal with, and of course $\mathcal{A}$ is a measure-zero subset of $\mathcal{%
\bar{A}}.$ So the proof of Theorem \ref{t.5.2} does not make direct use of
this calculation. Nevertheless we present it as motivation, with a precise
treatment of the domain issues, in Appendix A.

The main tool in the proof of Theorem \ref{t.5.2} is the Hermite expansion,
which for $L^{2}(\mathcal{\bar{A}},\tilde{P}_{s})$ takes the form of an
expansion in terms of multiple Wiener integrals, the so-called Wiener chaos
or homogeneous chaos expansion \cite{Ka,Ito}. The Segal-Bargmann transform $%
\tilde{S}_{s,t}$ has a very simple action on this expansion, given in
Theorem \ref{t.4.7} of Section \ref{s.4.2}.

Although it is natural from the standpoint of Yang-Mills theory to consider
functions of the holonomy $h\left( A\right) :=\theta _{1}\left( A\right) ,$
it makes sense to apply $\tilde{S}_{s,t}$ to arbitrary functions of the
parallel transport $\theta _{\tau },$ $\tau \in \left[ 0,1\right] .$ Results
similar to Theorem \ref{t.5.2} hold, described in Section \ref{s.6}.

Finally, let us mention the paper \cite{AHS}, which considers a sort of
Segal-Bargmann transform in the context of two-dimensional \textit{Euclidean}
Yang-Mills theory. That paper is not so much concerned with constructing the
theory as with understanding the structure of the Euclidean Yang-Mills
measure. Despite a superficial similarity, there is no overlap of results
between \cite{AHS} and the present paper.

\textit{Acknowledgments}. Our work on this problem was motivated by the
results of N.P. Landsman and K.K. Wren \cite{LW,W1,W2}. (See also \cite{L}.)
Although they take a different approach to the quantization, their results
strongly suggest a connection between Yang-Mills theory on a cylinder and
the generalized Segal-Bargmann transform for $K.$ See Section \ref{s.2} for
more on the relationship of our results to theirs. We thank both Landsman
and Wren for valuable scientific discussions.

In the case $s=t,$ a result very close to Theorem \ref{t.5.2} was proved by
L. Gross and P. Malliavin \cite{GM}. We thank Gross for helpful comments on
an earlier form of this paper. In the case $s=t,$ Theorem \ref{t.5.2} itself
was proved in \cite{Sa2} and in \cite{HS}. We thank A. Sengupta for his
insights into Theorem \ref{t.5.2}.

The first author is grateful to David Brydges for clarifying to him the role
of the heat equation in Wick ordering and thanks l'Institut Henri
Poincar\'{e} and l'\'{E}cole Normal Sup\'{e}riure for their hospitality.

We thank Nolan Wallach for valuable discussions concerning the energy
representation, and we thank the referee for helpful comments.

\section{The Yang-Mills interpretation\label{s.2}}

This section explains the motivation for, and the desired interpretation of,
the results of the paper. It may be skipped without a loss of understanding
of the statements.

The Segal-Bargmann transform was developed independently in the early 1960's
by Segal \cite{S1,S2,S3} in the infinite-dimensional context of scalar
quantum field theories and by Bargmann \cite{B} in the finite-dimensional
context of quantum mechanics on $\mathbb{R}^{n}.$ The paper \cite{H1}
introduced an analog of the Segal-Bargmann transform in the context of
quantum mechanics on a compact Lie group. A natural next step is to attempt
to combine the compact group with the field theory in order to obtain a
transform in the context of quantum gauge theories. One such transform has
already been obtained by Ashtekar, \textit{et al.} \cite{A}, with
application to quantum gravity.

This paper considers the canonical quantization of Yang-Mills theory in the
simplest non-trivial case, namely that of a space-time cylinder. We consider
first briefly the classical Yang-Mills theory. (See also \cite{L,RR}.) Let $K
$ be a connected compact Lie group (the structure group) together with an Ad-%
$K$-invariant inner product $\langle \cdot ,\cdot \rangle $ on its Lie
algebra $\frak{k}.$ We work in the temporal gauge, in which case the
configuration space for the classical Yang-Mills theory is the space of $%
\frak{k}$-valued 1-forms on the spatial circle. More precisely, let $%
\mathcal{A}$ denote the space of square-integrable $\frak{k}$-valued
1-forms, which can be identified with $L^{2}\left( \left[ 0,1\right] ;\frak{k%
}\right) ,$ where the circle is $\left[ 0,1\right] $ with ends identified.
The phase space of the system is then $\mathcal{A}_{\mathbb{C}}=\mathcal{A}+i%
\mathcal{A}.$ The dynamical part of the Yang-Mills equations (e.g., 
\cite[Eq. (2)]{Di}) may be expressed in Hamiltonian form, with the
Hamiltonian function on $\mathcal{A}_{\mathbb{C}}$ given by 
\begin{equation}
H\left( A+iP\right) =\frac{1}{2}\left\| P\right\| ^{2}=\frac{1}{2}%
\int_{0}^{1}\left| P_{\tau }\right| ^{2}\,d\tau .  \label{e.2.1}
\end{equation}
Note that since our spatial manifold is one-dimensional, the curvature term
which usually appears in the Hamiltonian is zero.

There is also a constraint part to the Yang-Mills equation (e.g., \cite[Eq.
(1)]{Di}), namely, 
\begin{equation}
\frac{dP_{\tau }}{d\tau }+\left[ A_{\tau },P_{\tau }\right] =0,\quad \tau
\in \left[ 0,1\right] .  \label{e.constraint}
\end{equation}
or equivalently, 
\begin{equation*}
J_{h}\left( A,P\right) :=-\left\langle h^{\prime }+\left[ A,h\right]
,P\right\rangle _{\mathcal{A}}=0
\end{equation*}
for all $h\in \mathcal{C}^{1}\left( S^{1}\rightarrow \frak{k}\right) .$ The
set of points $\left( A,P\right) $ satisfying this constraint is preserved
under the time evolution generated by (\ref{e.2.1}).

Now let $\mathcal{G}$ be the \textbf{gauge group}, namely, the group of maps
of the spatial circle into $K.$ This acts on $\mathcal{A}$ by 
\begin{equation*}
\left( g\cdot A\right) _{\tau }=g_{\tau }A_{\tau }g_{\tau }^{-1}-\frac{dg}{%
d\tau }g_{\tau }^{-1}
\end{equation*}
and on $\mathcal{A}_{\mathbb{C}}$ by 
\begin{equation*}
\left( g\cdot \left( A,P\right) \right) _{\tau }=\left( \left( g\cdot
A\right) _{\tau },g_{\tau }P_{\tau }g_{\tau }^{-1}\right) .
\end{equation*}
The gauge action preserves both the dynamics and the constraint. The
function $J_{h}$ on $\mathcal{A}_{\mathbb{C}}$ is the Hamiltonian generator
of the action of the one-parameter subgroup $e^{th}$ in $\mathcal{G},$ where 
$\left( e^{th}\right) _{\tau }=e^{th_{\tau }}.$ That is, $J$ is the moment
mapping for the action of $\mathcal{G}$ \cite[Sect. IV.3.6]{L}.

The parallel transport $\theta _{\tau }\left( A\right) $ of a connection $%
A\in \mathcal{A},$ is the solution to the $K$-valued differential equation 
\begin{equation}
\frac{d\theta }{d\tau }=\theta _{\tau }A_{\tau },\quad \theta _{0}=e,
\label{e.2.2}
\end{equation}
and $\theta $ transforms under gauge transformations as 
\begin{equation}
\theta _{\tau }\left( g\cdot A\right) =g_{0}\theta _{\tau }\left( A\right)
g_{\tau }^{-1},\quad g\in \mathcal{G}.  \label{e.gtheta}
\end{equation}
The \textbf{holonomy }of $A$ is the parallel transport around the circle: $%
h\left( A\right) :=\theta _{1}\left( A\right) .$ (In the interests of
consistency with \cite{G,GM,HS} we have put $\theta _{\tau }$ to the left of 
$A_{\tau }$ in (\ref{e.2.2}). Although this is the reverse of the usual
definition of parallel transport, it makes little difference. Theorem \ref
{t.5.2} would be unchanged with the other definition and Theorem \ref{t.6.3}
would require just the reversal of $x$ and $g$ in (\ref{e.6.2}).)

If we \textit{formally} apply the usual canonical quantization procedure to
this classical Yang-Mills theory, we find that the quantum mechanical
Hilbert space is $L^{2}\left( \mathcal{A},\mathcal{D}A\right) $ and the
Hamiltonian operator corresponding to the classical Hamiltonian (\ref{e.2.1}%
) is $-\Delta _{\mathcal{A}}/2.\ $Here $\mathcal{D}A$ is the \textit{%
non-existent} Lebesgue measure on $\mathcal{A}$ and $\Delta _{\mathcal{A}}$
is the Laplacian operator, that is, the sum of squares of derivatives in the
directions of an orthonormal basis. The quantum operator corresponding to
the function $J_{h}$ is the vector field $\hat{J}_{h}$ given by 
\begin{eqnarray*}
\hat{J}_{h}F\left( A\right) &=&i\left. \frac{d}{dt}\right| _{t=0}F\left(
A+t\left( h^{\prime }+\left[ A,h\right] \right) \right) \\
&=&-i\left. \frac{d}{dt}\right| _{t=0}F\left( e^{th}\cdot A\right) .
\end{eqnarray*}
Thus the quantum analog of the constraint equation is to require that $F\in
L^{2}\left( \mathcal{A},\mathcal{D}A\right) $ be $\mathcal{G}$-invariant.
(More precisely, this is true if $\mathcal{G}$ is connected, i.e., if $K$ is
simply connected. We will simply assume $\mathcal{G}$-invariance even if $K$
is not simply connected, and will not address here the issue of ``$\theta $%
-angles.'' See \cite{LW,W1,L}.)

We consider at first the \textbf{based gauge group} $\mathcal{G}_{0}$%
\begin{equation*}
\mathcal{G}_{0}=\left\{ g\in \mathcal{G}\left| g_{0}=g_{1}=e\right. \right\}
.
\end{equation*}
It is not hard to verify using (\ref{e.gtheta}) (see \cite{L}) that two
connections are $\mathcal{G}_{0}$-equivalent if and only if they have the
same holonomy. Thus, the $\mathcal{G}_{0}$-invariant functions are precisely
the functions of the form $\phi \left( h\left( A\right) \right) ,$ where $%
\phi $ is a function on $K.$ For invariance under the full gauge group, $%
\phi $ would be required to be a class function.

Now let $\Delta _{K}$ denote the Laplacian (quadratic Casimir) operator on $%
K $ associated to the chosen invariant inner product on $\frak{k}.$

\begin{claim}[Main Idea]
\label{main.2.1}Consider a function on $\mathcal{A}$ of the form $\phi
\left( h\left( A\right) \right) ,$ where $\phi $ is a function on $K.$ Then 
\begin{equation*}
\Delta _{\mathcal{A}}\phi \left( h\left( A\right) \right) =\left( \Delta
_{K}\phi \right) \left( h\left( A\right) \right) .
\end{equation*}
\end{claim}

That is, on functions of the holonomy, the Laplacian for the space of
connections should reduce to the Laplacian on the structure group. This idea
is not new. It is stated without proof in \cite[pp.166,169]{Wi}, and a
rigorous result in this direction is given in \cite[Lem. 3.2]{Di}. (See the
end of this section.) The challenge is not so much to prove the result but
to give it a rigorous interpretation. (See also \cite{Ra}, where reduction
is done before quantization.)

One approach is to approximate the non-existent Lebesgue measure by a
Gaussian measure $\tilde{P}_{s}$ with large variance $s,$ where ``large''
means that at the appropriate point in our calculations $s$ will tend to
infinity. The measure $\tilde{P}_{s}$ does not exist on $\mathcal{A}$
itself, but does exist on a certain space $\mathcal{\bar{A}}$ of generalized
connections. We then take $\mathcal{G}_{0}$ to be the group of \textbf{%
finite-energy} maps $g:\left[ 0,1\right] \rightarrow K$ satisfying $%
g_{0}=g_{1}=e,$ where finite energy means that $g$ is absolutely continuous
and $\int_{0}^{1}\left| g_{\tau }^{-1}\,dg/d\tau \right| ^{2}\,d\tau <\infty
.$ The action of $\mathcal{G}_{0}$ on $\mathcal{A}$ may be extended to an
action on $\mathcal{\bar{A}},$ and this action leaves $\tilde{P}_{s}$
quasi-invariant. We consider the Hilbert space $L^{2}(\mathcal{\bar{A}},%
\tilde{P}_{s})$ and define the gauge-trivial subspace to be: 
\begin{equation}
L^{2}(\mathcal{\bar{A}},\tilde{P}_{s})^{\mathcal{G}_{0}}=\left\{ f\in L^{2}(%
\mathcal{\bar{A}},\tilde{P}_{s})\left| \forall g\in \mathcal{G}_{0},\,\text{%
{}}f\left( g^{-1}\cdot A\right) =f\left( A\right) \,\text{a.e.}\right.
\right\} .  \label{e.2.3}
\end{equation}
Note that the map $f\left( A\right) \rightarrow f\left( g^{-1}\cdot A\right) 
$ is not unitary, since $\mathcal{G}_{0}$ leaves $\tilde{P}_{s}$
quasi-invariant but not invariant. We are deliberately not unitarizing the
action of $\mathcal{G}_{0}$ as in \cite{LW}; the point of letting $%
s\rightarrow \infty $ is to avoid having to do so. The following result
shows clearly our motivation for not unitarizing.

\begin{theorem}
\label{t.2.2}Let $U\left( g\right) $ be the unitary gauge action, as for
example in \cite{Di}. If $f\in L^{2}(\mathcal{\bar{A}},\tilde{P}_{s})$ and $%
U\left( g\right) f=f$ for all $g\in \mathcal{G}_{0},$ then $f=0.$
\end{theorem}

The corresponding results in dimensions 3+1 and higher (and in certain
(2+1)-dimensional cases) is a consequence of the irreducibility of the
energy representation \cite{Wa,AKT,GGV}, at least for the case when $K$ is
semisimple. In the one-dimensional case considered here, the energy
representation is reducible, so a different proof is needed, and is given in 
\cite{DH}. Defining the gauge-trivial subspace in terms of the un-unitarized
action as in (\ref{e.2.3}) gives a non-zero Hilbert space, as we shall see
momentarily. Unitarity is recovered, at least formally, in the $s\rightarrow
\infty $ limit. (However, in cases where the energy representation is
irreducible, the space defined in (\ref{e.2.3}) contains only the constants.
So our approach will not work without modification in high dimensions.)

The parallel transport map $\theta ,$ and so also the holonomy, may be
``extended'' from $\mathcal{A}$ to $\mathcal{\bar{A}}$ by replacing the
differential equation (\ref{e.2.2}) with a \textit{stochastic} differential
equation, the It\^{o} map (Section \ref{s.5}). A deep theorem of Gross
asserts that the elements of $L^{2}(\mathcal{\bar{A}},\tilde{P}_{s})^{%
\mathcal{G}_{0}}$ are precisely functions of the holonomy. (See also \cite
{Sa1}.) The reason this is not obvious is that although we have enlarged the
space of connections by replacing $\mathcal{A}$ with $\mathcal{\bar{A}}$, we
cannot unduly enlarge the group of gauge transformations without losing
quasi-invariance, without which (\ref{e.2.3}) does not make sense. As a
result, two connections in $\mathcal{\bar{A}}$ with the same holonomy need
not be gauge-equivalent. (If $K$ is simply connected, then the
characterization of $L^{2}(\mathcal{\bar{A}},\tilde{P}_{s})^{\mathcal{G}%
_{0}} $ is obtained by composing \cite[Thm. 2.5]{G} with the It\^{o} map.
The general case is easily reduced to the simply connected case. Note that
our $\mathcal{G}_{0}$ corresponds to $\hat{K}_{0}$ (\textit{not} $K_{0}$) in
the notation of Gross.)

We are back, then, to the matter of computing the Laplacian on functions of
the holonomy. Unfortunately, while $\Delta _{\mathcal{A}}$ is densely
defined in $L^{2}(\mathcal{\bar{A}},\tilde{P}_{s})$ (say on smooth,
compactly supported cylinder functions), it is not closable. So it is not
clear what it means to apply the Laplacian to a function of the holonomy. We
consider, then, a variant of the Segal-Bargmann transform. The transform
involves the heat operator $e^{t\Delta _{\mathcal{A}}/2},$ instead of the $%
\Delta _{\mathcal{A}}$ itself. More precisely, the Segal-Bargmann transform
consists of $e^{t\Delta _{\mathcal{A}}/2}$ followed by analytic
continuation. The transform maps from $L^{2}(\mathcal{\bar{A}},\tilde{P}%
_{s}) $ onto a certain $L^{2}$ space of holomorphic functions on a space $%
\mathcal{\bar{A}}_{\mathbb{C}}$ of complexified connections, rather than
from $L^{2}(\mathcal{\bar{A}},\tilde{P}_{s})$ to itself. Although the
Segal-Bargmann transform is defined initially only on cylinder functions, it
is an isometric map and so extends by continuity to all of $L^{2}(\mathcal{%
\bar{A}},\tilde{P}_{s}).$ In particular, it makes sense to apply the
Segal-Bargmann transform to functions of the holonomy.

Our main result is Theorem \ref{t.5.2}, described already in the
introduction. It asserts that for functions of the holonomy the
Segal-Bargmann transform (roughly, the heat operator $e^{t\Delta _{\mathcal{A%
}}/2},$ followed by analytic continuation) becomes the heat operator $%
e^{t\Delta _{K}/2}$ for the structure group $K$, followed by analytic
continuation. This holds for each fixed $s,$ not just in the $s\rightarrow
\infty $ limit. Thus Theorem \ref{t.5.2} gives a rigorous meaning to the
Main Idea in Claim \ref{main.2.1}.

Now, the gauge-trivial subspace, which consists of functions of the form $%
\phi \left( h\left( A\right) \right) ,$ may be identified with $L^{2}\left(
K,\rho _{s}\right) ,$ where $\rho _{s}$ is the distribution of $h\left(
A\right) $ with respect to $\tilde{P}_{s}.$ Similarly, the space of
functions of the form $\Phi \left( h_{\mathbb{C}}\left( C\right) \right) $
may be identified with $L^{2}\left( K_{\mathbb{C}},\mu _{s,t}\right) ,$
where $\mu _{s,t}$ is the distribution of $h_{\mathbb{C}}\left( C\right) $
with respect to the relevant Gaussian measure on $\mathcal{\bar{A}}.$ So
restricting the Segal-Bargmann transform to the gauge-trivial subspace gives
an isometric map from $L^{2}\left( K,\rho _{s}\right) $ into the holomorphic
subspace of $L^{2}\left( K_{\mathbb{C}},\mu _{s,t}\right) ,$ given by $%
e^{t\Delta _{K}/2}$ followed by analytic continuation. A finite-dimensional
argument shows that this transform maps \textit{onto} the holomorphic
subspace. This gives a unitary Segal-Bargmann-type transform (Theorem \ref
{t.5.3}) for $K,$ $B_{s,t}:L^{2}\left( K,\rho _{s}\right) \rightarrow 
\mathcal{H}L^{2}\left( K_{\mathbb{C}},\mu _{s,t}\right) $ given by 
\begin{equation*}
B_{s,t}f=\text{ analytic continuation of }e^{t\Delta _{K}/2}f.
\end{equation*}
Here $\mathcal{H}L^{2}$ denotes the space of square-integrable holomorphic
functions. This unitary transform was previously obtained in \cite{H1} for
the case $s=t$ and the limiting case $s\rightarrow \infty ,$ which we now
discuss.

Note that the formula for $B_{s,t}$ depends only on $t;$ the $s$-dependence
is only in the measures. Hence it makes sense to let $s$ tend to infinity.
In this limit $\rho _{s}$ converges to normalized Haar measure on $K$ and $%
\mu _{s,t}$ converges to a certain $K$-invariant measure $\nu _{t}$ on $K_{%
\mathbb{C}}.$ So taking the Segal-Bargmann transform for $\mathcal{\bar{A}},$
restricting to the gauge-trivial subspace, and taking the large variance
limit yields a unitary transform $C_{t},$ mapping $L^{2}\left( K,\text{ Haar}%
\right) $ onto the holomorphic subspace of $L^{2}\left( K_{\mathbb{C}},\nu
_{t}\right) .$ The transform is given, as always, by the time $t$ heat
operator followed by analytic continuation. Here $t$ is an arbitrary
positive parameter, which is to be interpreted physically as Planck's
constant. See Theorem \ref{t.5.5}.

We arrive, then, at the expected conclusion: the physical Hilbert space for
quantized Yang-Mills on a space-time cylinder is $L^{2}\left( K,\,\text{Haar}%
\right) $ and the Hamiltonian operator is $-\Delta _{K}/2.$ As a bonus, we
obtain a natural Segal-Bargmann transform $C_{t}$ for the physical Hilbert
space. Let us mention two other matters in passing. First, for invariance
under the full gauge group $\mathcal{G}$ we would restrict attention to the
Ad-invariant subspace of $L^{2}\left( K,\,\text{Haar}\right) .$ Second, the
Wilson loop operators naturally act as multiplication operators in $L^{2}(%
\mathcal{\bar{A}},\tilde{P}_{s})$ and so also in $L^{2}\left( K,\,\text{Haar}%
\right) .$

Let briefly compare our approach to others. Landsman and Wren \cite
{LW,W1,W2,L} use a method called Rieffel induction, in which gauge symmetry
is implemented by means of a certain integral over the gauge group. Under
this integration the classical coherent states for $\mathcal{A}$ map to the
coherent states that are associated to the transform $C_{t}$ described above
(the $s\rightarrow \infty $ limit of $B_{s,t}$). However, it is not clear
how to derive by this method the relevant measure $\nu _{t}$ on $K_{\mathbb{C%
}},$ and the computation of the reduced Hamiltonian \cite{W1} is complicated.

Dimock \cite{Di} adds a mass term to the Hamiltonian, which makes it
self-adjoint in $L^{2}(\mathcal{\bar{A}},\tilde{P}_{s}).$ However, because
the mass term destroys gauge-invariance, Dimock obtains a result like the
Main Idea only in the $s\rightarrow \infty $ limit \cite[Lem. 3.2]{Di}.

The Euclidean method for Yang-Mills on a cylinder constructs a probability
measure directly on connections modulo gauge transformations. Let $h^{t}$
denote the holonomy around the spatial circle at time $t.$ Then it can be
shown that: 1) for each $t,$ $h^{t}$ is distributed as Haar measure on $K,$
and 2) $h^{t}$ is a $K$-valued Brownian motion. Thinking in terms of the
temporal gauge, it is then reasonable to take as the time-zero Hilbert space 
$L^{2}\left( K,\,\text{Haar}\right) ,$ and (since the infinitesimal
generator of Brownian motion on $K$ is $\Delta _{K}/2$) to take as the
Hamiltonian $-\Delta _{K}/2.$ Since the Euclidean Yang-Mills measure does
not exist on the space of connections, but only on connections modulo gauge
transformations, the Euclidean method does not directly address the
relationship between $\Delta _{\mathcal{A}}$ and $\Delta _{K}.$

Finally in Appendix A, $\Delta _{\mathcal{A}}$ is considered as an operator
acting on functions on $\mathcal{A}$ rather than on $\mathcal{\bar{A}}.$
While functions of the form $\phi \left( h\left( A\right) \right) ,$ with $%
\phi $ smooth on $K,$ are differentiable on $\mathcal{A},$ the Hessian of
such a function is not in general trace-class. Hence, it is not possible to
define $\Delta _{\mathcal{A}}\phi \left( h\left( A\right) \right) $ as the
trace of the Hessian of $\phi \left( h\left( A\right) \right) .$ This
problem is circumvented by computing the trace by a two-step procedure--see
Definition \ref{d.8.5}. With this definition, we prove a rigorous version of
the Main Idea in Claim \ref{main.2.1} (Theorem \ref{t.8.11}). Since $%
\mathcal{A}$ is a set of $\tilde{P}_{s}$-measure zero, these results do not
bear directly on Theorem \ref{t.5.2}.

\section{Segal-Bargmann for $\mathbb{R}^{d}$\label{s.3}}

We consider a variant of the classical Segal-Bargmann transform that depends
on two parameters, one of which we wish to let tend to infinity. See Section 
\ref{s.2} for motivation. This is in contrast to the conventional version of
the transform, which has only one parameter (or none, depending on the
author). However, in the $\mathbb{R}^{d}$ case, this two-parameter transform
is not truly new, but can be reduced to the classical one-parameter version
by elementary changes of variable. This reduction is described in \cite{H5}.
In Section \ref{s.3.1} we describe the transform itself. In Section \ref
{s.3.2} we describe Hermite expansions on both the domain and range of the
transform, and we describe the action of the transform on these expansions.
Hermite expansions play a key role in the proof of our main result, Theorem 
\ref{t.5.2} in Section \ref{s.5}.

\subsection{The Transform for $\mathbb{R}^{d}$\label{s.3.1}}

Let $\Delta $ be the standard Laplacian on $\mathbb{R}^{d}$ and $P_{s}$ be
the associated Gaussian measure. Explicitly for $s>0,$ $dP_{s}\left(
x\right) =P_{s}\left( x\right) \,dx,$ where 
\begin{equation*}
P_{s}\left( x\right) =\left( 2\pi s\right) ^{-d/2}e^{-x^{2}/2s}.
\end{equation*}
Here $x=\left( x_{1},\cdots ,x_{d}\right) ,$ $x^{2}=x_{1}^{2}+\cdots
+x_{d}^{2},$ and $dx$ is standard Lebesgue measure on $\mathbb{R}^{d}.$ Note
that the function $P_{s}\left( x\right) $ admits an analytic continuation to 
$\mathbb{C}^{d},$ denoted $P_{s}(z).$ Now for any number $t$ with $t<2s$
(i.e., $s>t/2$) define a map 
\begin{equation*}
S_{s,t}:L^{2}\left( \mathbb{R}^{d},P_{s}\right) \rightarrow \mathcal{H}%
\left( \mathbb{C}^{d}\right)
\end{equation*}
by 
\begin{equation}
S_{s,t}f\left( z\right) =\int_{\mathbb{R}^{d}}P_{t}\left( z-x\right) f\left(
x\right) \,dx,\quad z\in \mathbb{C}^{d},  \label{e.3.1}
\end{equation}
where $\mathcal{H}\left( \mathbb{C}^{d}\right) $ denotes the space of
holomorphic functions on $\mathbb{C}^{d}.$ The integral is well defined
since for $t<2s,P_{t}\left( z-x\right) /P_{s}\left( x\right) $ is in $L^{2}(%
\mathbb{R}^{d},P_{s}\left( x\right) ).$ Using Morera's theorem one may show
that $S_{s,t}f$ is indeed holomorphic.

Since $P_{t}$ is just the fundamental solution at the identity of the heat
equation $du/dt=\frac{1}{2}\Delta u,$ $S_{s,t}f$ may be expressed as 
\begin{equation}
S_{s,t}f=\text{ analytic continuation of }e^{t\Delta /2}f.  \label{e.3.2}
\end{equation}
Here $e^{t\Delta /2}$ is to be interpreted as the usual contraction
semigroup on $L^{2}\left( \mathbb{R}^{d},dx\right) ,$ extended by continuity
to $L^{2}\left( \mathbb{R}^{d},P_{s}\right) .$

\begin{definition}
\label{d.3.1}For $s>t/2,$ let $A_{s,t}$ be the constant-coefficient elliptic
differential operator on $\mathbb{C}^{d}$ given by 
\begin{equation*}
A_{s,t}=\left( s-\frac{t}{2}\right) \sum_{k=1}^{d}\frac{\partial ^{2}}{%
\partial x_{k}^{2}}+\frac{t}{2}\sum_{k=1}^{d}\frac{\partial ^{2}}{\partial
y_{k}^{2}}.
\end{equation*}
Let $M_{s,t}$ denote the Gaussian measure given by $e^{A_{s,t}/2}\left(
\delta _{0}\right) .$ Explicitly $dM_{s,t}=M_{s,t}(z)\,dz,$ where $dz$ is
the standard Lebesgue measure on $\mathbb{C}^{d}$ and 
\begin{equation*}
M_{s,t}\left( z\right) =\left( \pi r\right) ^{-d/2}\left( \pi t\right)
^{-d/2}e^{-x^{2}/r}e^{-y^{2}/t}
\end{equation*}
Here $r=2\left( s-t/2\right) .$
\end{definition}

The Gaussian measures $P_{s}$ and $M_{s,t}$ may also be described by their
Fourier transforms: 
\begin{eqnarray}
\int_{\mathbb{R}^{d}}\exp \left( i\lambda \cdot x\right) \,dP_{s}\left(
x\right) &=&\exp \left( -\frac{s}{2}\lambda ^{2}\right)  \notag \\
\int_{\mathbb{C}^{d}}\exp (i\lambda \cdot x+i\alpha \cdot y)\,dM_{s,t}(z)
&=&\exp (-\frac{1}{4}(r\lambda ^{2}+t\alpha ^{2}))  \label{e.3.3}
\end{eqnarray}
for all $\lambda $ and $\alpha $ in $\mathbb{R}^{d}.$

Let $\mathcal{H}L^{2}\left( \mathbb{C}^{d},M_{s,t}\right) $ denote the
Hilbert space of holomorphic functions on $\mathbb{C}^{d}$ which are
square-integrable with respect to $M_{s,t}.$

\begin{theorem}[Extended Segal--Bargmann Transform]
\label{t.3.2}For all $s$ and $t$ with $s>t/2>0,$ the map $S_{s,t}$ defined
in (\ref{e.3.1}) is an isometric isomorphism of $L^{2}\left( \mathbb{R}%
^{d},P_{s}\right) $ onto $\mathcal{H}L^{2}\left( \mathbb{C}%
^{d},M_{s,t}\right) .$ The standard case is $s=t.$
\end{theorem}

We will prove this in Section \ref{s.3.2}, using Hermite expansions. The
surjectivity of $S_{s,t}$ is proved by showing that the holomorphic
polynomials are dense in $\mathcal{H}L^{2}\left( \mathbb{C}%
^{d},M_{s,t}\right) ,$ which is item 4 of Theorem \ref{t.3.6} below.

\subsection{Action on the Hermite Expansion\label{s.3.2}}

The classical Segal-Bargmann transform for $\mathbb{R}^{d}$ takes the
Hermite expansion of a function on $\mathbb{R}^{d}$ to the Taylor expansion
of the corresponding holomorphic function on $\mathbb{R}^{d},$ and this
property determines the transform \cite{B}. Our variant of the
Segal-Bargmann transform also has a simple action on the Hermite expansion,
which reduces to the above result when $s=t.$ Since we require Hermite
expansions on both $\mathbb{R}^{d}$ and $\mathbb{C}^{d},$ we will prove
abstract results which cover both cases simultaneously.

Let $V$ \ be a real finite-dimensional vector space and $L$ be a constant
coefficient pure second order elliptic operator on $V,$ i.e., $%
L=\sum_{i,j=1}^{N}g_{ij}\partial ^{2}/\partial x_{i}\partial x_{j}$ where $%
N=\dim \left( V\right) ,$ $\{x_{i}\}_{i=1}^{N}$ are linear coordinates on $%
V, $ and $\{g_{ij}\}$ is a positive-definite symmetric matrix. For $v,w\in
V, $ let $g\left( v,w\right) =\sum_{i,j=1}^{N}g^{ij}x_{i}\left( v\right)
x_{j}\left( w\right) ,$ where $\{g^{ij}\}$ denotes the matrix inverse of $%
\{g_{ij}\}.$ Then $g$ is an inner product on $V$ which is naturally induced
by $L.$ Indeed, if $g^{\ast }$ denotes the dual inner product of $V^{\ast }$
and $\alpha ,\beta \in V^{\ast },$ then $g^{\ast }\left( \alpha ,\beta
\right) =\frac{1}{2}L\left( \alpha \beta \right) .$ If $\{e_{i}\}_{i=1}^{N}$
is an orthonormal basis for $\left( V,g\right) ,$ then $L=\sum_{i=1}^{N}%
\partial _{i}^{2}$ where $\partial _{i}=\partial _{e_{i}}.$ This follows
from the observation that 
\begin{equation*}
\frac{1}{2}\sum_{i=1}^{N}\partial _{i}^{2}(\alpha \beta
)=\sum_{i=1}^{N}\alpha (e_{i})\beta (e_{i})=g^{\ast }(\alpha ,\beta )=\frac{1%
}{2}L(\alpha \beta ).
\end{equation*}

\begin{definition}[Heat Kernel Measure]
\label{d.3.3}For a pair $V$ and $L$ be as above, we associate the Gaussian
measures 
\begin{equation*}
dQ_{t}\left( v\right) =\left( \frac{1}{2\pi t}\right) ^{N/2}\exp \left( -%
\frac{1}{2t}g\left( v,v\right) \right) \,dv\quad \forall t>0,
\end{equation*}
where $dv$ denotes Lebesgue measure on $V$ normalized so that unit cube in $V
$ relative to $g$ has unit volume. We will abbreviate $Q_{1}$ by $Q.$ For
any measurable function $f$ on $V$ and $v\in V,$ let 
\begin{equation}
e^{tL/2}f(v)=\int_{V}f\left( v-w\right) \,dQ_{t}\left( w\right) 
\label{e.3.4}
\end{equation}
whenever the integral exists.
\end{definition}

The measures $Q_{t}$ may also be described by their Fourier transforms,
namely $Q_{t}$ is the unique measure on $V$ such that 
\begin{equation*}
\int_{V}e^{i\lambda \left( w\right) }\,dQ_{t}\left( w\right) =\exp \left( -%
\frac{tg^{\ast }(\lambda ,\lambda )}{2}\right) =\exp \left( -t\frac{L\left(
\lambda ^{2}\right) }{4}\right) .
\end{equation*}
for all $\lambda \in V^{\ast }.$

Given a reasonable function $f$ on $V$ (say continuous and exponentially
bounded), it is well known and easily checked that $u(t,v):=e^{tL/2}f(v)$ is
a solution to the heat equation $\partial u(t,v)/\partial t=\frac{1}{2}%
Lu(t,v)$ such that $\lim_{t\searrow 0}u(t,v)=f(v).$ It is also easily
checked that if $f$ is a polynomial function of $v,$ then $e^{tL/2}f$ may be
computed by the finite Taylor series expansion: 
\begin{equation}
e^{tL/2}f=\sum_{k=0}^{\infty }\left( \frac{tL}{2}\right) ^{k}f.
\label{e.3.5}
\end{equation}
The above sum in finite since $L^{k}f=0$ whenever $2k$ is greater than the
degree of $f.$ On polynomials, (\ref{e.3.5}) defines $e^{tL/2}f$ for all $%
t\in \mathbb{R}$ in such a way that $e^{-tL/2}$ is the inverse of $e^{tL/2}.$

\begin{definition}
\label{d.3.4}The\textbf{\ n}$^{th}$\textbf{\ level Hermite subspace of}\emph{%
\ }$L^{2}\left( V,Q\right) $ is the space $\mathcal{F}_{n}\left( L\right)
=e^{-L/2}\mathcal{P}_{n}(V),$ where $\mathcal{P}_{n}(V)$ denotes the space
of homogeneous polynomials of degree $n$ on $V.$
\end{definition}

The following result is well-known. We include a proof for completeness and
so that some calculations will be available for later use.

\begin{proposition}
\label{p.3.5}Let $V$ and $L$ be as above. Then

\begin{enumerate}
\item  $L^{2}\left( V,Q\right) $ is the orthogonal Hilbert space direct sum
of the subspaces $\mathcal{F}_{n}\left( L\right) $ for $n=0,1,2\ldots .$

\item  $\mathcal{F}_{n}\left( L\right) $ is the set of all polynomials on $V$
of degree $n$ which are orthogonal to all polynomials of degree at most $n-1.
$

\item  For every $f\in L^{2}\left( V,Q\right) ,$ $e^{L/2}f$ is a well
defined, real-analytic function on $V.$ Moreover, if the ``Hermite''
expansion of $f\in L^{2}\left( V,Q\right) $ is $f=\sum_{n=0}^{\infty }f_{n}$
with $f_{n}\in \mathcal{F}_{n}\left( L\right) ,$ then $f_{n}=e^{-L/2}p_{n},$
where $p_{n}(v)=\frac{1}{n!}(\partial _{v}^{n}e^{L/2}f)(0)$ and ($\partial
_{v}^{n}f)(0)=\frac{d^{n}}{dt^{n}}f(tv)|_{t=0}$. We will write this
succinctly as 
\begin{equation}
f(v)=\sum_{n=0}^{\infty }\frac{1}{n!}e^{-L/2}(\partial _{v}^{n}e^{L/2}f)(0).
\label{e.3.6}
\end{equation}
\end{enumerate}
\end{proposition}

\textit{Proof. }Let $\{e_{i}\}_{i=1}^{N}$ be an orthonormal basis for $(V,g)$
so that $L=\sum_{i=1}^{N}\partial _{i}^{2},$ where $\partial _{i}=\partial
_{e_{i}}.$ For functions $p,q$ on $V$ let $(p,q)=\int_{V}\bar{p}%
(v)q(v)\,dQ(v)$ be the $L^{2}$ inner product. Taking $p$ and $q$ to be
polynomials on $V\ $and $v=0$ in (\ref{e.3.4}), we find, using the fact that 
$Q$ is even, that 
\begin{equation*}
(e^{-L/2}p,e^{-L/2}q)=e^{L/2}(\overline{e^{-L/2}p}\,\cdot e^{-L/2}q)|_{0}.
\end{equation*}
Since $e^{tL/2}(\overline{e^{-tL/2}p}\,e^{-tL/2}q)$ is a polynomial in $%
(t,v),$ it follows by Taylor's theorem that 
\begin{equation*}
e^{L/2}\left( \overline{e^{-L/2}p}\,\cdot e^{-L/2}q\right)
=\sum_{n=0}^{\infty }\frac{1}{n!}\left. \frac{d^{n}}{dt^{n}}\right|
_{t=0}e^{tL/2}(e^{-tL/2}\bar{p}\,\cdot e^{-tL/2}q).
\end{equation*}
Using the product rule repeatedly shows that 
\begin{align*}
& \frac{d}{dt}e^{tL/2}(e^{-tL/2}\bar{p}\cdot e^{-tL/2}q) \\
& =e^{tL/2}\left( \frac{L}{2}(e^{-tL/2}\bar{p}\cdot e^{-tL/2}q)-(\frac{L}{2}%
e^{-tL/2}\bar{p}\cdot e^{-tL/2}q)-(e^{-tL/2}\bar{p}\cdot \frac{L}{2}%
e^{-tL/2}q)\right) \\
& =e^{tL/2}\left( \sum_{i=1}^{N}\partial _{i}e^{-tL/2}\bar{p}\cdot \partial
_{i}e^{-tL/2}q)\right) .
\end{align*}
This equation may now be used inductively to show 
\begin{equation*}
\left. \frac{d^{n}}{dt^{n}}\right| _{t=0}e^{tL/2}(e^{-tL/2}\bar{p}\cdot
e^{-tL/2}q).=\sum_{i_{1},i_{2},\ldots i_{n}=1}^{N}\partial _{i_{1}}\partial
_{i_{2}}\cdots \partial _{i_{n}}\bar{p}\cdot \partial _{i_{1}}\partial
_{i_{2}}\cdots \partial _{i_{n}}q.
\end{equation*}
Combining the previous four displayed equations shows that 
\begin{equation}
(e^{-L/2}p,e^{-L/2}q)=\sum_{n=0}^{\infty }\frac{1}{n!}\sum_{i_{1},i_{2},%
\ldots i_{n}=1}^{N}\left( \partial _{i_{1}}\partial _{i_{2}}\cdots \partial
_{i_{n}}\bar{p}(v)\cdot \partial _{i_{1}}\partial _{i_{2}}\cdots \partial
_{i_{n}}q(v)\right) |_{v=0},  \label{e.3.7}
\end{equation}
for all polynomials $p$ and $q$ on $V.$

Using (\ref{e.3.7}) we may prove items 1 and 2 as follows. Notice that $%
\mathcal{F}_{n}(L)$ consists of polynomials of degree $n$ and that $\oplus
_{k=0}^{n-1}\mathcal{F}_{k}(L)$ consists of \textit{all} polynomials on $V$
of degree $n-1$ or less. By (\ref{e.3.7}), it is easily seen that if $p\in 
\mathcal{P}_{n}(V)$ and $q\in \mathcal{P}_{m}(V)$ with $m\neq n,$ then $%
(e^{-L/2}p,e^{-L/2}q)=0.$ Hence $\mathcal{F}_{n}(L)$ is orthogonal to $%
\oplus _{k=0}^{n-1}\mathcal{F}_{k}(L)$. Since polynomials are dense in $%
L^{2}(V,Q),$ these observations immediately imply the first two items of the
theorem.

For item 3, suppose for the moment that $f$ is a polynomial on $V.$ By
Taylor's theorem applied to $e^{L/2}f,$ 
\begin{equation*}
e^{L/2}f=\sum_{n=0}^{\infty }\frac{1}{n!}(\partial _{v}^{n}e^{L/2}f)(0).
\end{equation*}
Applying $e^{-L/2}$ to both sides of this equation then proves (\ref{e.3.6})
when $f$ is a polynomial.

For general $f\in L^{2}\left( V,Q\right) ,$ we must first show that $%
e^{L/2}f $ is defined and smooth. Letting $q(v)=dQ(v)/dv,$ we may write 
\begin{equation}
e^{L/2}f(v)=\int_{V}f(v-w)q(w)\,dw=\int_{V}f(w)\frac{q(v-w)}{q(w)}q(w)\,dw.
\label{e.3.8}
\end{equation}
Since $q(v-w)/q(w)=\exp (-\frac{1}{2}g(v,v)+g(v,w))\in L^{2}\left(
V,Q(dw)\right) ,$ it follows that $f(v-w)q(w)$ is integrable and hence $%
e^{L/2}f$ is defined. More generally one may show $\sup_{v\in K}\left|
\partial ^{\alpha }q(v-w)/q\left( w\right) \right| \in L^{2}\left(
V,Q(dw)\right) $ for all compact sets $K\subset V.$ Hence $e^{L/2}f(v)$ is
smooth and 
\begin{equation}
\partial ^{\alpha }e^{L/2}f(v)=\int_{V}f(w)\frac{\partial ^{\alpha }q(v-w)}{%
q(w)}q(w)dw.  \label{e.3.9}
\end{equation}
So for each integer $n\geq 0,$ let $(P_{n}f)(v)=\frac{1}{n!}%
e^{-L/2}(\partial _{v}^{n}e^{L/2}f)(0)\in \mathcal{F}_{n}(L).$ Because of (%
\ref{e.3.9}), $P_{n}:L^{2}\left( V,Q\right) \rightarrow \mathcal{F}_{n}(L)$
is a well defined continuous linear map. Moreover, $P_{n}f$ is the same as
orthogonal projection onto $\mathcal{F}_{n}(L)$ when $f$ is a polynomial.
Hence it follows by density of polynomials in $L^{2}\left( V,Q\right) $ that 
$P_{n}$ is orthogonal projection onto $\mathcal{F}_{n}(L).$ This proves (\ref
{e.3.6}) for general $f\in L^{2}\left( V,Q\right) .$ \qed

We will need the following holomorphic version of Proposition \ref{p.3.5}.

\begin{theorem}
\label{t.3.6}Suppose that $V$ is a real vector space and $L$ be a pure
second order constant coefficient differential operator on $C^{\infty }(V).$
Also assume that $V$ is equipped with a complex structure $J,$ i.e., $%
J:V\rightarrow V$ is a linear map such that $J^{2}=-I.$ Using $J,$ $V$ may
considered to be a complex vectors space by defining $iv=Jv$ for $v\in V.$
Let $\mathcal{H}(V)\mathcal{\ }$denote the space of holomorphic functions on 
$V$ and let $\mathcal{H}L^{2}\left( V,Q\right) =\mathcal{H}(V)\cap
L^{2}\left( V,Q\right) $ be the space of $L^{2}$ holomorphic functions. Let $%
\mathcal{HF}_{n}(L)=\mathcal{H}\left( V\right) \cap \mathcal{F}_{n}(L).$ Then

\begin{enumerate}
\item  $\mathcal{H}L^{2}\left( V,Q\right) $ is the orthogonal Hilbert space
direct sum of the subspaces $\mathcal{HF}_{n}\left( L\right) $ for $%
n=0,1,2\ldots .$

\item  $\mathcal{HF}_{n}\left( L\right) $ is the set of all holomorphic
polynomials on $V$ of degree $n$ or less which are orthogonal to all
holomorphic polynomials on $V$ of degree $n-1$ or less.

\item  Let $f=\sum_{n=0}^{\infty }f_{n}$ be the Hermite expansion of $f\in 
\mathcal{H}L^{2}\left( V,Q\right) .$ Then $f_{n}\in \mathcal{HF}_{n}(L)$ for 
$n=0,1,2,\ldots .$

\item  The holomorphic polynomials on $V$ are dense in $\mathcal{H}%
L^{2}\left( V,Q\right) .$
\end{enumerate}
\end{theorem}

\textit{Proof}. Since $\mathcal{F}_{m}\left( L\right) $ and $\mathcal{F}%
_{n}\left( L\right) $ are orthogonal for $m\neq n,$ $\mathcal{HF}_{n}\left(
L\right) $ and $\mathcal{HF}_{n}\left( L\right) $ are also clearly
orthogonal for $m\neq n.$

Now for $f\in L^{2}\left( V,Q\right) $ we have already seen that $e^{L/2}f$
is a smooth function. If $f$ is also holomorphic, then $e^{L/2}f$ is
holomorphic. To see this it suffices to show, for each $u,v\in V,$ that $%
e^{L/2}f\left( u+zv\right) $ is holomorphic as a function of $z\in \mathbb{C}%
.$ This is easily done using Morera's Theorem and the fact that $f$ is
holomorphic. We omit the details.

Hence if $f\in \mathcal{H}L^{2}\left( V,Q\right) ,$ then $p_{n}(v):=\frac{1}{%
n!}(\partial _{v}^{n}e^{L/2}f)(0)$ is a holomorphic polynomial that is
homogeneous of degree $n.$ Since $L$ preserves the space of holomorphic
functions, it follows that $f_{n}=e^{-L/2}p_{n}$ is both holomorphic and in $%
\mathcal{F}_{n}(L),$ i.e., $f_{n}\in \mathcal{HF}_{n}(L).$ Hence we have
proved items 1, 3, and 4 of the Theorem. Finally, for item 2, if $p$ is a
holomorphic polynomial of degree less than or equal to $n$, then 
\begin{equation*}
p(v)=\sum_{k=0}^{n}\frac{1}{k!}e^{-L/2}(\partial _{v}^{n}e^{L/2}p)(0).
\end{equation*}
Since $\oplus _{k=0}^{n-1}\mathcal{HF}_{k}(L)$ is the collection $\mathcal{H}%
_{n-1}$ of holomorphic polynomials of degree less than or equal to $n-1,$ it
follows $p$ is orthogonal to $\mathcal{H}_{n-1}$ if and only if $p(v)=\frac{1%
}{n!}e^{-L/2}(\partial _{v}^{n}e^{L/2}p)(0)$ which is equivalent to $p$
being in $\mathcal{HF}_{n}(L).$ \qed

We now apply our results in two cases: $V=\mathbb{R}^{d}$ and $L=s\Delta ,$
and $V=\mathbb{C}^{d}$ and $L=A_{s,t}.$

\begin{definition}
\label{d.3.7}Let $\mathcal{F}_{n,s}\left( \mathbb{R}^{d}\right) =\mathcal{F}%
_{n}(s\Delta )\subset $\emph{\ }$L^{2}\left( \mathbb{R}^{d},P_{s}\right) $
and $\mathcal{F}_{n,s,t}\left( \mathbb{C}^{d}\right) =\mathcal{F}_{n}\left(
A_{s,t}\right) \subset L^{2}\left( \mathbb{C}^{d},M_{s,t}\right) .$ Let $%
\mathcal{HF}_{n,s,t}\left( \mathbb{C}^{d}\right) $ denote the holomorphic
polynomials in $\mathcal{F}_{n,s,t}\left( \mathbb{C}^{d}\right) .$
\end{definition}

\begin{theorem}
\label{t.3.8}The transform $S_{s,t}$ in (\ref{e.3.1}) takes $\mathcal{F}%
_{n,s}\left( \mathbb{R}^{d}\right) $ onto $\mathcal{HF}_{n,s,t}\left( 
\mathbb{C}^{d}\right) .$ Specifically, let $p$ be a homogeneous polynomial
of degree $n$ on $\mathbb{R}^{d}$ and let $p_{\mathbb{C}}$ be its analytic
continuation to $\mathbb{C}^{d}.$ Then 
\begin{equation}
S_{s,t}\left( e^{-s\Delta /2}p\right) =e^{-A_{s,t}/2}\left( p_{\mathbb{C}%
}\right) .  \label{e.3.10}
\end{equation}
\end{theorem}

Note that if $s=t,$ then the operator $A_{s,t}$ is zero on all holomorphic
functions. So when $s=t$ the holomorphic subspace of $\mathcal{F}%
_{n,s,t}\left( \mathbb{C}^{d}\right) $ is precisely the space of holomorphic
polynomials which are homogeneous of degree $n.$ In that case, the transform 
$S_{t,t}$ takes the Hermite expansion of $f\in L^{2}\left( \mathbb{R}%
^{d},P_{t}\right) $ to the Taylor expansion of $S_{t,t}f.$ If $s\neq t,$
then by Theorem \ref{t.3.6} and Theorem \ref{t.3.8}, $S_{s,t}$ takes the
Hermite expansion of $f\in L^{2}\left( \mathbb{R}^{d},P_{s}\right) $ to an $%
L^{2}$-convergent expansion of $S_{s,t}f$ in terms of non-homogeneous
holomorphic polynomials. For $s\neq t$ it is not clear (to us) whether the
Taylor series of a function in $\mathcal{H}L^{2}\left( \mathbb{C}%
^{d},M_{s,t}\right) $ is always $L^{2}$-convergent.

\textit{Proof. }By the definition of $S_{s,t},$ 
\begin{equation*}
S_{s,t}\left( e^{-s\Delta /2}p\right) =(e^{t\Delta /2}e^{-s\Delta /2}p)_{%
\mathbb{C}}=(e^{(t-s)\Delta /2}p)_{\mathbb{C}}.
\end{equation*}
On the other hand since $p_{\mathbb{C}}$ is holomorphic, $\partial p_{%
\mathbb{C}}/\partial y_{k}=i\partial p_{\mathbb{C}}/\partial x_{k}$ and
hence 
\begin{equation*}
A_{s,t}p_{\mathbb{C}}=\left( s-\frac{t}{2}-\frac{t}{2}\right) \sum_{k=1}^{d}%
\frac{\partial ^{2}}{\partial x_{k}^{2}}p_{\mathbb{C}}=\left( s-t\right)
\sum_{k=1}^{d}\frac{\partial ^{2}}{\partial x_{k}^{2}}p_{\mathbb{C}%
}=-(t-s)(\Delta p)_{\mathbb{C}}.
\end{equation*}
Therefore, $e^{-A_{s,t}/2}\left( p_{\mathbb{C}}\right) =(e^{(t-s)\Delta
/2}p)_{\mathbb{C}}=S_{s,t}\left( e^{-s\Delta /2}p\right) .$\qed

\textit{Proof of Theorem \ref{t.3.2}}. If $p$ and $q$ are polynomials on $%
\mathbb{R}^{d},$ then 
\begin{eqnarray*}
A_{s,t}(\overline{p_{\mathbb{C}}}q_{\mathbb{C}})-(A_{s,t}\overline{p_{%
\mathbb{C}}})q_{\mathbb{C}}-\overline{p_{\mathbb{C}}}(A_{s,t}q_{\mathbb{C}})
&=&2\left( s-\frac{t}{2}\right) \sum_{k=1}^{d}\overline{\frac{\partial p_{%
\mathbb{C}}}{\partial x_{k}}}\frac{\partial q_{\mathbb{C}}}{\partial x_{k}}+%
\frac{t}{2}\sum_{k=1}^{d}\overline{\frac{\partial p_{\mathbb{C}}}{\partial
y_{k}}}\frac{\partial q_{\mathbb{C}}}{\partial y_{k}} \\
&=&2\left( s-\frac{t}{2}\right) \sum_{k=1}^{d}\overline{\frac{\partial p_{%
\mathbb{C}}}{\partial x_{k}}}\frac{\partial q_{\mathbb{C}}}{\partial x_{k}}+%
\frac{t}{2}\sum_{k=1}^{d}\left( -i\overline{\frac{\partial p_{\mathbb{C}}}{%
\partial x_{k}}}\right) \left( i\frac{\partial q_{\mathbb{C}}}{\partial x_{k}%
}\right) \\
&=&2s\sum_{k=1}^{d}\overline{\frac{\partial p_{\mathbb{C}}}{\partial x_{k}}}%
\frac{\partial q_{\mathbb{C}}}{\partial x_{k}}.
\end{eqnarray*}
This formula and computations similar to those used to prove (\ref{e.3.7})
show that 
\begin{align}
& (e^{-A_{s,t}/2}p_{\mathbb{C}},e^{-A_{s,t}/2}q_{\mathbb{C}%
})_{L^{2}(M_{s,t})}  \notag \\
& =e^{A_{s,t}/2}(\overline{e^{-A_{s,t}/2}p_{\mathbb{C}}}\cdot
e^{-A_{s,t}/2}q_{\mathbb{C}})|_{z=0}  \notag \\
& =\sum_{n=0}^{\infty }\frac{s^{n}}{n!}\sum_{k_{1},k_{2},\ldots
k_{,n}=1}^{d}\left. \overline{\partial _{x_{k_{1}}}\partial
_{x_{k_{2}}}\cdots \partial _{x_{k_{n}}}p_{\mathbb{C}}}\cdot \partial
_{x_{k_{1}}}\partial _{x_{k_{2}}}\cdots \partial _{x_{k_{n}}}q_{\mathbb{C}%
}\right| _{z=0} \\
& =\sum_{n=0}^{\infty }\frac{s^{n}}{n!}\sum_{k_{1},k_{2},\ldots
k_{,n}=1}^{d}\left. \overline{\partial _{x_{k_{1}}}\partial
_{x_{k_{2}}}\cdots \partial _{x_{k_{n}}}p}\cdot \partial
_{x_{k_{1}}}\partial _{x_{k_{2}}}\cdots \partial _{x_{k_{n}}}q\right| _{x=0}
\label{e.3.11} \\
& =e^{s\Delta /2}(\overline{e^{-s\Delta /2}p}\cdot e^{-s\Delta /2}q)\left(
0\right) =(e^{-s\Delta /2}p,e^{-s\Delta /2}q)_{L^{2}(P_{s})}.  \notag
\end{align}
Here $\partial _{x_{i}}=\partial /\partial x_{i}.$ In light of (\ref{e.3.10}%
), which holds by linearity for all polynomials, this shows that $S_{s,t}$
is isometric on polynomials.

Since polynomials are dense in $L^{2}\left( \mathbb{R}^{d},P_{s}\right) $
and the linear functionals $f\in L^{2}\left( \mathbb{R}^{d},P_{s}\right)
\rightarrow \left( S_{s,t}f\right) (z)\in \mathbb{C}$ are continuous for
each $z\in \mathbb{C}^{d},$ it follows that $S_{s,t}$ is isometric on all of 
$L^{2}\left( \mathbb{R}^{d},P_{s}\right) .$ Note that $e^{-s\Delta /2}$ is
isometric and hence injective and hence invertible on the space of
polynomials of degree at most $n.$ This plus the fact that every holomorphic
polynomial on $\mathbb{C}^{d}$ is the analytic continuation of its
restriction to $\mathbb{R}^{d}$ shows that every holomorphic polynomial is
in the image of $S_{s,t}.$ But Item 4 of Theorem \ref{t.3.6} asserts that
the holomorphic polynomials are dense in $\mathcal{H}L^{2}\left( \mathbb{C}%
^{d},M_{s,t}\right) \ $and therefore $S_{s,t}$ is surjective. \qed

\section{Segal-Bargmann for the Wiener space\label{s.4}}

Because it is formulated in terms of Gaussian measures, our variant of the
Segal-Bargmann transform admits an infinite-dimensional ($d\rightarrow
\infty $) limit. While this could be formulated in terms of an arbitrary
abstract Wiener space, we will for concreteness consider only the classical
Wiener space case relevant to this paper. Thus $\mathbb{R}^{d}$ will be
replaced by an infinite-dimensional space of Lie algebra-valued generalized
functions on $\left[ 0,1\right] $, with a white noise measure. By
integrating once, this space may be identified with the space of continuous
Lie algebra-valued functions on $\left[ 0,1\right] $ with a Wiener measure.
Similarly, $\mathbb{R}^{d}$ will be replaced by a space of generalized
functions with values in the complex Lie algebra (with a white noise
measure), which may be identified with the space of continuous functions
with values in the complex Lie algebra (with a Wiener measure).

\subsection{The Transform for the Wiener Space\label{s.4.1}}

Let $K$ be a compact connected Lie group. Fix once and for all an Ad-$K$%
-invariant inner product $\left\langle \cdot ,\cdot \right\rangle $ on the
Lie algebra $\frak{k}$ of $K.$ Let $K_{\mathbb{C}}$ be the complexification
of $K$ in the sense of \cite{Ho,H1}, and let $\frak{k}_{\mathbb{C}}=\frak{k}%
+i\frak{k}$ be the Lie algebra of $K_{\mathbb{C}}.$

We then consider the space of connections on the spatial circle. These are
Lie algebra-valued 1-forms, which can be identified with Lie algebra-valued
functions on the interval $\left[ 0,1\right] ,$ where the circle is this
interval with ends identified. Specifically, let 
\begin{equation}
\mathcal{A}=L^{2}\left( \left[ 0,1\right] ;\frak{k}\right) ,  \label{e.4.1}
\end{equation}
where the norm is computed using Lebesgue measure on $\left[ 0,1\right] $
and the inner product on $\frak{k}.$ We need also a larger space $\mathcal{%
\bar{A}},$ which may be taken to be 
\begin{equation}
\mathcal{\bar{A}}=\{A=\frac{da_{\tau }}{d\tau }\left| a\in W(\frak{k}%
)\right. \},  \label{e.4.2}
\end{equation}
a subspace of $\frak{k}$-valued distributions. Here $W(\frak{k)}$ denotes
the set of continuous paths $a$ from $[0,1]$ to $\frak{k}$ such that $%
a_{0}=0,$ and $\frac{da_{\tau }}{d\tau }$ denotes the distributional
derivative of $a.$ For each $A\in \mathcal{\bar{A}},$ the function $a\in
W\left( \frak{k}\right) $ is unique, and so may be thought of as a function
of $A.$ We will write, suggestively, 
\begin{equation}
a_{\tau }\left( A\right) =\int_{0}^{\tau }A_{\sigma }\,d\sigma .
\label{e.4.3}
\end{equation}
Note that we are reversing convention by using the lowercase letter $a$ for
the anti-derivative of $A.$ We can make $\mathcal{\bar{A}}$ into a Banach
space whose norm is the supremum norm on $a.$

We similarly define 
\begin{equation*}
\mathcal{A}_{\mathbb{C}}=L^{2}\left( \left[ 0,1\right] ;\frak{k}_{\mathbb{C}%
}\right)
\end{equation*}
using Lebesgue measure on $\left[ 0,1\right] $ and the sesquilinear
extension of the inner product from $\frak{k}$ to $\frak{k}_{\mathbb{C}}$;
and 
\begin{equation*}
\mathcal{\bar{A}}_{\mathbb{C}}=\{C=\frac{dc_{\tau }}{d\tau }\left| c\in W(%
\frak{k}_{\mathbb{C}})\right. \},
\end{equation*}
where $W\left( \frak{k}_{\mathbb{C}}\right) $ is defined analogously to $%
W\left( \frak{k}\right) .$ As in the real case, $c$ is unique and we will
write 
\begin{equation}
c_{\tau }\left( C\right) =\int_{0}^{\tau }C_{\sigma }\,d\sigma .
\label{e.4.4}
\end{equation}

\begin{definition}
\label{d.4.1}Let $\tilde{P}_{s}$ denote the unique Gaussian measure on $%
\mathcal{\bar{A}}$ such that for all continuous linear functionals $\phi $
on $\mathcal{\bar{A}}$%
\begin{equation*}
\int_{\mathcal{\bar{A}}}e^{i\phi \left( A\right) }\,d\tilde{P}_{s}\left(
A\right) =\exp \left( -\frac{s}{2}\left\| \phi \right\| ^{2}\right) ,
\end{equation*}
where $\left\| \phi \right\| $ denotes the norm of $\phi $ as a linear
functional on $\mathcal{A}$.

Let $\tilde{M}_{s,t}$ denote the unique Gaussian measure on $\mathcal{\bar{A}%
}_{\mathbb{C}}$ such that for all continuous linear functionals $\phi $ and $%
\psi $ on $\mathcal{\bar{A}}$%
\begin{equation*}
\int_{\mathcal{\bar{A}}_{\mathbb{C}}}e^{i\phi \left( A\right) +i\psi \left(
B\right) }\,d\tilde{M}_{s,t}\left( A+iB\right) =\exp \left( -\frac{1}{4}%
(r\left\| \phi \right\| ^{2}+t\left\| \psi \right\| ^{2})\right) ,
\end{equation*}
where $\left\| \phi \right\| $ and $\left\| \psi \right\| $ denote norms as
linear functionals on $\mathcal{A}$ and $r=2\left( s-t/2\right) .$
\end{definition}

These measures have the formal expressions: 
\begin{eqnarray*}
d\tilde{P}_{s}\left( A\right) &=&\frac{1}{Z_{1}}\,e^{-\left\| A\right\|
^{2}/2s}\,\mathcal{D}A\quad \text{and} \\
d\tilde{M}_{s,t}\left( A+iB\right) &=&\frac{1}{Z_{2}}e^{-\left\| A\right\|
^{2}/r-\left\| B\right\| ^{2}/t}\,\mathcal{D}A\,\mathcal{D}B.
\end{eqnarray*}
Here $\left\| \cdot \right\| $ is the $L^{2}$ -- norm for $\mathcal{A}$, $%
\mathcal{D}A$ and $\mathcal{D}B$ refer to the (non-existent) Lebesgue
measure on $\mathcal{A},$ and $Z_{1},Z_{2}$ are ``normalization constants.''
Note that the measure $\tilde{P}_{s}$ may be thought of as the heat kernel
measure at the origin, that is, the fundamental solution at the origin of
the equation $du/dt=\frac{1}{2}\Delta _{\mathcal{A}}u,$ where $\Delta _{%
\mathcal{A}}$ is the sum of squares of derivatives in the directions of an
orthonormal basis for $\mathcal{A}.$

The space $\mathcal{A}$ is the Cameron-Martin subspace for the Gaussian
measure space $(\mathcal{\bar{A}},\tilde{P}_{s})$ and is a set of $\tilde{P}%
_{s}$-measure zero. Similarly, $\mathcal{A}_{\mathbb{C}}$ is the
Cameron-Martin subspace for $(\mathcal{\bar{A}}_{\mathbb{C}},\tilde{M}%
_{s,t}) $ and is a set of $\tilde{M}_{s,t}$-measure zero.

The measure $\tilde{P}_{s}$ is the law of a scaled $\frak{k}$-valued white
noise on $\left[ 0,1\right] .$ This is equivalent to saying that if $A$ is
distributed as $\tilde{P}_{s}$ then $a_{\tau }\left( A\right) $ (defined in (%
\ref{e.4.3})) is a scaled $\frak{k}$-valued Brownian motion. Specifically,
if $\left\{ X^{k}\right\} $ is an orthonormal basis for $\frak{k},$ then $%
a_{\tau }^{k}:=\left\langle X^{k},a_{\tau }\right\rangle $ are real-valued
Brownian motions satisfying 
\begin{equation}
E\left\{ a_{\sigma }^{k}a_{\tau }^{l}\right\} =s\min \left\{ \sigma ,\tau
\right\} \delta _{kl}.  \label{e.4.5}
\end{equation}
Similarly, the measure $\tilde{M}_{s,t}$ is the law of a scaled $\frak{k}_{%
\mathbb{C}}$-valued white noise on $\left[ 0,1\right] .$ Let $c$ be as in (%
\ref{e.4.4}), and decompose $c$ as $c_{\tau }=\text{Re}c_{\tau }+i\text{Im}%
c_{\tau },$ with $\text{Re}c_{\tau }$ and $\text{Im}c_{\tau }$ taking values
in $\frak{k}.$ Then $\left\langle X^{k},\text{Re}c_{\tau }\right\rangle $
and $\left\langle X^{k},\text{Im}c_{\tau }\right\rangle $ are independent
real-valued Brownian motions satisfying: 
\begin{eqnarray}
E\left\{ \langle X^{k},\text{Re}c_{\sigma }\rangle \left\langle X^{l},\text{%
Re}c_{\tau }\right\rangle \right\} &=&\left( s-\frac{t}{2}\right) \min
\left\{ \sigma ,\tau \right\} \delta _{kl}  \notag \\
E\left\{ \left\langle X^{k},\text{Im}c_{\sigma }\right\rangle \left\langle
X^{l},\text{Im}c_{\tau }\right\rangle \right\} &=&\frac{t}{2}\min \left\{
\sigma ,\tau \right\} \delta _{kl}.  \label{e.4.6}
\end{eqnarray}

We now define the two-parameter version of the Segal-Bargmann transform for $%
L^{2}(\mathcal{\bar{A}},\tilde{P}_{s}).$ The reader should keep in mind that
we are ``trying'' to work on the space $\mathcal{A}$, with the larger space $%
\mathcal{\bar{A}}$ introduced as a technical necessity.

\begin{definition}
\label{d.4.2}Let $\left\{ e_{1},\cdots ,e_{d}\right\} $ be a finite
orthonormal set in the real Hilbert space $\mathcal{A}$ with the property
that each linear functional $\left\langle e_{j},\cdot \right\rangle $
extends continuously to $\mathcal{\bar{A}}.$ Each linear functional $%
\left\langle e_{j},\cdot \right\rangle $ then has a unique complex-linear
extension from $\mathcal{\bar{A}}$ to $\mathcal{\bar{A}}_{\mathbb{C}}.$ A 
\textbf{cylinder function} on $\mathcal{\bar{A}}$ is a function that can be
expressed in the form

\begin{equation}
f\left( A\right) =\phi \left( \left\langle e_{1},A\right\rangle ,\cdots
\left\langle e_{d},A\right\rangle \right) ,  \label{e.4.7}
\end{equation}
where $\phi $ is a measurable function on $\mathbb{R}^{d}$ and $\left\{
e_{1},\cdots ,e_{d}\right\} $ is a orthonormal basis as above. A \textbf{%
holomorphic cylinder function} on $\mathcal{\bar{A}}_{\mathbb{C}}$ is a
function of the form 
\begin{equation*}
F\left( C\right) =\Phi \left( \left\langle e_{1},C\right\rangle ,\cdots
\left\langle e_{d},C\right\rangle \right) ,
\end{equation*}
where $\Phi $ is a holomorphic function on $\mathbb{C}^{d}.$ The \textbf{%
holomorphic subspace} of $L^{2}(\mathcal{\bar{A}}_{\mathbb{C}},\tilde{M}%
_{s,t}),$ denoted $\mathcal{H}L^{2}(\mathcal{\bar{A}}_{\mathbb{C}},\tilde{M}%
_{s,t}),$ is the $L^{2}$ closure of the $L^{2}$ holomorphic cylinder
functions.
\end{definition}

The transform $\tilde{S}_{s,t}$ will be defined in Theorem \ref{t.4.3} below
so as to coincide with the finite dimensional transform $S_{s,t}$ acting on
cylinder functions, and then extending by continuity to all of $L^{2}(%
\mathcal{\bar{A}},\tilde{P}_{s})$. In the standard case ($s=t$) one can and
often does define the transform differently (e.g., \cite{BSZ,GM}), with the
range Hilbert space being a certain space of holomorphic functions on $%
\mathcal{A}_{\mathbb{C}}$ rather than on $\mathcal{\bar{A}}_{\mathbb{C}}.$
Since the necessary dimension-independent pointwise bounds hold only when $%
s=t,$ this approach does not work when $s\neq t.$ Formally, $\tilde{S}%
_{s,t}f $ is the analytic continuation of $e^{t\Delta _{\mathcal{A}}/2}f;$
this description may be taken fairly literally when $f$ is a cylinder
function.

\begin{theorem}
\label{t.4.3}Fix $s$ and $t$ with $s>t/2>0.$ There exists a unique isometric
map $\tilde{S}_{s,t}$ of $L^{2}(\mathcal{\bar{A}},\tilde{P}_{s})$ onto $%
\mathcal{H}L^{2}(\mathcal{\bar{A}}_{\mathbb{C}},\tilde{M}_{s,t})$ such that
for all $f\in L^{2}(\mathcal{\bar{A}},\tilde{P}_{s})$ of the form 
\begin{equation*}
f\left( A\right) =\phi \left( \left\langle e_{1},A\right\rangle ,\cdots
,\left\langle e_{d},A\right\rangle \right) 
\end{equation*}
with $\left\{ e_{1},\cdots ,e_{d}\right\} $ as in Definition \ref{d.4.2}, $%
\tilde{S}_{s,t}f$ is given by 
\begin{equation*}
\tilde{S}_{s,t}f\left( C\right) =\left( S_{s,t}\phi \right) \left(
\left\langle e_{1},C\right\rangle ,\cdots ,\left\langle e_{d},C\right\rangle
\right) .
\end{equation*}
\end{theorem}

\textit{Proof}. We want to define $\tilde{S}_{s,t}$ to coincide with $%
S_{s,t} $ on cylinder functions. The fact that $\tilde{S}_{s,t}f$ is well
defined independent of how $f$ is represented as a cylinder function is a
consequence of the two observations: 1) the measure $P_{t}$ on $\mathbb{R}%
^{d}$ is rotationally-invariant, and 2) the $\left( d+k\right) $-dimensional
measure $P_{t}$ factors as the product of the corresponding $d$-dimensional
and $k$-dimensional measures.

Now isometricity on cylinder functions follows immediately from Theorem \ref
{t.3.2}. Since cylinder functions are dense, $\tilde{S}_{s,t}$ has a unique
isometric extension to $L^{2}(\mathcal{\bar{A}},\tilde{P}_{s}).$ The
surjectivity in Theorem \ref{t.3.2} shows that every $L^{2}$ holomorphic
cylinder function is in the image of $\tilde{S}_{s,t}.$ Since by definition
the $L^{2}$ holomorphic cylinder functions are dense in the holomorphic
subspace, we conclude that $\tilde{S}_{s,t}$ maps onto $\mathcal{H}L^{2}(%
\mathcal{\bar{A}}_{\mathbb{C}},\tilde{M}_{s,t}).$ \qed

\begin{definition}
\label{d.4.4}Let $\Delta _{\mathcal{A}}$ be the unique operator on cylinder
functions such that 
\begin{equation*}
\Delta _{\mathcal{A}}f\left( A\right) =\left( \Delta \phi \right) \left(
\left\langle e_{1},A\right\rangle ,\cdots ,\left\langle e_{d},A\right\rangle
\right) 
\end{equation*}
when $f$ is a cylinder function as in (\ref{e.4.7}) of Definition \ref{d.4.2}%
. The domain of $\Delta _{\mathcal{A}}$ is taking to be the set of those
cylinder functions for which $\phi $ is smooth and both $\phi $ and $\Delta
\phi $ are in $L^{2}\left( \mathbb{R}^{d},P_{s}\right) .$
\end{definition}

One checks as in the proof of Theorem \ref{t.4.3} that $\Delta _{\mathcal{A}%
} $ is well defined, independent of how $f$ is represented as a cylinder
function. Although it is densely defined, $\Delta _{\mathcal{A}}$ is a
non-closable operator.

\subsection{Action on the Hermite expansion\label{s.4.2}}

We now turn to the infinite-dimensional version of the Hermite expansion,
and the action of $\tilde{S}_{s,t}$ on it.

\begin{definition}
\label{d.4.5}The \textbf{n}$^{th}$\textbf{\ level Hermite subspace of} $%
L^{2}(\mathcal{\bar{A}},\tilde{P}_{s}),$ denoted $\mathcal{F}_{n,s}\left( 
\mathcal{\bar{A}}\right) ,$ is the $L^{2}$ closure of the space of functions
of the form 
\begin{equation*}
f\left( A\right) =\phi \left( \left\langle e_{1},A\right\rangle ,\cdots
,\left\langle e_{d},A\right\rangle \right) 
\end{equation*}
where $\left\{ e_{1},\cdots ,e_{d}\right\} $ is as in Definition \ref{d.4.2}
and where $\phi \in \mathcal{F}_{n,s}\left( \mathbb{R}^{d}\right) .$ The 
\textbf{n}$^{th}$\textbf{\ level holomorphic Hermite subspace of} $L^{2}(%
\mathcal{\bar{A}}_{\mathbb{C}},\tilde{M}_{s,t}),$ denoted $\mathcal{HF}%
_{n,s,t}\left( \mathcal{\bar{A}}_{\mathbb{C}}\right) ,$ is the $L^{2}$
closure of the space of functions of the form 
\begin{equation*}
F\left( C\right) =\Phi \left( \left\langle e_{1},C\right\rangle ,\cdots
,\left\langle e_{d},C\right\rangle \right) 
\end{equation*}
where $\Phi $ is in $\mathcal{HF}_{n,s,t}\left( \mathbb{C}^{d}\right) $ as
defined in Definition \ref{d.3.7}.
\end{definition}

Recall that $a_{\tau }=\int_{0}^{\tau }A_{\sigma }\,d\sigma $ is a scaled $%
\frak{k}$-valued Brownian motion whose components with respect to an
orthonormal basis $\left\{ X^{k}\right\} $ for $\frak{k}$ are denoted $%
a_{\tau }^{k}.$ Now consider the $n$-simplex 
\begin{equation*}
\Delta _{n}=\left\{ \left( \tau _{1},\cdots ,\tau _{n}\right) \in \mathbb{R}%
^{n}\left| 0\leq \tau _{1}\leq \tau _{2}\leq \cdots \leq \tau _{n}\right.
\leq 1\right\} .
\end{equation*}
Let $\mathbf{H}$\textbf{$=$}$\mathbf{\{}H_{k_{1},\cdots
,k_{n}}|k_{i}=1,\dots ,\dim \frak{k}\}$ be a collection of square-integrable
complex-valued functions on $\Delta _{n}$ and let 
\begin{equation*}
\sum_{k_{1},\cdots ,k_{n}=1}^{\dim \frak{k}}\int_{\Delta
_{n}}H_{k_{1},\cdots ,k_{n}}\left( \tau _{1},\cdots ,\tau _{n}\right)
\,da_{\tau _{1}}^{k_{1}}\cdots \,da_{\tau _{n}}^{k_{n}}
\end{equation*}
denote the multiple Wiener integral of $\mathbf{H}$ relative to $a,$ see 
\cite{Ito} or Definitions \ref{d.4.12} and \ref{d.4.12} below.

Similarly, $c_{\tau }=\int_{0}^{\tau }C_{\sigma }\,d\sigma $ for $C\in (%
\mathcal{\bar{A}}_{\mathbb{C}},\tilde{M}_{s,t})$ is a $\frak{k}_{\mathbb{C}}$%
-valued Brownian motion. Regard the orthonormal basis $\left\{ X_{k}\right\} 
$ for $\frak{k}$ as a basis of $\frak{k}_{\mathbb{C}}$ as a complex vector
space and let $c_{\tau }^{k}$ be the corresponding complex-valued components
of $c_{\tau }$. If $\mathbf{H}$\textbf{$=$}$\mathbf{\{}H_{k_{1},\cdots
,k_{n}}\}$ as above let 
\begin{equation*}
\sum_{k_{1},\cdots ,k_{n}=1}^{\dim \frak{k}}\int_{\Delta
_{n}}H_{k_{1},\cdots ,k_{n}}\left( \tau _{1},\cdots ,\tau _{n}\right)
\,dc_{\tau _{1}}^{k_{1}}\cdots \,dc_{\tau _{n}}^{k_{n}}
\end{equation*}
denote the multiple Wiener integral of $\mathbf{H}$ with respect to $c.$ By
expanding in terms of the real and imaginary parts of $c_{\tau }^{k},$ we
could express this as an integral in terms of independent real-valued
Brownian motions. Note that this integral is a formally holomorphic function
of $c$ (and hence of $C$) since it depends only on the complex increments of 
$c.$

\begin{proposition}
\label{p.4.6}The Hilbert space $L^{2}(\mathcal{\bar{A}},\tilde{P}_{s})$ is
the orthogonal direct sum of the subspaces $\mathcal{F}_{n,s}\left( \mathcal{%
\bar{A}}\right) .$ The Hilbert space $\mathcal{\ H}L^{2}(\mathcal{\bar{A}}_{%
\mathbb{C}},\tilde{M}_{s,t})$ is the orthogonal direct sum of the subspaces $%
\mathcal{HF}_{n,s,t}\left( \mathcal{\bar{A}}_{\mathbb{C}}\right) .$

A function $f\in L^{2}(\mathcal{\bar{A}},\tilde{P}_{s})$ is in $\mathcal{F}%
_{n,s}\left( \mathcal{\bar{A}}\right) $ if and only if there exist
square-integrable complex-valued functions $H_{k_{1},\cdots ,k_{n}}$ on $%
\Delta _{n}$ such that 
\begin{equation*}
f\left( A\right) =\sum_{k_{1},\cdots ,k_{n}=1}^{\dim \frak{k}}\int_{\Delta
_{n}}H_{k_{1},\cdots ,k_{n}}\left( \tau _{1},\cdots ,\tau _{n}\right)
\,da_{\tau _{1}}^{k_{1}}\cdots \,da_{\tau _{n}}^{k_{n}}.
\end{equation*}
A function $F\in \mathcal{H}L^{2}(\mathcal{\bar{A}}_{\mathbb{C}},\tilde{M}%
_{s,t})$ is in $\mathcal{HF}_{n,s,t}\left( \mathcal{\bar{A}}_{\mathbb{C}%
}\right) $ if and only if there exist square-integrable complex-valued
functions $H_{k_{1},\cdots ,k_{n}}$ on $\Delta _{n}$ such that 
\begin{equation*}
F\left( C\right) =\sum_{k_{1},\cdots ,k_{n}=1}^{\dim \frak{k}}\int_{\Delta
_{n}}H_{k_{1},\cdots ,k_{n}}\left( \tau _{1},\cdots ,\tau _{n}\right)
\,dc_{\tau _{1}}^{k_{1}}\cdots \,dc_{\tau _{n}}^{k_{n}}.
\end{equation*}
Here as usual $a_{\tau }=\int_{0}^{\tau }A_{\sigma }\,d\sigma $ and $c_{\tau
}=\int_{0}^{\tau }C_{\sigma }\,d\sigma ,$ and in either case the $H$'s are
unique up to a set of measure zero in $\Delta _{n}.$
\end{proposition}

The expansion of a function $f\in L^{2}(\mathcal{\bar{A}},\tilde{P}_{s})$
into a sum over $n$ of such stochastic integrals is called the Wiener chaos
expansion and goes back to Kakutani \cite{Ka} and It\^{o} \cite{Ito}.
Nevertheless, we will give a proof of this result (after Lemma \ref{l.4.11}
below) to emphasize the relation of this result to those in Section \ref{s.3}%
. The transform $\tilde{S}_{s,t}$ has the following simple action on the
Wiener chaos expansion, which will be used in the proof of Theorem \ref
{t.5.2} in Section \ref{s.5}.

\begin{theorem}
\label{t.4.7}The transform $\tilde{S}_{s,t}$ takes $\mathcal{F}_{n,s}\left( 
\mathcal{\bar{A}}\right) $ onto $\mathcal{HF}_{n,s,t}\left( \mathcal{\bar{A}}%
_{\mathbb{C}}\right) .$ Specifically, if $f\in L^{2}(\mathcal{\bar{A}},%
\tilde{P}_{s})$ is of the form 
\begin{equation*}
f=\sum_{k_{1},\cdots ,k_{n}=1}^{\dim K}\int_{\Delta _{n}}H_{k_{1},\cdots
,k_{n}}\left( \tau _{1},\cdots ,\tau _{n}\right) \,da_{\tau
_{1}}^{k_{1}}\cdots \,da_{\tau _{n}}^{k_{n}}
\end{equation*}
then $\tilde{S}_{s,t}f$ is given by 
\begin{equation*}
\tilde{S}_{s,t}f=\sum_{k_{1},\cdots ,k_{n}=1}^{\dim K}\int_{\Delta
_{n}}H_{k_{1},\cdots ,k_{n}}\left( \tau _{1},\cdots ,\tau _{n}\right)
\,dc_{\tau _{1}}^{k_{1}}\cdots \,dc_{\tau _{n}}^{k_{n}}.
\end{equation*}
\end{theorem}

This result is the infinite-dimensional analog of Theorem \ref{t.4.3}, with $%
e^{-s\Delta /2}$ and $e^{-A_{s,t}/2}$ hidden in the definition of the
stochastic integrals. The proof will be given at the end of this section.

We use a slightly unorthodox definition of the multiple Wiener integral,
which emphasizes the role of the heat equation. Equation (\ref{e.4.12})
below shows that our definition agrees with the usual one. The symmetric
group $S_{n}$ acts on the complex Hilbert space $L^{2}\left( \left[ 0,1%
\right] ^{n};\frak{k}_{\mathbb{C}}^{\otimes n}\right) $ by $\left( \sigma
\cdot f\right) \left( x_{1},x_{2},\dots ,x_{n}\right) :=\sigma f\left(
x_{\sigma 1},x_{\sigma 2},\dots ,x_{\sigma n}\right) ,$ where $\sigma f$
denotes action of $S_{n}$ on $\frak{k}_{\mathbb{C}}^{\otimes n}$ determined
by $\sigma (\xi _{1}\otimes \dots \otimes \xi _{n})=\xi _{\sigma
^{-1}1}\otimes \dots \otimes \xi _{\sigma ^{-1}n}.$ In particular, if $%
f=f_{1}\otimes \dots \otimes f_{n}$ with $f_{i}\in L^{2}\left( \left[ 0,1%
\right] ;\frak{k}_{\mathbb{C}}\right) ,$ then $\sigma \cdot f=f_{\sigma
^{-1}1}\otimes \dots \otimes f_{\sigma ^{-1}n}.$ The symmetric subspace,
denoted $\mathcal{S}L^{2}\left( \left[ 0,1\right] ^{n};\frak{k}_{\mathbb{C}%
}^{\otimes n}\right) ,$ is the space of those $f\in L^{2}\left( \left[ 0,1%
\right] ^{n};\frak{k}_{\mathbb{C}}^{\otimes n}\right) $ for which $\sigma
\cdot f=f$ for all $\sigma \in S_{n}.$ A function $f\in \mathcal{S}%
L^{2}\left( \left[ 0,1\right] ^{n};\frak{k}_{\mathbb{C}}^{\otimes n}\right) $
is determined by its restriction to $\Delta _{n}.$ The restriction is in $%
L^{2}\left( \Delta _{n};\frak{k}_{\mathbb{C}}^{\otimes n}\right) $ and every
element of $L^{2}\left( \Delta _{n};\frak{k}_{\mathbb{C}}^{\otimes n}\right) 
$ arises as such a restriction. For a symmetric $f,$ its norm-squared over $%
\left[ 0,1\right] ^{n}$ is $n!$ times its norm-squared over $\Delta _{n}.$
Finally, note that given functions $H_{k_{1},\cdots ,k_{n}}\in L^{2}\left(
\Delta _{n}\right) $ there is a unique $f\in \mathcal{S}L^{2}\left( \left[
0,1\right] ^{n};\frak{k}_{\mathbb{C}}^{\otimes n}\right) $ such that 
\begin{equation*}
\left( f\left( \tau _{1},\cdots ,\tau _{n}\right) ,X_{k_{1}}\otimes \cdots
\otimes X_{k_{n}}\right) =H_{k_{1},\cdots ,k_{n}}\left( \tau _{1},\cdots
,\tau _{n}\right) \text{ for }\left( \tau _{1},\cdots ,\tau _{n}\right) \in
\Delta _{n},
\end{equation*}
i.e. $f|_{\Delta _{n}}=\sum H_{k_{1},\cdots ,k_{n}}X_{k_{1}}\otimes \cdots
\otimes X_{k_{n}}.$ Here $\left( \cdot ,\cdot \right) $ refers to the
bilinear extension of the inner product from $\frak{k}$ to $\frak{k}_{%
\mathbb{C}}.$

\begin{definition}
\label{d.4.8}Let $\mathcal{E}$ denote the subspace of $L^{2}\left( \left[ 0,1%
\right] ^{n};\frak{k}_{\mathbb{C}}^{\otimes n}\right) $ consisting of finite
linear combinations of functions of the form 
\begin{equation}
f\left( \tau _{1},\cdots ,\tau _{n}\right) =f_{1}\left( \tau _{1}\right)
\otimes \cdots \otimes f_{n}\left( \tau _{n}\right)   \label{e.4.8}
\end{equation}
where $f_{i}\in L^{2}\left( \left[ 0,1\right] ;\frak{k}_{\mathbb{C}}\right) $
is of finite variation. As above, $f$ will be denoted by $f_{1}\otimes
\cdots \otimes f_{n}$. .The\textbf{\ multiple Stratonovich integral} is the
linear map $\mathrm{Strat}_{n,s}:\mathcal{E}\rightarrow L^{2}(\mathcal{%
\bar{A}},\tilde{P}_{s})$ determined by 
\begin{equation}
\mathrm{Strat}_{n,s}\left( f_{1}\otimes \cdots \otimes f_{n}\right) =\frac{1%
}{n!}\prod_{i=1}^{n}\int_{0}^{1}\left( f_{i}\left( \tau \right) ,da_{\tau
}\right) .  \label{e.4.9}
\end{equation}
Here $\left( \cdot ,\cdot \right) $ refers to the bilinear extension of the
inner product from $\frak{k}$ to $\frak{k}_{\mathbb{C}}.$
\end{definition}

\begin{remark}
\label{r.4.9}Since $f_{i}$ is assumed to be of finite variation, the
integrals in (\ref{e.4.9}) make sense for all continuous functions $a$ as
Stiltjies integrals. Moreover each $\left( f_{i}\left( \tau \right)
,da_{\tau }\right) $ is a continuous linear function of $a.$ Using these
remarks $\mathrm{Strat}_{n,s}\left( f\right) $ for $f\in \mathcal{E}$ is
completely determined by its values on $\mathcal{A}$ which are: 
\begin{equation}
\mathrm{Strat}_{n,s}\left( f\right) (A)=\frac{1}{n!}\left( f,A\otimes \dots
\otimes A\right) _{L^{2}\left( \left[ 0,1\right] ^{n};\frak{k}_{\mathbb{C}%
}^{\otimes n}\right) }\quad \text{for all \quad }A\in \mathcal{A}.
\label{e.4.10}
\end{equation}
Thus the right side of (\ref{e.4.10}) is a cylinder function, where the $%
\phi $ in Definition \ref{d.4.2} is a homogeneous polynomial of degree $n,$
and every such cylinder function arises in this way.
\end{remark}

We will be interested in this map just on the symmetric subspace of $%
\mathcal{E},$ denoted $\mathcal{E}_{\mathcal{S}}.$

\begin{definition}
\label{d.4.10}The\textbf{\ multiple It\^{o} integral} is the bounded linear
map $I_{n,s}$ from $\mathcal{S}L^{2}\left( \left[ 0,1\right] ^{n};\frak{k}_{%
\mathbb{C}}^{\otimes n}\right) $ to $L^{2}(\mathcal{\bar{A}},\tilde{P}_{s})$
which is determined uniquely by 
\begin{equation}
I_{n,s}\left( f\right) =e^{-s\Delta _{\mathcal{A}}/2}\mathrm{Strat}%
_{n,s}\left( f\right) \text{ for all }f\in \mathcal{E}_{\mathcal{S}}.
\label{e.4.11}
\end{equation}
The fact that there exists such a bounded linear operator satisfying this
equation is a consequence of Lemma \ref{l.4.11} below.
\end{definition}

As an example, for $i=1,2,,\ldots ,n,$ let $f_{i}=X_{k_{i}}1_{\left[
l_{i},m_{i}\right] }\in L^{2}\left( \left[ 0,1\right] ;\frak{k}_{\mathbb{C}%
}\right) $ where $l_{1},\cdots ,l_{n}$ and $m_{1},\cdots ,m_{n}$ are real
numbers such that $0\leq l_{i}<m_{i}\leq l_{i+1}\leq 1.$ Because of the
conditions on the $l$'s and $m$'s$,$ $f:=\sum_{\sigma \in S_{n}}f_{\sigma
1}\otimes \dots \otimes f_{\sigma n}$ is the unique functions in $\mathcal{E}%
_{\mathcal{S}}$ such that 
\begin{equation*}
f|_{\Delta _{n}}\left( \tau _{1},\cdots ,\tau _{n}\right) =\left(
X_{k_{1}}1_{\left[ l_{1},m_{1}\right] }\left( \tau _{1}\right) \right)
\otimes \cdots \otimes \left( X_{k_{n}}1_{\left[ l_{n},m_{n}\right] }\left(
\tau _{n}\right) \right) .
\end{equation*}
By Definition \ref{d.4.8}, 
\begin{equation*}
\mathrm{Strat}_{n,s}\left( f\right) =\prod_{i=1}^{n}\int_{0}^{1}\left(
f_{i}\left( \tau \right) ,da_{\tau }\right) =\prod_{i=1}^{n}\left(
a_{m_{i}}^{k_{i}}-a_{l_{i}}^{k_{i}}\right)
\end{equation*}
Again the assumptions on the $l$'s and $m$'s imply that $\left\{
f_{1},f_{2},\dots ,f_{n}\right\} \subset L^{2}\left( \left[ 0,1\right] ;%
\frak{k}_{\mathbb{C}}\right) $ is an orthonormal set and hence that $\Delta
_{\mathcal{A}}\mathrm{Strat}_{n,s}\left( f\right) =0.$ Therefore the
multiple Wiener integral coincides with the multiple Stratonovich integral,
i.e., 
\begin{equation}
I_{n,s}\left( f\right) =\prod_{i=1}^{n}\left(
a_{m_{i}}^{k_{i}}-a_{l_{i}}^{k_{i}}\right) .  \label{e.4.12}
\end{equation}
This expression agrees with any other reasonable definition of the multiple
Wiener integral. For more on multiple Stratonovich integrals and there
relationship to multiple It\^{o} integrals, see \cite{HM,JK} and the
references therein.

\begin{lemma}
\label{l.4.11}For $f\in \mathcal{E}_{\mathcal{S}},$%
\begin{equation*}
\left\| I_{n,s}\left( f\right) \right\| _{L^{2}(\mathcal{\bar{A}},\tilde{P}%
_{s})}^{2}=\frac{s^{n}}{n!}\left\| f\right\| _{L^{2}\left( \left[ 0,1\right]
^{n};\frak{k}_{\mathbb{C}}^{\otimes n}\right) }^{2}=s^{n}\left\| f\right\|
_{L^{2}\left( \Delta _{n};\frak{k}_{\mathbb{C}}^{\otimes n}\right) .}^{2}
\end{equation*}
\end{lemma}

\textit{Proof of Lemma \ref{l.4.11}. }Choose an orthonormal set $\left\{
e_{1},\cdots ,e_{d}\right\} \subset \mathcal{A}$ with each $e_{i}$ of finite
variation and such that 
\begin{equation}
f\in \text{span}\left\{ e_{i_{1}}\otimes \dots \otimes e_{i_{n}}\right\}
_{i_{j}=1}^{d}.  \label{e.4.13}
\end{equation}
Then there is a homogeneous polynomial $p$ of degree $n$ on $\mathbb{R}^{d}$
such that $\mathrm{Strat}_{n,s}\left( f\right) (A)=p(\langle e_{1},A\rangle
,\dots ,\langle e_{d},A\rangle ).$ Since the distribution (under $\tilde{P}%
_{s}$) of $(\langle e_{1},A\rangle ,\dots ,\langle e_{d},A\rangle )$ is the
Gaussian measure $e^{s\Delta _{\mathbb{R}^{d}}/2}\delta _{0},$ (\ref{e.3.7})
with $L=s\Delta _{\mathbb{R}^{d}}$ implies that 
\begin{equation}
\left\| I_{n,s}\left( f\right) \right\| _{L^{2}(\mathcal{\bar{A}},\tilde{P}%
_{s})}^{2}=\frac{s^{n}}{n!}\sum_{i_{1},i_{2},\ldots i_{n}=1}^{d}\left|
\left( \partial _{i_{1}}\partial _{i_{2}}\cdots \partial _{i_{n}}p\right)
(0)\right| ^{2},.  \label{e.4.14}
\end{equation}
where $\partial _{i}=\partial /\partial x_{i}$ and the factor $s^{n}$
results from the fact that $L=s\Delta _{\mathbb{R}^{d}}$ rather than $\Delta
_{\mathbb{R}^{d}}.$ By (\ref{e.4.10}) of Remark \ref{r.4.9}, 
\begin{equation*}
p(\langle e_{1},A\rangle ,\dots ,\langle e_{d},A\rangle )=\frac{1}{n!}%
\langle f,A\otimes \dots \otimes A\rangle _{L^{2}\left( \left[ 0,1\right]
^{n};\frak{k}_{\mathbb{C}}^{\otimes n}\right) }.
\end{equation*}
This equation and the chain rule gives, 
\begin{eqnarray}
\left( \partial _{i_{1}}\partial _{i_{2}}\cdots \partial _{i_{n}}p\right)
(0) &=&\frac{1}{n!}\partial _{e_{i_{1}}}\partial _{e_{i_{2}}}\cdots \partial
_{e_{i_{n}}}\langle f,A\otimes \dots \otimes A\rangle _{L^{2}\left( \left[
0,1\right] ^{n};\frak{k}_{\mathbb{C}}^{\otimes n}\right) }|_{A=0}  \notag \\
&=&\frac{1}{n!}\sum_{\sigma \in S_{n}}\langle f,e_{i_{\sigma 1}}\otimes
\dots \otimes e_{i_{\sigma n}}\rangle _{L^{2}\left( \left[ 0,1\right] ^{n};%
\frak{k}_{\mathbb{C}}^{\otimes n}\right) }  \notag \\
&=&\langle f,e_{i_{1}}\otimes \dots \otimes e_{i_{n}}\rangle _{L^{2}\left( 
\left[ 0,1\right] ^{n};\frak{k}_{\mathbb{C}}^{\otimes n}\right) },
\label{e.4.15}
\end{eqnarray}
where in the last equality we have used $f\in \mathcal{E}_{\mathcal{S}}.$
Combining (\ref{e.4.13})--(\ref{e.4.15}) proves the Lemma. \qed

We may now define the integrals that appear in Proposition \ref{p.4.6}.

\begin{definition}
\label{d.4.12}For $f\in L^{2}\left( \Delta _{n}\right) ,$ let 
\begin{equation*}
\int_{\Delta _{n}}f\,da_{\tau _{1}}^{k_{1}}\cdots da_{\tau
_{n}}^{k_{n}}=I_{n,s}\left( g\right) ,
\end{equation*}
where $g$ is the symmetric extension to $\left[ 0,1\right] ^{n}$ of the
function $f\cdot X_{k_{1}}\otimes \cdots \otimes X_{k_{n}}$ in $L^{2}\left(
\Delta _{n};\frak{k}_{\mathbb{C}}^{\otimes n}\right) .$
\end{definition}

\bigskip

\textit{Proof of Proposition \ref{p.4.6}.} Identifying $\mathcal{S}%
L^{2}\left( \left[ 0,1\right] ^{n};\frak{k}_{\mathbb{C}}^{\otimes n}\right) $
with $L^{2}\left( \Delta _{n};\frak{k}_{\mathbb{C}}^{\otimes n}\right) $ by
restriction, the It\^{o} integral is an isometric map of $L^{2}\left( \Delta
_{n};\frak{k}_{\mathbb{C}}^{\otimes n}\right) $ into $L^{2}(\mathcal{\bar{A}}%
,\tilde{P}_{s}).$ From the proof of Lemma \ref{l.4.11}, the image of $%
L^{2}\left( \Delta _{n};\frak{k}_{\mathbb{C}}^{\otimes n}\right) $ in $L^{2}(%
\mathcal{\bar{A}},\tilde{P}_{s})$ is precisely $\mathcal{F}_{n,s}\left( 
\mathcal{\bar{A}}\right) $ in Definition \ref{d.4.5}. This gives the
characterization of $\mathcal{F}_{n,s}\left( \mathcal{\bar{A}}\right) $
given in the proposition. That the spaces $\mathcal{F}_{n,s}\left( \mathcal{%
\bar{A}}\right) $ are orthogonal and that their sum is all of $\ L^{2}(%
\mathcal{\bar{A}},\tilde{P}_{s})$ follow from the corresponding results
(Proposition \ref{p.3.5}) for the spaces $\mathcal{F}_{n,s}\left( \mathbb{R}%
^{d}\right) $ in $L^{2}\left( \mathbb{R}^{d},P_{s}\right) $ and the density
of cylinder functions.

We now turn to the complex case. As in the real case we think of the
integrand as an element of $L^{2}\left( \Delta _{n};\frak{k}_{\mathbb{C}%
}^{\otimes n}\right) ,$ which we then identify with the symmetric subspace
of $L^{2}\left( \left[ 0,1\right] ^{n};\frak{k}_{\mathbb{C}}^{\otimes
n}\right) .$ Continuing the notation of Definition \ref{d.4.4}, let $%
\widetilde{A}_{s,t}$ be the unique operator on cylinder functions such that $%
\widetilde{A}_{s,t}f\left( A\right) =\left( A_{s,t}\phi \right) \left(
\left\langle e_{1},A\right\rangle ,\cdots ,\left\langle e_{d},A\right\rangle
\right) .$ Define the complex Stratonovich integral by analogy to (\ref
{e.4.9}) to be 
\begin{equation}
\mathrm{Strat}_{n,s,t}\left( f_{1}\otimes \cdots \otimes f_{n}\right) =\frac{%
1}{n!}\prod_{i=1}^{n}\int_{0}^{1}\left( f_{i}\left( \tau \right) ,dc_{\tau
}\right)  \label{e.4.20}
\end{equation}
and the complex It\^{o} integral to be 
\begin{equation}
I_{n,s,t}\left( f\right) =e^{-\widetilde{A}_{s,t}/2}\mathrm{Strat}%
_{n,s,t}\left( f\right) .  \label{e.4.21}
\end{equation}

The analog of Lemma \ref{l.4.11} in the complex case is: 
\begin{equation*}
\left\| I_{n,s,t}\left( f\right) \right\| _{L^{2}(\mathcal{\bar{A}}_{\mathbb{%
C}},\tilde{M}_{s,t})}^{2}=\frac{s^{n}}{n!}\left\| f\right\| _{L^{2}\left( %
\left[ 0,1\right] ^{n};\frak{k}_{\mathbb{C}}^{\otimes n}\right)
}^{2}=s^{n}\left\| f\right\| _{L^{2}\left( \Delta _{n};\frak{k}_{\mathbb{C}%
}^{\otimes n}\right) .}^{2}
\end{equation*}
The proof is the same provided that (\ref{e.3.11}) is used in place of (\ref
{e.3.7}). Alternatively, this equation is a consequence of Theorem \ref
{t.3.8} and the isometricity of $S_{s,t}$ in Theorem \ref{t.3.2}. The rest
of the proof is the same as the real case. \qed

\textit{Proof of Theorem \ref{t.4.7}}. Identify the integrand as above with
an element of $\mathcal{S}L^{2}\left( \left[ 0,1\right] ^{n};\frak{k}_{%
\mathbb{C}}^{\otimes n}\right) .$ By the continuity of the transform and the
integrals is suffices to prove the result on the dense subspace $\mathcal{E}%
_{\mathcal{S}}.$ But comparing (\ref{e.4.9}) and (\ref{e.4.11}) to (\ref
{e.4.20}) and (\ref{e.4.21}) and using Theorem \ref{t.3.8} gives the result
on $\mathcal{E}_{\mathcal{S}}.$ \qed

\begin{remark}
\label{r.4.13}In Section \ref{s.5} we will need to know that, at least in
certain cases, the \textit{multiple} Wiener integral coincides with the 
\textit{iterated} It\^{o} integral. It is easily seen that the two coincide
for nice integrands, as in (\ref{e.4.12}). Moreover, using repeatedly the
isometry property of the one-dimensional It\^{o} integral shows that the
iterated It\^{o} integral has the same isometry property (Lemma \ref{l.4.11}%
) as the multiple Wiener integral. It follows that the multiple Wiener and
iterated It\^{o} integrals coincide, \textit{provided} that the iterated It%
\^{o} integral makes sense. For us it is enough to have this for integrands
which are constant on a set of the form $0\leq u\leq \tau _{1}\leq \cdots
\leq \tau _{n}\leq v$ and zero elsewhere (Lemmas \ref{l.5.7} and \ref{l.5.8}%
), in which case there is no difficulty.
\end{remark}

\section{Functions of the holonomy\label{s.5}}

\subsection{Statements\label{s.5.1}}

Recall that $a_{\tau }(A):=\int_{0}^{\tau }A_{\sigma }\,d\sigma $ and $%
c_{\tau }(A):=\int_{0}^{\tau }C_{\sigma }\,d\sigma $ are Brownian motions
with values in $\frak{k}$ and $\frak{k}_{\mathbb{C}}$ respectively.

\begin{definition}[It\^{o} Maps]
\label{d.5.1}Let $\theta _{\tau }$ and $\theta _{\tau }^{\mathbb{C}}$ denote
the solutions to the Stratonovich stochastic differential equations 
\begin{eqnarray}
d\theta _{\tau }=\theta _{\tau }\circ da_{\tau }\text{ \ with }\theta
_{0}=e\in K,  \label{e.5.1} \\
d\theta _{\tau }^{\mathbb{C}}=\theta _{\tau }^{\mathbb{C}}\circ dc_{\tau }%
\text{ \ with }\theta _{0}^{\mathbb{C}}=e\in K_{\mathbb{C}}.  \label{e.5.2}
\end{eqnarray}
We define the holonomies $h\left( A\right) $ and $h_{\mathbb{C}}\left(
C\right) $ by 
\begin{eqnarray*}
h\left( A\right) =\theta _{1}\left( A\right)  \\
h_{\mathbb{C}}\left( C\right) =\theta _{1}^{\mathbb{C}}\left( C\right) .
\end{eqnarray*}
Notice that $\theta $ and $h$ are defined on $(\mathcal{\bar{A}},\tilde{P}%
_{s}),$ and that $\theta ^{\mathbb{C}}$ and $h_{\mathbb{C}}$ are defined on $%
(\mathcal{\bar{A}}_{\mathbb{C}},\tilde{M}_{s,t}).$
\end{definition}

Recall that $\theta _{\tau }\left( A\right) $ and $\theta _{\tau }^{\mathbb{C%
}}\left( C\right) $ are to be interpreted as the parallel transport from $0$
to $\tau $ of the generalized connections $A$ and $C,$ respectively. The
meaning of the stochastic differential equations (\ref{e.5.1}) and (\ref
{e.5.2}) is described in detail in Section \ref{s.5.3}. We are now ready to
state our main result.

\begin{theorem}
\label{t.5.2}Fix $s$ and $t$ with $s>t/2>0.$ Suppose $f\in L^{2}(\mathcal{%
\bar{A}},\tilde{P}_{s})$ is of the form 
\begin{equation*}
f\left( A\right) =\phi \left( h\left( A\right) \right) ,
\end{equation*}
where $\phi $ is a function on $K.$ Then there exists a unique holomorphic
function $\Phi $ on $K_{\mathbb{C}}$ such that 
\begin{equation*}
\tilde{S}_{s,t}f\left( C\right) =\Phi \left( h_{\mathbb{C}}\left( C\right)
\right) .
\end{equation*}
The function $\Phi $ is determined by the condition that 
\begin{equation*}
\left. \Phi \right| _{K}=e^{t\Delta _{K}/2}\phi .
\end{equation*}
\end{theorem}

Here $\Delta _{K}$ is the Laplace-Beltrami operator on $K$ associated to the
bi-invariant Riemannian metric that agrees at the identity with the chosen
inner product on $\frak{k}.$ The meaning of $e^{t\Delta _{K}/2}$ is
discussed following Theorem \ref{t.5.3}.

Observe that the space of functions $f\in L^{2}(\mathcal{\bar{A}},\tilde{P}%
_{s})$ of the form $\phi \circ h$ may be identified with $L^{2}\left( K,\rho
_{s}\right) ,$ where $\rho _{s}$ is the distribution of $h$ with respect to $%
\tilde{P}_{s}.$ It is known that $\rho _{s}$ coincides with the heat kernel
measure on $K;$ thus the Hilbert space $L^{2}\left( K,\rho _{s}\right) $
coincides with the one considered in \cite{H1}. Similarly, the space of
functions $F\in L^{2}(\mathcal{\bar{A}}_{\mathbb{C}},\tilde{M}_{s,t})$ of
the form $F=\Phi \circ \theta _{\mathbb{C}},$ with $\Phi $ a
not-necessarily-holomorphic function on $K_{\mathbb{C}},$ may be identified
with $L^{2}\left( K_{\mathbb{C}},\mu _{s,t}\right) ,$ where $\mu _{s,t}$ is
the distribution of $h_{\mathbb{C}}$ with respect to $\tilde{M}_{s,t},$
which is a certain heat kernel measure on $K_{\mathbb{C}}$. So if we apply
the isometric transform $\tilde{S}_{s,t}$ to functions of the form $\phi
\circ h,$ then we obtain an isometric map of $L^{2}\left( K,\rho _{s}\right) 
$ into the holomorphic subspace of $L^{2}\left( K_{\mathbb{C}},\mu
_{s,t}\right) .$ The proof of surjectivity in Theorem 2 of \cite{H1} applies
essentially without change to show that this map is onto the holomorphic
subspace. The following theorem summarizes these observations.

\begin{theorem}
\label{t.5.3}For all $s$ and $t$ with $s>t/2>0,$ the map 
\begin{equation*}
\phi \rightarrow \text{ analytic continuation of }e^{t\Delta _{K}/2}\phi 
\end{equation*}
is an isometric isomorphism of $L^{2}\left( K,\rho _{s}\right) $ onto the
space of holomorphic functions in $L^{2}\left( K_{\mathbb{C}},\mu
_{s,t}\right) .$
\end{theorem}

This result was proved in \cite{H1} for the case $s=t.$ Part of the theorem
is that $e^{t\Delta _{K}/2}\phi $ always has a unique analytic continuation
to $K_{\mathbb{C}}.$ Note that the measure $\rho _{s}$ has a density with
respect to Haar measure which is strictly positive and continuous, and
therefore bounded and bounded away from zero by compactness. This means that 
$L^{2}\left( K,\rho _{s}\right) $ is the same space of functions as $%
L^{2}\left( K,\text{\thinspace Haar}\right) .$ So $e^{t\Delta _{K}/2}$ is to
be interpreted as the standard contraction semigroup on $L^{2}\left( K,\,%
\text{Haar}\right) .$

If we restrict the transform $\tilde{S}_{s,t}$ to functions of the holonomy,
then it makes sense to allow $s$ to tend to infinity. See Section \ref{s.2}
for a discussion of why this limit is natural from the point of view of
Yang-Mills theory.

\begin{theorem}
\label{t.5.4}Normalize Haar measure $dx$ on the compact group $K$ to have
mass one. Then $f\in L^{2}\left( K,\rho _{s}\right) $ if and only if $f\in
L^{2}\left( K,dx\right) ,$ and 
\begin{equation*}
\left\| f\right\| _{L^{2}\left( K,dx\right) }=\lim_{s\rightarrow \infty
}\left\| f\right\| _{L^{2}\left( K,\rho _{s}\right) }.
\end{equation*}
The measure 
\begin{equation*}
d\nu _{s,t}\left( g\right) =\int_{K}d\mu _{s,t}\left( gx\right) \,dx,\quad
g\in K_{\mathbb{C}}
\end{equation*}
is independent of $s$ and will be denoted $\nu _{t}\left( g\right) .$ For
all $F\in \mathcal{H}\left( K_{\mathbb{C}}\right) ,$ $F\in L^{2}\left( K_{%
\mathbb{C}},\mu _{s,t}\right) $ if and only if $F\in L^{2}\left( K_{\mathbb{C%
}},\nu _{t}\right) ,$ and 
\begin{equation*}
\left\| F\right\| _{L^{2}\left( K_{\mathbb{C}},\nu _{t}\right)
}=\lim_{s\rightarrow \infty }\left\| F\right\| _{L^{2}\left( K_{\mathbb{C}%
},\mu _{s,t}\right) }.
\end{equation*}
Thus the map 
\begin{equation*}
f\rightarrow \text{ analytic continuation }\left( e^{t\Delta _{K}/2}f\right) 
\end{equation*}
is an isometric isomorphism of $L^{2}\left( K,dx\right) $ onto $\mathcal{H}%
L^{2}\left( K_{\mathbb{C}},\nu _{t}\right) .$
\end{theorem}

The last isometric isomorphism, with domain $L^{2}\left( K,dx\right) ,$ was
obtained in \cite[Thm. 2]{H1}, and was denoted $C_{t}.$

Recall that the transform $\tilde{S}_{s,t}$ maps into the holomorphic
subspace of $L^{2}(\mathcal{\bar{A}}_{\mathbb{C}},\tilde{M}_{s,t}).$ Theorem 
\ref{t.5.2} together with the ``onto'' part of Theorem \ref{t.5.3} gives the
following.

\begin{theorem}
\label{t.5.5}Suppose $\Phi $ is a holomorphic function on $K_{\mathbb{C}}$
such that $\Phi \circ h_{\mathbb{C}}$ is in $L^{2}(\mathcal{\bar{A}}_{%
\mathbb{C}},\tilde{M}_{s,t}).$ Then $\Phi \circ h_{\mathbb{C}}$ is in $%
\mathcal{H}L^{2}(\mathcal{\bar{A}}_{\mathbb{C}},\tilde{M}_{s,t}).$
\end{theorem}

It would be desirable to have a direct proof of this result. See the
discussion in Section 2.5 of \cite{HS}, which contains the $s=t$ case of
Theorem \ref{t.5.5}. (Note that there is a gap in one of the two proofs of
this result in \cite{HS}. The paper \cite{DHu} will close this gap. See the
discussion at the end of Section \ref{s.7}.) A more general version of
Theorem \ref{t.5.5} is given in Theorem \ref{t.6.3} of Section \ref{s.6}.

\subsection{Heuristics\label{s.5.2}}

We now give a simple heuristic argument for Theorem \ref{t.5.2} based on the
following proposition. The proof will be given in the next subsection. We
are grateful to Ambar Sengupta for showing us the significance of
Proposition \ref{p.5.6}. See also Appendix A for another ``explanation'' of
Theorem \ref{t.5.2}.

\begin{proposition}
\label{p.5.6}If $A\in \mathcal{\bar{A}}$ is distributed according to the
measure $\tilde{P}_{s}$ and $B$ is a fixed element of $\mathcal{A},$ then $%
\theta \!\left( A+B\right) $ has the same distribution as $\theta \left(
A\right) \theta \left( B\right) .$
\end{proposition}

\textit{Proof.} Direct calculation shows that for $A$ and $B$ smooth, 
\begin{equation*}
\theta _{\tau }\left( A\right) \theta _{\tau }\left( B\right) =\theta
\!\left( \text{Ad}\theta _{\tau }\left( B\right) ^{-1}\left( A_{\tau
}\right) +B_{\tau }\right) .
\end{equation*}
Standard stochastic techniques show that this remains true almost surely if $%
A$ is a white noise and $B$ is in $\mathcal{A}.$ But the white noise measure 
$\tilde{P}_{s}$ is invariant under the pointwise adjoint action, which is
just a ``rotation'' of $\mathcal{A}.$ \qed

Using Proposition \ref{p.5.6}, we may formally calculate $e^{t\Delta _{%
\mathcal{A}}/2}f.$ The measure $\tilde{P}_{t}$ is the fundamental solution
at the origin of the heat equation, which means that $e^{t\Delta _{\mathcal{A%
}}/2}$ should be given by convolution with $\tilde{P}_{t}.$ Similarly, $\rho
_{t}$ is the fundamental solution at the identity of the heat equation on $%
K. $ So if $f\left( A\right) =\phi \left( \theta \!_{1}\left( A\right)
\right) $ then 
\begin{eqnarray*}
e^{t\Delta _{\mathcal{A}}/2}f\left( B\right) &=&\int_{\mathcal{\bar{A}}}\phi
\left( \theta _{1}\left( A+B\right) \right) \,d\tilde{P}_{t}\left( A\right)
\\
&=&\int_{\mathcal{\bar{A}}}\phi \left( \theta _{1}\left( A\right) \theta
_{1}\left( B\right) \right) \,d\tilde{P}_{t}\left( A\right) \\
&=&\int_{K}\phi \left( x\theta _{1}\left( B\right) \right) \,d\rho
_{t}\left( x\right) \\
&=&e^{t\Delta _{K}/2}\phi \left( \theta _{1}\left( B\right) \right) .
\end{eqnarray*}
We have used the proposition between the first and second lines. Assuming
this is valid for $B\in \mathcal{\bar{A}}$ and then analytically continuing
formally to $\mathcal{\bar{A}}_{\mathbb{C}}$ we obtain Theorem \ref{t.5.2}.

Some rigorous variant of this argument is used in \cite{GM,Sa2,HS,AHS}, all
of which consider only the $s=t$ case. In that case it is possible to work
with holomorphic functions on $\mathcal{A}_{\mathbb{C}}$ instead of $%
\mathcal{\bar{A}}_{\mathbb{C}},$ so that the above argument is essentially
rigorous. The results of \cite{GM} are stated only on $\mathcal{A}_{\mathbb{C%
}},$ while the other papers work first on $\mathcal{A}_{\mathbb{C}}$ and
then extend to $\mathcal{\bar{A}}_{\mathbb{C}}.$ However, this approach does
not work when $s\neq t,$ since the pointwise bounds needed to obtain
(everywhere-defined) holomorphic functions on $\mathcal{A}_{\mathbb{C}}$ do
not hold when $s\neq t.$

\subsection{Proofs\label{s.5.3}}

Let us explain more precisely Definition \ref{d.5.1} of the It\^{o} maps. A
continuous $K$--valued semi-martingale $\theta _{\tau }$ satisfies (\ref
{e.5.1}) if and only if for all $f\in C^{\infty }([0,1]\times K),$ 
\begin{equation}
f(\tau ,\theta _{\tau })=f(0,e)+\int_{0}^{\tau }\frac{\partial f}{\partial
\sigma }\left( \sigma ,\theta _{\sigma }\right) \,d\sigma +\sum_{k=1}^{\dim 
\frak{k}}\int_{0}^{\tau }X_{k}f\left( \sigma ,\theta _{\sigma }\right) \circ
da_{\sigma }^{k},  \label{e.5.3}
\end{equation}
where $a_{\tau }=\sum_{k=1}^{\dim \frak{k}}a_{\tau }^{k}X_{k}.$ Here $%
\left\{ X_{k}\right\} _{k=1}^{\dim \frak{k}}$ is an orthonormal basis for $%
\frak{k},$ with each $X_{k}$ viewed as a left-invariant vector field on $K.$

Noting that $\{a_{\cdot }^{k}\}_{k=1}^{\dim \frak{k}}$ are independent
Brownian motions with variance $s,$ (\ref{e.5.3}) may be written in It\^{o}
form as 
\begin{eqnarray}
f(\tau ,\theta _{\tau }) &=&f(0,e)+\int_{0}^{\tau }\left( \frac{\partial f}{%
\partial \sigma }(\sigma ,\theta _{\sigma })+\frac{s}{2}\Delta _{K}f(\sigma
,\theta _{\sigma })\right) \,d\sigma  \notag \\
&&+\sum_{k=1}^{\dim \frak{k}}\int_{0}^{\tau }X_{k}f(\sigma ,\theta _{\sigma
})\,da_{\sigma }^{k}.  \label{e.5.4}
\end{eqnarray}
where $\Delta _{K}=\sum X_{k}^{2}.$

Write $c_{\tau }=\sum_{k=1}^{\dim \frak{k}}a_{\tau
}^{k}X_{k}+\sum_{k=1}^{\dim \frak{k}}b_{\tau }^{k}Y_{k},$ where $%
Y_{k}=JX_{k}.$ \textbf{Warning: }we are using the same letter $a$ for both
the process in $\frak{k}$ and the real part of the process in $\frak{k}_{%
\mathbb{C}},$ which do not even have the same distribution. The context
should make it clear whether we are in the real or the complex setting. A
continuous $K_{\mathbb{C}}$--valued semi-martingale $\theta _{\tau }^{%
\mathbb{C}}$ is said to solve (\ref{e.5.2}) provided that for all $u\in
C^{\infty }\left( [0,1]\times K_{\mathbb{C}}\right) ,$ 
\begin{eqnarray}
u(\tau ,\theta _{\tau }^{\mathbb{C}}) &=&u(0,e)+\int_{0}^{\tau }\frac{%
\partial u}{\partial \sigma }(\sigma ,\theta _{\sigma }^{\mathbb{C}%
})\,d\sigma  \notag \\
&&+\sum_{k=1}^{\dim \frak{k}}\int_{0}^{\tau }X_{k}u(\sigma ,\theta _{\sigma
}^{\mathbb{C}})\circ da_{\sigma }^{k}+\sum_{k=1}^{\dim \frak{k}%
}\int_{0}^{\tau }Y_{k}u(\sigma ,\theta _{\sigma }^{\mathbb{C}})\circ
db_{\sigma }^{k}.  \label{e.5.5}
\end{eqnarray}
Noting that $\{a_{\cdot }^{k}\}_{k=1}^{\dim \frak{k}}$ and $\{b_{\cdot
}^{k}\}_{k=1}^{\dim \frak{k}}$ are independent Brownian motions with
variances $\left( s-t/2\right) $ and $t/2,$ respectively, (\ref{e.5.5}) may
be written in It\^{o} form as 
\begin{eqnarray}
u\left( \tau ,\theta _{\tau }^{\mathbb{C}}\right) &=&u\left( 0,e\right)
+\int_{0}^{\tau }\left( \frac{\partial u}{\partial \sigma }\left( \sigma
,\theta _{\sigma }^{\mathbb{C}}\right) +\frac{1}{2}A_{s,t}^{K_{\mathbb{C}%
}}u\left( \sigma ,\theta _{\sigma }^{\mathbb{C}}\right) \right) d\sigma 
\notag \\
&&+\sum_{k=1}^{\dim \frak{k}}\int_{0}^{\tau }X_{k}u\left( \theta _{\sigma }^{%
\mathbb{C}}\right) \,da_{\sigma }^{k}+\sum_{k=1}^{\dim \frak{k}%
}\int_{0}^{\tau }Y_{k}u\left( \theta _{\sigma }^{\mathbb{C}}\right)
\,db_{\sigma }^{k}.  \label{e.5.6}
\end{eqnarray}
where $A_{s,t}^{K_{\mathbb{C}}}$ is defined, by analogy to $A_{s,t},$ to be $%
\left( s-t/2\right) \Sigma X_{k}^{2}+\frac{t}{2}\Sigma Y_{k}^{2}.$ For
existence and uniqueness of solutions to (\ref{e.5.5}) and (\ref{e.5.6})
see, for example, Elworthy \cite{El}, Emery \cite{Em}, Ikeda and Watanabe 
\cite{IW}, or Kunita \cite[Theorem 4.8.7]{Ku}.

We now begin working toward the proof of Theorem \ref{t.5.2}. For use in
Section \ref{s.6}, we will actually compute $\tilde{S}_{s,t}$ on functions
of the form $\phi (\theta _{u}^{-1}\theta _{v}),$ for two fixed times $u$
and $v.$ The transformed function is then $\Phi \left( \left( \theta _{u}^{%
\mathbb{C}}\right) ^{-1}\theta _{v}^{\mathbb{C}}\right) ,$ where $\Phi $ is
holomorphic on $K_{\mathbb{C}}$ and where $\left. \Phi \right|
_{K}=e^{\left( v-u\right) t\Delta _{K}/2}\phi .$ Theorem \ref{t.5.2} is the
special case $u=0,$ $v=1.$

The following Lemma is essentially a special case of the results of
Veretennikov and Krylov\cite{VK}.

\begin{lemma}
\label{l.5.7}Suppose that $0\leq u<v\leq 1,$ and $\phi $ is a measurable
function on $K$ such that $\phi (\theta _{u}^{-1}\theta _{v})\in L^{2}(%
\mathcal{\bar{A}},\tilde{P}_{s}).$ Then the Wiener Chaos expansion of $\phi
(\theta _{u}^{-1}\theta _{v})$ is 
\begin{equation}
\phi (\theta _{u}^{-1}\theta _{v})=\sum_{n=0}^{\infty }\sum_{k_{1},\cdots
,k_{n}=1}^{\dim \frak{k}}\int_{\Delta _{n}(u,v)}\alpha _{k_{1},\cdots
,k_{n}}\,da_{\tau _{1}}^{k_{1}}\cdots \,da_{\tau _{n}}^{k_{n}},
\label{e.5.7}
\end{equation}
where 
\begin{equation}
\alpha _{k_{1},\cdots ,k_{n}}=\left( X_{k_{1}}\cdots
X_{k_{n}}e^{(v-u)s\Delta _{K}/2}\phi \right) (e)  \label{e.5.8}
\end{equation}
and $\Delta _{n}(u,v)=\{(\tau _{1},\dots ,\tau _{n})\left| u\leq \tau
_{1}\leq \tau _{2}\leq \cdots \leq \tau _{n}\leq v\right. \}.$
\end{lemma}

\textit{Proof. }To simplify notation, let $\xi _{\tau }=\theta
_{u}^{-1}\theta _{\tau }$ for $u\leq \tau \leq 1.$ Let us first assume that $%
\phi (x)=\langle h,\pi (x)w\rangle ,$ where $\pi :K\rightarrow End(W)$ is a
finite-dimensional representation of $K$, $h\in W^{\ast }$ and $w\in W.$ Set 
\begin{equation}
f(\tau ,x)=\left( e^{(v-\tau )s\Delta _{K}/2}\phi \right) \left( x\right)
=\left\langle h,\pi (x)e^{(v-\tau )s\pi \left( \Delta _{K}\right)
/2}w\right\rangle ,  \label{e.5.9}
\end{equation}
where $\pi \left( \Delta _{K}\right) =\sum_{k=1}^{\dim \frak{k}}\pi
(X_{k})^{2}$ and by abuse of notation we are writing $\pi (X)$ for $\frac{d}{%
dt}|_{0}\pi (e^{tX})$ when $X\in \frak{k.}$ The second equality in (\ref
{e.5.9}) follows from uniqueness of solutions to the heat equation. (See
also \cite{H1}). Notice that $f$ solves the backward heat equation $\partial
f(\tau ,x)/\partial \tau +s\Delta f(\tau ,x)/2=0$ with $f(v,x)=\phi (x).$
Therefore by the analog of (\ref{e.5.4}) for $\xi _{\tau },$ 
\begin{equation*}
f(\tau ,\xi _{\tau })=f(u,e)+\sum_{k=1}^{\dim \frak{k}}\int_{u}^{\tau
}X_{k}f\left( \sigma ,\xi _{\sigma }\right) \,da_{\sigma }^{k}
\end{equation*}
Since $\Delta $ and $X_{k}$ commute, it follows that $X_{k}f$ also satisfies
the backward heat equation. Hence, the previous equation applies with $f$
replaced by $X_{k}f,$ namely 
\begin{equation*}
X_{k}f\left( \tau ,\xi _{\tau }\right) =X_{k}f\left( u,e\right)
+\sum_{l=1}^{\dim \frak{k}}\int_{u}^{\tau }X_{l}X_{k}f\left( r,\xi
_{r}\right) \,da_{r}^{l}.
\end{equation*}
Combining the two previous equations gives 
\begin{equation*}
f\left( \tau ,\xi _{\tau }\right) =f(u,e)+\sum_{k=1}^{\dim \frak{k}%
}\int_{u}^{\tau }X_{k}f(u,e)\,da_{\sigma }^{k}+\sum_{k,l=1}^{\dim \frak{k}%
}\int_{u}^{\tau }\left( \int_{u}^{\sigma }X_{l}X_{k}f\left( r,\xi
_{r}\right) \,da_{r}^{l}\right) \,da_{\sigma }^{k}.
\end{equation*}

Iterating this procedure gives (Remark \ref{r.4.13}) 
\begin{eqnarray*}
\phi (\theta _{u}^{-1}\theta _{v}) &=&\phi (\xi _{v})=f(v,\xi _{v}) \\
&=&\sum_{n=0}^{N}\sum_{k_{1},\cdots ,k_{n}=1}^{\dim \frak{k}}\int_{\Delta
_{n}(u,v)}\alpha _{k_{1},\cdots ,k_{n}}\,da_{\tau _{1}}^{k_{1}}\cdots
\,da_{\tau _{n}}^{k_{n}}+R_{N}(u,v),
\end{eqnarray*}
where $\alpha _{k_{1},\cdots ,k_{n}}$ is defined by (\ref{e.5.8}) and where 
\begin{equation*}
R_{N}(u,v)=\sum_{k_{1},\cdots ,k_{N+1}=1}^{\dim \frak{k}}\int_{u}^{v}%
\int_{u}^{\tau _{N+1}}\cdots \int_{u}^{\tau _{2}}X_{k_{1}}\cdots
X_{k_{n}}f(\tau _{1},\xi _{\tau _{1}})\,da_{\tau _{1}}\cdots \,da_{\tau
_{N+1}}.
\end{equation*}
Using the isometry property of iterated It\^{o} integral together with the
assumption that $\phi $ is a matrix entry of a finite-dimensional
representation one shows that 
\begin{equation*}
\left\| R_{N}\left( u,v\right) \right\| _{L^{2}(\mathcal{\bar{A}},\tilde{P}%
_{s})}^{2}\leq \frac{C^{N+1}}{\left( N+1\right) !}\rightarrow 0\text{ as }%
N\rightarrow \infty .
\end{equation*}
Therefore, (\ref{e.5.7}) holds when $\phi (x)$ is a linear combination of
matrix elements of finite-dimensional representations of $K.$

We now consider general $\phi .$ Note that the distribution of $\theta
_{u}^{-1}\theta _{v}$ is the same as that of $\theta _{v-u},$ namely, the
heat kernel measure $\rho _{\left( v-u\right) s}.$ This measure has a smooth
strictly positive density with respect to Haar measure on $K,$ which by
compactness is bounded and bounded away from zero. Thus by the Peter-Weyl
theorem, there exist functions $\phi _{n}$ which are finite linear
combinations of matrix entries such that $\phi _{n}\rightarrow \phi $ in $%
L^{2}\left( K,\rho _{\left( u-v\right) s}\right) $ and thus $\phi
_{n}(\theta _{u}^{-1}\theta _{v})\rightarrow \phi (\theta _{u}^{-1}\theta
_{v})$ in $L^{2}(\mathcal{\bar{A}},\tilde{P}_{s}).$ The smoothness of the
heat kernel shows that the map $\phi \rightarrow \left( X_{k_{1}}\cdots
X_{k_{n}}e^{(v-\tau )s\Delta _{K}/2}\phi \right) \left( e\right) $ is a
continuous linear functional on $L^{2}\left( K,\rho _{\left( v-u\right)
s}\right) .$ So passing to the limit gives the lemma in general. \qed

We have the following holomorphic analog of the previous lemma. Recall that $%
A_{s,t}^{K_{\mathbb{C}}}=\left( s-t/2\right) \Sigma X_{k}^{2}+\frac{t}{2}%
\Sigma Y_{k}^{2},$ where $Y_{k}=JX_{k}.$

\begin{lemma}
\label{l.5.8}Suppose that $\Phi $ is a holomorphic function on $K_{\mathbb{C}%
}$ which is a finite linear combination of matrix entries. Then the
holomorphic Wiener chaos expansion of $\Phi \left( \left( \theta _{u}^{%
\mathbb{C}}\right) ^{-1}\theta _{v}^{\mathbb{C}}\right) $ is 
\begin{equation}
\Phi \left( \left( \theta _{u}^{\mathbb{C}}\right) ^{-1}\theta _{v}^{\mathbb{%
C}}\right) =\sum_{n=0}^{\infty }\sum_{k_{1},\cdots ,k_{n}=1}^{\dim \frak{k}%
}\int_{\Delta _{n}(u,v)}\beta _{k_{1},\cdots ,k_{n}}\,dc_{\tau
_{1}}^{k_{1}}\cdots \,dc_{\tau _{n}}^{k_{n}},  \label{e.5.10}
\end{equation}
where 
\begin{equation}
\beta _{k_{1},\cdots ,k_{n}}=\left( X_{k_{1}}\cdots
X_{k_{n}}e^{(v-u)A_{s,t}^{K_{\mathbb{C}}}/2}\Phi \right) (e)  \label{e.5.11}
\end{equation}
and $\Delta _{n}(u,v)=\{(\tau _{1},\dots ,\tau _{n})\left| u\leq \tau
_{1}\leq \tau _{2}\leq \cdots \leq \tau _{n}\leq v\right. \}.$
\end{lemma}

\textit{Proof.} The argument is very similar to the preceding one. If $\Phi
\left( g\right) =\left\langle h,\pi \left( g\right) w\right\rangle ,$ where $%
\pi $ is a finite-dimensional holomorphic representation of $K_{\mathbb{C}},$
then we set 
\begin{equation*}
u\left( \tau ,g\right) =e^{\left( v-\tau \right) A_{s,t}^{K_{\mathbb{C}%
}}/2}\Phi \left( g\right) =\left\langle h,\pi \left( g\right) e^{\left(
v-\tau \right) \pi \left( A_{s,t}^{K_{\mathbb{C}}}\right) }w\right\rangle .
\end{equation*}
Again it may be verified that the second and third expressions are equal
(see for example \cite{H1}), with the second interpreted as convolution
against the relevant heat kernel. Then $u\left( \tau ,g\right) $ is
holomorphic in $g$ for each $\tau .$ Thus $Y_{k}u=iX_{k}u,$ and the last two
terms in (\ref{e.5.6}) combine into one term involving integration against $%
da_{\tau }+idb_{t}=dc_{\tau }.$ Iteration then proceeds as in the real case.
The remainder estimate is similar as well after using the standard fact that
the $L^{2}$ norm of $\pi \left( \left( \theta _{u}^{\mathbb{C}}\right)
^{-1}\theta _{v}^{\mathbb{C}}\right) $ is bounded uniformly for $0\leq u\leq
v\leq 1.$ The lemma in fact holds for all holomorphic $\Phi $ for which $%
\Phi \left( \left( \theta _{u}^{\mathbb{C}}\right) ^{-1}\theta _{v}^{\mathbb{%
C}}\right) $ is square-integrable, but we will not require this.\qed

\bigskip

\textit{Proof of Theorem \ref{t.5.2}}. Let $\phi \in L^{2}(K)$ and $f=\phi
(\theta _{u}^{-1}\theta _{v})\in L^{2}(\mathcal{\bar{A}},\tilde{P}_{s}).$ We
will show that there exists a unique holomorphic function $\Phi $ such that $%
\tilde{S}_{s,t}f\left( C\right) =\Phi \left( \left( \theta _{u}^{\mathbb{C}%
}\right) ^{-1}\theta _{v}^{\mathbb{C}}\right) $ and such that $\left. \Phi
\right| _{K}=e^{\left( v-u\right) t\Delta _{K}/2}\phi .$ Theorem \ref{t.5.2}
is the special case, $u=0,$ $v=1.$

By standard density arguments it suffices to prove the theorem in the case
where $\phi (x)=\langle h,\pi (x)w\rangle $ with $\pi :K_{\mathbb{C}%
}\rightarrow End(W)$ being a finite-dimensional holomorphic representation
of $K_{\mathbb{C}}$, $h\in W^{\ast }$ and $w\in W.$ In this case the
holomorphic function in the statement of Theorem \ref{t.5.2} is 
\begin{equation*}
\Phi (g)=(e^{\left( v-u\right) t\Delta _{K}/2}\phi )(g)=\left\langle h,\pi
(g)e^{\left( v-u\right) t\pi (\Delta _{K})/2}w\right\rangle .
\end{equation*}

By Lemma \ref{l.5.7} and Theorem \ref{t.4.7}, 
\begin{equation*}
\tilde{S}_{s,t}(\phi (\theta _{u}^{-1}\theta _{v}))=\sum_{n=0}^{\infty
}\sum_{k_{1},\cdots ,k_{n}=1}^{\dim \frak{k}}\int_{\Delta _{n}(u,v)}\alpha
_{k_{1},\cdots ,k_{n}}\,dc_{\tau _{1}}^{k_{1}}\cdots \,dc_{\tau
_{n}}^{k_{n}},
\end{equation*}
where $\alpha _{k_{1},\cdots ,k_{n}}=\left( X_{k_{1}}\cdots
X_{k_{n}}e^{(v-u)s\Delta _{K}/2}\phi \right) (e).$ Lemma \ref{l.5.8} will
now give the theorem, provided that the coefficients $\beta _{k_{1},\cdots
,k_{n}}$ in Lemma \ref{l.5.8} with $\Phi $ as above coincide with the
coefficients $\alpha _{k_{1},\cdots ,k_{n}}.$ So we require that 
\begin{equation}
\left( X_{k_{1}}\cdots X_{k_{n}}e^{(v-u)s\Delta _{K}/2}\right) \phi
(e)=\left( X_{k_{1}}\cdots X_{k_{n}}e^{(v-u)A_{s,t}/2}e^{\left( v-u\right)
t\Delta _{K}/2}\phi \right) (e).  \label{e.5.12}
\end{equation}
But $\Phi =e^{\left( v-u\right) t\Delta _{K}/2}\phi $ is holomorphic, and $%
A_{s,t}\Phi =\left( s-t\right) \Delta _{K}\Phi $ on holomorphic functions,
so (\ref{e.5.12}) holds. \qed

\textit{Proof of Theorem \ref{t.5.4}}. Since we are assuming that $K$ is
compact, $\rho _{s}\left( x\right) $ will be bounded and bounded away from
zero. Thus the $L^{2}$ norm with respect to $\rho _{s}\left( x\right) dx$ is
finite if and only if the $L^{2}$ norm with respect to Haar measure is
finite. It is an easy and standard result that $\rho _{s}\left( x\right) $
converges uniformly to the constant function 1, which establishes the first
limit in the theorem.

Now, let $\delta _{K}$ denote Haar measure on $K,$ viewed as a measure on $%
K_{\mathbb{C}}.$ Then since $A_{s,t}$ is a left-invariant operator, we have
formally 
\begin{equation*}
\nu _{s,t}=e^{A_{s,t}}\left( \delta _{K}\right) .
\end{equation*}
But the two terms in the definition of $A_{s,t}$ commute, so 
\begin{equation*}
\nu _{s,t}=e^{t/2\sum JX_{k}^{2}}e^{\left( s-t/2\right) \sum
X_{k}^{2}}\left( \delta _{K}\right) .
\end{equation*}
Since $\delta _{K}$ is $K$-invariant, the exponential involving $\sum
X_{k}^{2}$ has no effect, and the $s$-dependence vanishes.

The equivalence of square-integrability with respect to $\mu _{s,t}$ and $%
\nu _{t}$ is implied by the ``averaging lemma'' \cite[Lem. 11]{H1}. This is
stated in \cite{H1} for the case $s=t,$ but the same proof applies in
general. Using the commutativity of $\sum X_{k}^{2}$ and $\sum \left(
JX_{k}\right) ^{2},$ 
\begin{equation*}
\mu _{s,t}=e^{\left( s-t\right) \sum X_{k}^{2}}\,e^{t/2\sum
JX_{k}^{2}}\,e^{t/2\sum X_{k}^{2}}\left( \delta _{e}\right) \text{\quad }%
\forall \quad s>t.
\end{equation*}
Thus 
\begin{equation*}
\mu _{s,t}\left( g\right) =\int_{K}\mu _{t,t}\left( gx^{-1}\right) \rho
_{s-t}\left( x\right) \,dx,
\end{equation*}
from which it follows that $\lim_{s\rightarrow \infty }\mu _{s,t}\left(
g\right) =\nu _{t}\left( g\right) $ for all $g.$ Furthermore, applying the
averaging lemma to $\mu _{t,t}$ we see that for all $s>t,$ $\mu _{s,t}\left(
g\right) $ is dominated by a constant (independent of $s$) times $\nu
_{t}\left( g\right) .$ So Dominated Convergence gives the second limit in
the theorem. The methods of \cite{H1} are sufficient to make all of this
rigorous. \qed

\section{General functions of the parallel transport \label{s.6}}

Recall that $\theta $ and $\theta ^{\mathbb{C}}$ are the It\^{o} maps
satisfying the stochastic differential equations (\ref{e.5.1}) and (\ref
{e.5.2}) of Definition \ref{d.5.1}.

\begin{definition}
\label{d.6.1}Let $W\left( K\right) $ denote the group of continuous path $x$
with values in $K,$ with time interval $\left[ 0,1\right] $ and satisfying $%
x_{0}=e.$ Define $W\left( K_{\mathbb{C}}\right) $ similarly. Let $\tilde{\rho%
}_{s}$ be the \textbf{Wiener measure on} $W\left( K\right) ,$ that is, the
law of the process $\theta _{\cdot }\left( A\right) ,$ where $A$ is
distributed as $\tilde{P}_{s}.$ Similarly let $\tilde{\mu}_{s,t}$ be the 
\textbf{Wiener measure on} $W\left( K_{\mathbb{C}}\right) ,$ the law of the
process $\theta _{\cdot }^{\mathbb{C}}\left( C\right) ,$ where $C$ is
distributed as $\tilde{M}_{s,t}.$

For each partition $\mathcal{P}=\{0=\tau _{0}<\tau _{1}<\cdots <\tau _{n}=1\}
$ of $[0,1],$ let $\theta _{\mathcal{P}}=\left( \theta _{\tau _{1}},\cdots
,\theta _{\tau _{n}}\right) ,$ $K^{\mathcal{P}}=K^{n}$ and $\rho _{s}^{%
\mathcal{P}}$ denote the law of $\theta _{\mathcal{P}},$ a probability
measure on $K^{\mathcal{P}}.$ Define $\theta _{\mathcal{P}}^{\mathbb{C}},$ $%
K_{\mathbb{C}}^{\mathcal{P}}$ and $\mu _{s,t}^{\mathcal{P}}$ similarly.
\end{definition}

As in Theorem \ref{t.5.3}, let $\rho _{s}$denote the measure $\rho _{s}^{%
\mathcal{P}}$ on $K,$ where $\mathcal{P}=\left\{ 0,1\right\} .$ This measure
has a smooth strictly positive density with respect to Haar measure, which
we also call $\rho _{s}.$ If $\mathbf{x}=\left( x_{1},\cdots ,x_{n}\right) $
is a typical element in $K^{\mathcal{P}},$ then (as is well known) 
\begin{equation}
d\rho _{s}^{\mathcal{P}}(\mathbf{x})=\prod_{i=1}^{n}\rho _{s\Delta _{i}\tau
}(x_{i-1}^{-1}x_{i})\,dx_{i},  \label{e.6.1}
\end{equation}
where $\Delta _{i}\tau =\tau _{i}-\tau _{i-1}.$ As for $\rho _{s},$ we will
also use $\rho _{s}^{\mathcal{P}}$ to denote the density on the right side
of (\ref{e.6.1}).

If $\mathcal{P}$ is a partition and $x\in W\left( K\right) $ and $g\in W(K_{%
\mathbb{C}}),$ let $x_{\mathcal{P}}=\left( x_{\tau _{1}},\cdots ,x_{\tau
_{n}}\right) \in K^{\mathcal{P}}$ and $g_{\mathcal{P}}=\left( g_{\tau
_{1}},\cdots ,g_{\tau _{n}}\right) \in K_{\mathbb{C}}^{\mathcal{P}}.$

\begin{definition}
\label{d.6.2}Let $\mathcal{P}=\{0=\tau _{0}<\tau _{1}<\cdots <\tau _{n}=1\}$
be a partition of $[0,1].$ A function $f\in L^{2}(W(K),\tilde{\rho}_{s})$ is
said to be a \textbf{cylinder function based on }$\mathcal{P}$ if $f$ is the
form $f(x)=\phi (x_{\mathcal{P}})$ for some measurable function $\phi :K^{%
\mathcal{P}}\rightarrow \mathbb{C}.$

Similarly, we say a function $F\in L^{2}(W(K_{\mathbb{C}}),\tilde{\mu}_{s,t})
$ is a \textbf{holomorphic cylinder function based on }$\mathcal{P}$
provided that $F$ is of the form $F(g)=\Phi (g_{\mathcal{P}}),$ where $\Phi $
is a holomorphic function on $K_{\mathbb{C}}^{\mathcal{P}}.$

The \textbf{holomorphic subspace of }$L^{2}\left( W\left( K_{\mathbb{C}%
}\right) ,\tilde{\mu}_{s,t}\right) ,$ denoted $\mathcal{H}L^{2}\left(
W\left( K_{\mathbb{C}}\right) ,\tilde{\mu}_{s,t}\right) ,$ is the $L^{2}$
closure of the $L^{2}$ holomorphic cylinder functions.
\end{definition}

\begin{theorem}
\label{t.6.3}There exists a unique isometric isomorphism $\tilde{B}%
_{s,t}:L^{2}\left( W\left( K\right) ,\tilde{\rho}_{s}\right) \rightarrow 
\mathcal{H}L^{2}\left( W\left( K_{\mathbb{C}}\right) ,\widetilde{\mu }%
_{s,t}\right) $ such that for all partitions $\mathcal{P}$ and all cylinder
functions 
\begin{equation*}
f\left( x\right) =\phi \left( x_{\mathcal{P}}\right) 
\end{equation*}
based on $\mathcal{P},$ $\tilde{B}_{s,t}$ is of the form 
\begin{equation*}
\tilde{B}_{s,t}f\left( g\right) =\Phi \left( g_{\mathcal{P}}\right) 
\end{equation*}
where $\Phi $ is holomorphic on $K_{\mathbb{C}}^{\mathcal{P}}$ which is
determined uniquely by the condition that 
\begin{equation}
\Phi (\mathbf{g})=\int_{K^{\mathcal{P}}}\rho _{t}^{\mathcal{P}}\left( 
\mathbf{gx}^{-1}\right) \phi (\mathbf{x})\,d\mathbf{x}  \label{e.6.2}
\end{equation}
for all $\mathbf{g}\in K^{\mathcal{P}}.$ If $F\in \mathcal{H}L^{2}\left(
W\left( K_{\mathbb{C}}\right) ,\tilde{\mu}_{s,t}\right) $ is a holomorphic
cylinder function, then $\tilde{B}_{s,t}^{-1}F$ is a cylinder function based
on the same partition.

The following diagram is well defined and commutative, and all maps are
one-to-one, onto, and isometric. 
\begin{equation*}
\begin{array}{ccc}
L^{2}(\mathcal{\bar{A}},\tilde{P}_{s}) & \overset{\tilde{S}_{s,t}}{%
\rightarrow } & \mathcal{H}L^{2}(\mathcal{\bar{A}}_{\mathbb{C}},\tilde{M}%
_{s,t}) \\ 
\uparrow \theta  &  & \uparrow \theta ^{\mathbb{C}} \\ 
L^{2}\left( W\left( K\right) ,\tilde{\rho}_{s}\right)  & \overset{\tilde{B}%
_{s,t}}{\rightarrow } & \mathcal{H}L^{2}\left( W\left( K_{\mathbb{C}}\right)
,\tilde{\mu}_{s,t}\right) ,
\end{array}
\end{equation*}
where $\theta $ and $\theta ^{\mathbb{C}}$ are being used here to denote the
unitary maps, $f\in L^{2}\left( W\left( K\right) ,\tilde{\rho}_{s}\right)
\rightarrow f\circ \theta \in L^{2}(\mathcal{\bar{A}},\tilde{P}_{s})$ and $%
F\in L^{2}\left( W\left( K_{\mathbb{C}}\right) ,\tilde{\mu}_{s,t}\right)
\rightarrow F\circ \theta ^{\mathbb{C}}\in L^{2}(\mathcal{\bar{A}}_{\mathbb{C%
}},\tilde{M}_{s,t}),$ respectively.
\end{theorem}

Note that the vertical arrow on the right side is not obviously well
defined, since composition with $\theta ^{\mathbb{C}}$ does not take
cylinder functions to cylinder functions. Theorems \ref{t.5.2} and \ref
{t.5.3} are special cases in which the partition is $\mathcal{P}=\left\{
0=\tau _{0}<\tau _{1}=1\right\} .$ The $s=t$ case of this theorem is part of
Theorem 17 of \cite{HS}. The theorem implies that if $f\in L^{2}(\mathcal{%
\bar{A}},\tilde{P}_{s})$ is a function of the parallel transport $\theta $
at a finite number of times $\tau _{1},\cdots ,\tau _{n},$ then $\tilde{S}%
_{s,t}f$ is a function of the complex parallel transport at the same times $%
\tau _{1},\cdots ,\tau _{n}.$

The heuristic argument for Theorem \ref{t.5.2} applies just as well when
applying $\tilde{S}_{s,t}$ to a function of the form $\phi \left( \theta
_{\tau _{1}}\left( A\right) ,\cdots \theta _{\tau _{n}}\left( A\right)
\right) ,$ and so provides a heuristic argument for the commutative diagram
in Theorem \ref{t.6.3}. Isometricity of $\tilde{B}_{s,t}$ would then follow
from the isometricity of $\tilde{S}_{s,t}.$ Appendix A provides another
heuristic argument for Theorem \ref{t.6.3}. See especially Example \ref
{ex.8.9} and Theorem \ref{t.8.11}. The actual proof of Theorem \ref{t.6.3}
will be by reduction to Theorem \ref{t.5.2}, using the following result.
(See also \cite[Prop. 3.3.1]{AHS}.)

\begin{proposition}[Factorization Proposition]
\label{p.6.4}For $0\leq l<m\leq 1,$ let $\mathcal{F}_{[l,m]}$ be the $\sigma 
$-algebra in $\mathcal{\bar{A}}$ generated by $a_{s}-a_{l},$ with $s\in %
\left[ l,m\right] .$ Suppose $l\in (0,1)$, $f\in L^{2}(\mathcal{\bar{A}},%
\mathcal{F}_{[0,l]},\tilde{P}_{s})$ and $g\in L^{2}(\mathcal{\bar{A}},%
\mathcal{F}_{[l,1]},\tilde{P}_{s})$. Then $fg\in L^{2}(\mathcal{\bar{A}},%
\tilde{P}_{s})$ and 
\begin{equation*}
\tilde{S}_{s,t}(fg)=\tilde{S}_{s,t}(f)\tilde{S}_{s,t}(g).
\end{equation*}
\end{proposition}

\textit{Proof.} First suppose that 
\begin{equation*}
f\left( A\right) =\phi \left( \left\langle e_{1},A\right\rangle ,\cdots
\left\langle e_{d},A\right\rangle \right) \text{ and }g\left( A\right) =\psi
\left( \left\langle u_{1},A\right\rangle ,\cdots \left\langle
u_{k},A\right\rangle \right)
\end{equation*}
where $\left\{ e_{1},\cdots ,e_{d}\right\} $ and $\left\{ u_{1},\cdots
,u_{k}\right\} $ are orthonormal subsets of $\mathcal{A}$ which are
contained in $\mathcal{A}\cap C^{\infty }(\left[ 0,c\right] ;\frak{k)}$ and $%
\mathcal{A}\cap C^{\infty }(\left[ c,1\right] ;\frak{k)}$ respectively. Then 
\begin{eqnarray*}
\langle e_{i},A\rangle &=&-\int_{0}^{c}\left\langle e_{i}^{\prime }(\tau
),a_{\tau }(A)\right\rangle \,d\tau , \\
\langle u_{i},A\rangle &=&-\int_{c}^{1}\left\langle u_{i}^{\prime }(\tau
),a_{\tau }(A)-a_{c}(A)\right\rangle \,d\tau .
\end{eqnarray*}
Approximating the integrals by Riemann sums, one shows that these
expressions are $\mathcal{F}_{[0,c]}$- and $\mathcal{F}_{[c,1]}$-measurable,
respectively. Therefore $f$ and $g$ are cylinder functions that are $%
\mathcal{F}_{[0,c]}$- and $\mathcal{F}_{[c,1]}$-measurable, respectively.
Since each $e_{i}$ is orthogonal to each $u_{j},$ $\left\{ e_{1},\cdots
,e_{d},u_{1},\cdots ,u_{k}\right\} $ is an orthonormal set and $\mathcal{F}%
_{[0,c]}$ and $\mathcal{F}_{[c,1]}$ are $\tilde{P}_{t}$ independent $\sigma $%
-fields. The heat kernel on $\mathbb{R}^{d+k}$ factors, so applying the
finite-dimensional transform $S_{s,t}$ (\ref{e.3.1}) to the function $\phi
\left( x\right) \psi \left( y\right) $ gives $S_{s,t}\left( \phi \right)
S_{s,t}\left( \psi \right) .$ Hence $\tilde{S}_{s,t}(fg)=\tilde{S}_{s,t}(f)%
\tilde{S}_{s,t}(g).$

For general $f\in L^{2}(\mathcal{\bar{A}}_{\mathbb{C}},\mathcal{F}_{[0,c]},%
\tilde{P}_{s})$ and $g\in L^{2}(\mathcal{\bar{A}}_{\mathbb{C}},\mathcal{F}%
_{[c,1]},\tilde{P}_{s}),$ choose cylinder functions $f_{n}$ and $g_{n}$ as
above such that $f_{n}\rightarrow f$ in $L^{2}(\mathcal{\bar{A}}_{\mathbb{C}%
},\mathcal{F}_{[0,c]},\tilde{P}_{s})$ and $g_{n}\rightarrow g$ in $L^{2}(%
\mathcal{\bar{A}}_{\mathbb{C}},\mathcal{F}_{[c,1]},\tilde{P}_{s}).$ Because
of the independence of $\mathcal{F}_{[0,c]}$ and $\mathcal{F}_{[c,1]}$, $%
f_{n}g_{n}\rightarrow fg$ in $L^{2}(\mathcal{\bar{A}}_{\mathbb{C}},\mathcal{F%
}_{[0,1]},\tilde{P}_{s}).$ Furthermore, it is easily seen that $\tilde{S}%
_{s,t}\left( f_{n}\right) $ and $\tilde{S}_{s,t}\left( g_{n}\right) $ are
measurable with respect to independent $\sigma $-algebras in $\mathcal{\bar{A%
}}_{\mathbb{C}}.$ Thus by the isometry property of $\tilde{S}_{s,t},$%
\begin{equation*}
\tilde{S}_{s,t}\left( fg\right) =\lim_{n\rightarrow \infty }\tilde{S}%
_{s,t}\left( f_{n}g_{n}\right) =\lim_{n\rightarrow \infty }\tilde{S}%
_{s,t}(f_{n})\tilde{S}_{s,t}(g_{n})=\tilde{S}_{s,t}(f)\tilde{S}_{s,t}\left(
g\right) .
\end{equation*}
\qed

\textit{Proof of Theorem \ref{t.6.3}}. Our strategy is to use Theorem \ref
{t.5.2} and the Factorization Proposition to compute $\tilde{S}_{s,t}$ on
functions of the form $\phi \left( \theta _{\tau _{1}},\cdots ,\theta _{\tau
_{n}}\right) .$ The result shows that $\tilde{S}_{s,t}\left( f\circ \theta
\right) =\left( \tilde{B}_{s,t}f\right) \circ \theta _{\mathbb{C}}$ and
hence that $\tilde{B}_{s,t}$ is isometric because $\tilde{S}_{s,t}$ is
isometric. A surjectivity argument for $\tilde{B}_{s,t}$ then establishes
the well-definedness and commutativity of the diagram.

It is easily seen that there is at most one isometric isomorphism having the
given action on cylinder functions. We will now prove the existence of $%
\tilde{B}_{s,t}$ and establish the commutative diagram in the Theorem.

Suppose $f\in L^{2}(\mathcal{\bar{A}},\tilde{P}_{s})$ is of the form 
\begin{equation*}
f=\psi _{1}\left( \theta _{\tau _{1}}\right) \psi _{2}\left( \theta _{\tau
_{1}}^{-1}\theta _{\tau _{2}}\right) \cdots \psi _{n}\left( \theta _{\tau
_{n-1}}^{-1}\theta _{\tau _{n}}\right)
\end{equation*}
where $\psi _{1},\cdots ,\psi _{n}$ are functions on $K.$ Then $\psi
_{i}\left( \theta _{\tau _{i-1}}^{-1}\theta _{\tau _{i}}\right) $ is $%
\mathcal{F}_{\left[ \tau _{i-1},\tau _{i}\right] }$-measurable. So by the
strong form of Theorem \ref{t.5.2} proved in Section \ref{s.5.3} and by the
Factorization Proposition \ref{p.6.4} (extended by induction to hold for
products of $n$ factors), 
\begin{equation}
\tilde{S}_{s,t}f=\Psi _{1}\left( \theta _{\tau _{1}}^{\mathbb{C}}\right)
\Psi _{2}\left( \left( \theta _{\tau _{1}}^{\mathbb{C}}\right) ^{-1}\theta
_{\tau _{2}}^{\mathbb{C}}\right) \cdots \Psi _{n}\left( \left( \theta _{\tau
_{n-1}}^{\mathbb{C}}\right) ^{-1}\theta _{\tau _{n}}^{\mathbb{C}}\right) ,
\label{e.6.3}
\end{equation}
where $\Psi _{i}$ is a holomorphic function on $K_{\mathbb{C}}$ whose
restriction to $K$ is $e^{\left( \tau _{i}-\tau _{i-1}\right) t\Delta
_{K}/2}\psi _{i}.$

Now suppose $f\in L^{2}(\mathcal{\bar{A}},\tilde{P}_{s})$ is any function of
the form 
\begin{equation}
f=\psi \left( \theta _{\tau _{1}},\theta _{\tau _{1}}^{-1}\theta _{\tau
_{2}},\cdots ,\theta _{\tau _{n-1}}^{-1}\theta _{\tau _{n}}\right)
\label{e.sf1}
\end{equation}
with $\psi \in L^{2}\left( K^{n},\rho _{s\Delta _{1}\tau }\times \cdots
\times \rho _{s\Delta _{n}\tau }\right) .$ Then we claim that 
\begin{equation}
\tilde{S}_{s,t}f=\Psi \left( \theta _{\tau _{1}}^{\mathbb{C}},\left( \theta
_{\tau _{1}}^{\mathbb{C}}\right) ^{-1}\theta _{\tau _{2}}^{\mathbb{C}%
},\cdots ,\left( \theta _{\tau _{n-1}}^{\mathbb{C}}\right) ^{-1}\theta
_{\tau _{n}}^{\mathbb{C}}\right)  \label{e.sf2}
\end{equation}
where $\Psi $ is the unique holomorphic function on $K_{\mathbb{C}}^{n}$
whose restriction to $K^{n}$ is given by 
\begin{equation}
\Psi \left( a_{1},\cdots ,a_{n}\right) =\int_{K}\cdots \int_{K}\rho
_{t\Delta _{1}\tau }\left( a_{1}b_{1}^{-1}\right) \cdots \rho _{t\Delta
_{n}\tau }\left( a_{n}b_{n}^{-1}\right) \psi \left( b_{1},\cdots
,b_{n}\right) \,db_{1}\,db_{2}\,\cdots \,db_{n}.  \label{e.sf3}
\end{equation}
If $\psi $ is a product function then this assertion is simply (\ref{e.6.3}%
); since linear combinations of product functions are dense in $L^{2}\left(
K^{n},\rho _{s\Delta _{1}\tau }\times \cdots \times \rho _{s\Delta \tau
_{n}}\right) $ the assertion holds in general. (Recall (\ref{e.6.1}).)

The equations (\ref{e.sf1})-(\ref{e.sf3}) express the action of $\tilde{S}%
_{s,t}$ on cylinder functions in terms of the ``incremental coordinates'' $%
\theta _{\tau _{i-1}}^{-1}\theta _{\tau _{i}}.$ We wish to have in addition
a formula for the action of $\tilde{S}_{s,t}$ on cylinder functions in terms
of the ``direct coordinates'' $\theta _{\tau _{1}},\cdots ,\theta _{\tau
_{n}}.$ (See the proof of Theorem 3 in \cite[Sect. 3.2]{HS}.) So suppose $f$
is any cylinder function based on the partion $\mathcal{P}$: 
\begin{equation}
f=\phi \left( \theta _{\tau _{1}},\cdots ,\theta _{\tau _{n}}\right)
\label{e.sf4}
\end{equation}
with $\phi \in L^{2}\left( K^{\mathcal{P}},\rho _{s}^{\mathcal{P}}\right) .$
Then we claim that 
\begin{equation}
\tilde{S}_{s,t}f=\Phi \left( \theta _{\tau _{1}}^{\mathbb{C}},\theta _{\tau
_{2}}^{\mathbb{C}},\cdots ,\theta _{\tau _{n}}^{\mathbb{C}}\right) ,
\label{e.sf5}
\end{equation}
where $\Phi $ is the unique holomorphic function on $K_{\mathbb{C}}^{%
\mathcal{P}}$ such that 
\begin{equation}
\Phi (\mathbf{g})=\int_{K^{\mathcal{P}}}\rho _{t}^{\mathcal{P}}\left( 
\mathbf{gx}^{-1}\right) \phi (\mathbf{x})\,d\mathbf{x},  \label{e.sf6}
\end{equation}
for all $\mathbf{g}\in K^{\mathcal{P}}.$ Explicitly, by (\ref{e.6.1}) this
means that 
\begin{equation}
\Phi \left( g_{1},\cdots ,g_{n}\right) =\int_{K^{n}}\left[
\prod_{i=1}^{n}\rho _{t\Delta _{i}\tau }\left( \left(
g_{i-1}x_{i-1}^{-1}\right) ^{-1}g_{i}x_{i}^{-1}\right) \right] \phi \left(
x_{1},\cdots ,x_{n}\right) \,dx_{1}\cdots dx_{n}.  \label{e.sf8}
\end{equation}
(Here $g_{0}=x_{0}=e.$)

To verify this, we note that $f$ can be expressed in the form (\ref{e.sf1})
with 
\begin{equation*}
\psi \left( a_{1},\cdots ,a_{n}\right) =\phi \left( a_{1},a_{1}a_{2},\cdots
,a_{1}a_{2}\cdots a_{n}\right) ,
\end{equation*}
in which case $\phi $ can be expressed in terms of $\psi $ by 
\begin{equation*}
\phi \left( x_{1},\cdots ,x_{n}\right) =\psi \left(
x_{1},x_{1}^{-1}x_{2},\cdots ,x_{n-1}^{-1}x_{n}\right) .
\end{equation*}
Thus $\tilde{S}_{s,t}f$ is given by (\ref{e.sf2}). But then $\tilde{S}%
_{s,t}f $ can be expressed in the form (\ref{e.sf5}), where 
\begin{equation*}
\Phi \left( g_{1},\cdots ,g_{n}\right) =\Psi \left(
g_{1},g_{1}^{-1}g_{2},\cdots ,g_{n-1}^{-1}g_{n}\right) ,
\end{equation*}
with $\Psi $ given in (\ref{e.sf3}). We need only verify the relationship
between $\Phi $ and $\phi .$ Putting together the definitions we get 
\begin{equation}
\Phi \left( g_{1},\cdots ,g_{n}\right) =\int_{K^{n}}\left[ \prod \rho
_{t\Delta _{i}\tau }\left( g_{i-1}^{-1}g_{i}b_{i}^{-1}\right) \right] \psi
\left( b_{1},\cdots ,b_{n}\right) \,db_{1}\cdots db_{n}.  \label{e.sf9}
\end{equation}
We then make successive changes of variable $\left( x_{1},\cdots
,x_{n}\right) =\left( b_{1},b_{1}b_{2},\cdots ,b_{1}b_{2}\cdots b_{n}\right)
,$ so that $b_{i}=x_{i-1}^{-1}x_{i}.$ Since the heat kernel on $K$ is a
class function, we have 
\begin{equation*}
\rho _{t\Delta _{i}\tau }\left( g_{i-1}^{-1}g_{i}b_{i}^{-1}\right) =\rho
_{t\Delta _{i}\tau }\left( g_{i-1}^{-1}g_{i}x_{i}^{-1}x_{i-1}\right) =\rho
_{t\Delta _{i}\tau }\left( \left( g_{i-1}x_{i-1}^{-1}\right)
^{-1}g_{i}x_{i}^{-1}\right) .
\end{equation*}
Thus (\ref{e.sf9}) agrees with (\ref{e.sf8}).

Now recall that $\theta $ is a measure-theoretic isomorphism of $(\mathcal{%
\bar{A}},\tilde{P}_{s})$ with $\left( W\left( K\right) ,\tilde{\rho}%
_{s}\right) $ and $\theta ^{\mathbb{C}}$ is an isomorphism of $(\mathcal{%
\bar{A}}_{\mathbb{C}},\tilde{M}_{s,t})$ with $\left( W\left( K_{\mathbb{C}%
}\right) ,\tilde{\mu}_{s,t}\right) .$ So let us now \textit{define} a
transform $\tilde{B}_{s,t}$ by 
\begin{equation*}
\tilde{B}_{s,t}f=\left[ \tilde{S}_{s,t}\left( f\circ \theta \right) \right]
\circ \left( \theta ^{\mathbb{C}}\right) ^{-1}.
\end{equation*}
This is a well-defined isometric map of $L^{2}\left( W\left( K\right) ,%
\tilde{\rho}_{s}\right) $ into $L^{2}\left( W\left( K_{\mathbb{C}}\right) ,%
\tilde{\mu}_{s,t}\right) .$ But (\ref{e.sf4})-(\ref{e.sf8}) tell us that $%
\tilde{B}_{s,t}$ takes cylinder functions to holomorphic cylinder functions.
Since cylinder functions are dense in $L^{2}\left( W\left( K\right) ,\tilde{%
\rho}_{s}\right) $ then by the definition of $\mathcal{H}L^{2}\left( W\left(
K_{\mathbb{C}}\right) ,\tilde{\mu}_{s,t}\right) ,$ $\tilde{B}_{s,t}$ maps
into the holomorphic subspace. Equations (\ref{e.sf4})-(\ref{e.sf8}) assert
that $\tilde{B}_{s,t}$ satisfies (\ref{e.6.2}) of the theorem. By the
definition of $\tilde{B}_{s,t},$ the diagram in the theorem would commute if
the $\mathcal{H}L^{2}$'s were replaced by $L^{2}$'s.

It remains then to show that composition with $\theta ^{\mathbb{C}}$ takes $%
\mathcal{H}L^{2}\left( W\left( K_{\mathbb{C}}\right) ,\tilde{\mu}%
_{s,t}\right) $ onto $\mathcal{H}L^{2}(\mathcal{\bar{A}}_{\mathbb{C}},\tilde{%
M}_{s,t}),$ and that $\tilde{B}_{s,t}$ maps \textit{onto} $\mathcal{H}%
L^{2}\left( W\left( K_{\mathbb{C}}\right) ,\tilde{\mu}_{s,t}\right) .$ We
address the second point first. We compute $\tilde{B}_{s,t}$ in
``incremental coordinates'' using (\ref{e.sf1})-(\ref{e.sf3}). So we are
applying the heat equation in the ``increments'' $x_{i-1}^{-1}x_{i}.$ Since
in incremental coordinates both $\rho _{s}^{\mathcal{P}}$ and $\mu _{s,t}^{%
\mathcal{P}}$ factor as product measures, the surjectivity argument in
Theorem \ref{t.5.3} (using the method of \cite{H1}) applies to show that
every $L^{2}$ holomorphic cylinder function $\Phi \left( g_{\mathcal{P}%
}\right) $ comes from an $L^{2}$ cylinder function $\phi \left( x_{\mathcal{P%
}}\right) .$ For the details of this argument see again the proof of Theorem
3 in \cite{HS}. Since holomorphic cylinder functions are dense by definition
in $\mathcal{H}L^{2}\left( W\left( K_{\mathbb{C}}\right) ,\tilde{\mu}%
_{s,t}\right) ,$ $\tilde{B}_{s,t}$ is surjective.

Now that we know $\tilde{B}_{s,t}$ is surjective, we may show that
composition with $\theta ^{\mathbb{C}}$ takes the holomorphic subspace onto
the holomorphic subspace. If $F\in \mathcal{H}L^{2}\left( W\left( K_{\mathbb{%
C}}\right) ,\tilde{\mu}_{s,t}\right) ,$ then letting $f=\tilde{B}%
_{s,t}^{-1}F,$ we have 
\begin{equation*}
F\circ \theta ^{\mathbb{C}}=\left( \tilde{B}_{s,t}f\right) \circ \theta ^{%
\mathbb{C}}=\tilde{S}_{s,t}\left( f\circ \theta \right) .
\end{equation*}
But $\tilde{S}_{s,t}$ maps into the holomorphic subspace, so $F\circ \theta
^{\mathbb{C}}\in \mathcal{H}L^{2}(\mathcal{\bar{A}}_{\mathbb{C}},\tilde{M}%
_{s,t}).$ A similar argument using the surjectivity of $\tilde{S}_{s,t}$
shows that if $F\in \mathcal{H}L^{2}(\mathcal{\bar{A}}_{\mathbb{C}},\tilde{M}%
_{s,t})$ then $F\circ \left( \theta ^{\mathbb{C}}\right) ^{-1}\in \mathcal{H}%
L^{2}(\mathcal{\bar{A}}_{\mathbb{C}},\tilde{M}_{s,t}).$ \qed

\section{Another proof of Theorem \ref{t.5.2}\label{s.7}}

In this section, we sketch another method for proving Theorem \ref{t.5.2}.
The idea is to approximate $h\left( A\right) $ by cylinder functions and to
compute $\tilde{S}_{s,t}$ by first principles. So for a partition $\mathcal{P%
}$ we make a piecewise-linear approximation $a^{\mathcal{P}}$ to the
Brownian motion $a,$ and then apply the deterministic It\^{o} map to $a^{%
\mathcal{P}}.$ This gives an approximation $h^{\mathcal{P}}\left( A\right) $
to the holonomy $h\left( A\right) ,$ given explicitly by 
\begin{equation}
h^{\mathcal{P}}\left( A\right) =e^{\Delta _{1}\left( a\right) }e^{\Delta
_{2}\left( a\right) }\cdots e^{\Delta _{N}\left( a\right) },  \label{e.7.1}
\end{equation}
where $\Delta _{i}\left( a\right) =a_{\tau _{i}}-a_{\tau _{i-1}}$ is the $i$%
th increment of $a.$ Standard approximation results (e.g., \cite[Thm. VI.7.2]
{IW}) show that for any finite-dimensional irreducible representation $\pi $
of $K,$%
\begin{equation*}
\lim_{\left| \mathcal{P}\right| \rightarrow 0}\pi \circ h^{\mathcal{P}}=\pi
\circ h
\end{equation*}
in $L^{2}(\mathcal{\bar{A}},\tilde{P}_{s})$ where $\left| \mathcal{P}\right| 
$ is the partition size. The proof relies on the fact that $\pi \circ \theta 
$ satisfies its own (matrix-valued) stochastic differential equation.

Meanwhile, applying the Segal-Bargmann transform to the (matrix-valued)
cylinder function $\pi \circ h^{\mathcal{P}}$ gives 
\begin{equation}
\tilde{S}_{s,t}\left( \pi \circ h^{\mathcal{P}}\right) \left( C\right)
=\int_{\mathcal{\bar{A}}}\pi \left( e^{\Delta _{1}\left( c+a\right) }\cdots
e^{\Delta _{N}\left( c+a\right) }\right) \,d\tilde{P}_{s}\left( A\right)
\label{e.7.2}
\end{equation}
where $\pi $ now refers to the holomorphic extension of $\pi $ to $K_{%
\mathbb{C}}.$ Unfortunately the authors have not been able to find in the
current literature similar $L^{2}$ convergence results which are applicable
when $K$ is replaced by $K_{\mathbb{C}}.$ This is because for a non-compact
group the vector fields entering in the stochastic differential equations
are not bounded as is required in all references known to the authors.
Nevertheless, it is possible to show \cite{DHu} by essentially standard
arguments that 
\begin{equation*}
\lim_{\left| \mathcal{P}\right| \rightarrow 0}\pi \left( e^{\Delta
_{1}\left( c+a\right) }\cdots e^{\Delta _{N}\left( c+a\right) }\right) =\pi
\left( h_{\mathbb{C}}\left( A+C\right) \right)
\end{equation*}
in $L^{2}\left( \mathcal{\bar{A}}\times \mathcal{\bar{A}}_{\mathbb{C}},%
\tilde{P}_{s}\times \tilde{M}_{s,t}\right) ,$ where $h_{\mathbb{C}}\left(
A+C\right) =g_{1},$ $g_{\tau }$ being the $K_{\mathbb{C}}$-valued solution
to the stochastic differential equation, $dg_{\tau }=g_{\tau }\circ d(a+c).$

Because conditional expectations are contractive, we may interchange the $%
\left| \mathcal{P}\right| \rightarrow 0$ limit with the integral in (\ref
{e.7.2}) to find that 
\begin{equation}
\tilde{S}_{s,t}\left( \pi \circ h\right) \left( C\right) =\int_{\mathcal{%
\bar{A}}}\pi \left( h_{\mathbb{C}}\left( A+C\right) \right) \,d\tilde{P}%
_{t}\left( A\right) .  \label{e.7.3}
\end{equation}
To compute this integral we write out in It\^{o} form the matrix-valued
s.d.e. satisfied by $\theta ^{\pi }:=\pi \left( \theta ^{\mathbb{C}}\left(
A+C\right) \right) $: 
\begin{equation}
\theta _{\tau }^{\pi }=I+\int_{0}^{\tau }\theta _{\sigma }^{\pi
}\,da_{\sigma }^{\pi }+\frac{t}{2}\int_{0}^{\tau }\theta _{\sigma }^{\pi
}\pi \left( \Delta _{K}\right) \,d\sigma +\int_{0}^{\tau }\theta _{\sigma
}^{\pi }\,dc_{\sigma }^{\pi }+\frac{s-t}{2}\int_{0}^{\tau }\theta _{\sigma
}^{\pi }\pi \left( \Delta _{K}\right) \,d\sigma .  \label{.7.4}
\end{equation}
Here $a_{\tau }^{\pi }=\pi \left( a_{\tau }\right) $ and $c_{\tau }^{\pi
}=\pi \left( c_{\tau }\right) ,$ where as above $\pi (\xi )=d\pi (e^{t\xi
})/dt|_{t=0}$ for $\xi \in \frak{k}_{\mathbb{C}}.$ (We are using $\pi $ for
both the representation on $K_{\mathbb{C}}$ and the induced representation
on $\frak{k}_{\mathbb{C}}.)$ We now wish to take the expectation in $A$ with 
$C$ fixed. So let $\bar{\theta}_{\tau }^{\pi }\left( C\right) =\int \theta
_{\tau }^{\pi }\left( A+C\right) \,d\tilde{P}_{t}\left( A\right) .$ By the
Martingale property of stochastic integrals, the term $\int_{0}^{\tau
}\theta _{\sigma }^{\pi }\,da_{\sigma }^{\pi }$ integrates to zero so that 
\begin{eqnarray}
\bar{\theta}_{\tau }^{\pi }\left( C\right) &=&I+\frac{t}{2}\int_{0}^{\tau }%
\bar{\theta}_{\sigma }^{\pi }\pi \left( \Delta _{K}\right) \,d\sigma
+\int_{0}^{\tau }\bar{\theta}_{\sigma }^{\pi }\,dc_{\sigma }^{\pi }+\frac{s-t%
}{2}\int_{0}^{\tau }\bar{\theta}_{\sigma }^{\pi }\pi \left( \Delta
_{K}\right) \,d\sigma  \notag \\
&=&I+\frac{t}{2}\int_{0}^{\tau }\bar{\theta}_{\sigma }^{\pi }\pi \left(
\Delta _{K}\right) \,d\sigma +\int_{0}^{\tau }\bar{\theta}_{\sigma }^{\pi
}\circ dc_{\sigma }^{\pi }.  \label{e.7.5}
\end{eqnarray}
Since $\pi $ is assumed irreducible, $\pi \left( \Delta _{K}\right) $ is
simply a multiple of the identity, and so it is easily verified that the
(unique) solution to (\ref{e.7.5}) is 
\begin{equation*}
\bar{\theta}_{\tau }^{\pi }=e^{\tau t\pi \left( \Delta _{K}\right) /2}\pi
\left( \theta _{\tau }^{\mathbb{C}}\left( C\right) \right) .
\end{equation*}
Putting $\tau =1$ and recalling (\ref{e.7.3}) we see that Theorem \ref{t.5.2}
holds for the matrix entries of $\pi .$ Using the Peter-Weyl theorem it
follows that Theorem \ref{t.5.2} holds in general.

Note that the second of two proofs of \cite[Lem. 24]{HS} mistakenly applies 
\cite{IW} even on $K_{\mathbb{C}};$ that proof is therefore incomplete. The
convergence results in \cite{DHu} correct this error.

\section{Appendix A: Laplacians on $\mathcal{A}$ and $H(K)$\label{s.8}}

Let $\langle \cdot ,\cdot \rangle $ denote the unique bi-invariant
Riemannian metric on $K$ which agrees with the given Ad-$K$-invariant inner
product on $\frak{k}$. For $v\in T\left( K\right) $ we will simply write $%
|v|^{2}$ for $\langle v,v\rangle .$ Let $H(K)$ denote the finite-energy
paths in $K.$ Explicitly, $H(K)$ is the collection of absolutely continuous
paths $x:[0,1]\rightarrow K$ such that $x(0)=e$ and $\int_{0}^{1}|\dot{x}%
(\tau )|^{2}d\tau <\infty .$ It is well known that $H(K)$ is a Hilbert Lie
group under pointwise multiplication and that the map 
\begin{equation*}
(x,h)\in H(K)\times H(\frak{k)}\rightarrow R_{x\ast }h\in T\left( H\left(
K\right) \right)
\end{equation*}
is a trivialization of the tangent bundle of $H(K).$ (We are using $%
R_{x}:H(K)\rightarrow H(K)$ to denote right multiplication by $x.)$ This
trivialization induces a right-invariant Riemannian metric $(\cdot ,\cdot )$
on $H(K)$ given explicitly by 
\begin{equation*}
\left( R_{x\ast }h,R_{x\ast }h\right) =\int_{0}^{1}\left\langle \dot{h}(\tau
),\dot{h}(\tau )\right\rangle \,d\tau \quad \forall x\in H(K)\text{ and }%
h\in H(\frak{k).}
\end{equation*}
The following theorem appears (in a disguised form) in Theorem 3.14 and
Lemma 3.15 of Gross \cite{G}.

\begin{theorem}
\label{t.8.1}Let $\theta :\mathcal{A}\rightarrow H(K)$ denote the
deterministic solution to (\ref{e.5.1}), i.e., $d\theta _{\tau }(A)/d\tau
=\theta _{\tau }(A)A_{\tau }$ with $\theta _{0}(A)=e.$ Then $\theta $ is an
isometric isomorphism of infinite-dimensional Riemannian manifolds. In
particular, $H(K)$ is flat.
\end{theorem}

\textit{Proof.} For simplicity of notation, we will assume, without loss of
generality since $K$ is compact, that $K$ is a matrix group. In order to
compute the differential of $\theta ,$ let $x(\tau ,s)=\theta _{\tau
}(A+sB), $ $x(\tau )=x(\tau ,0)=\theta _{\tau }(A)$ and $h(\tau )=x^{\prime
}(\tau ,0)x(\tau )^{-1}.$ To simplify the exposition, first assume that $A$
and $B$ are $C^{1}.$ Then by smooth dependence of ordinary differential
equations on parameters, $h$ is differentiable and satisfies 
\begin{eqnarray*}
\dot{h}(\tau ) &=&\dot{x}^{\prime }(\tau ,0)x(\tau )^{-1}-x^{\prime }(\tau
,0)A(\tau )x(\tau )^{-1} \\
&=&\left. \frac{d}{ds}\right| _{s=0}\left( x(\tau ,s)\left[ A(\tau )+sB(\tau
)\right] \right) x(\tau )^{-1}-x^{\prime }(\tau ,0)A(\tau )x(\tau )^{-1} \\
&=&Ad_{x(\tau )}B(\tau )=Ad_{\theta _{\tau }(A)}B(\tau ).
\end{eqnarray*}
Here $\cdot $ indicates a derivative with respect to $\tau $ and $^{\prime }$
a derivative with respect to $s.$ The above equation says that 
\begin{equation}
\theta _{\ast }B_{A}=R_{\theta (A)\ast }\int_{0}^{\cdot }Ad_{\theta _{\tau
}(A)}B(\tau )d\tau .  \label{e.8.1}
\end{equation}
and therefore 
\begin{eqnarray}
(\theta _{\ast }B_{A},\theta _{\ast }B_{A}) &=&(x^{\prime }(\cdot
,0),x^{\prime }(\cdot ,0))=\int_{0}^{1}\langle \dot{h}(\tau ),\dot{h}(\tau
)\rangle \,d\tau  \notag \\
&=&\int_{0}^{1}|Ad_{x(\tau )}B(\tau )|^{2}d\tau =\int_{0}^{1}|B(\tau
)|^{2}d\tau =(B,B)_{\mathcal{A}}.  \label{e.8.2}
\end{eqnarray}
Since the map $\theta :\mathcal{A}\rightarrow H(K)$ is smooth as a mapping
of infinite-dimensional Hilbert manifolds (see for example \cite{P} or \cite
{D1}), both (\ref{e.8.1}) and (\ref{e.8.2}) extend by continuity to $A,B\in 
\mathcal{A}.$\qed

\begin{definition}[Directional Derivatives]
\label{d.8.2}For a function $F:\mathcal{A}\rightarrow \mathbb{C}$ and $%
A,B\in \mathcal{A}$ let 
\begin{equation*}
\partial _{B}F(A)=\left. \frac{d}{dt}\right| _{0}F(A+tB)
\end{equation*}
provided the limit exists. For a function $f:H(\frak{k})\rightarrow \mathbb{C%
}$ and $x\in H(K\frak{)}$ and $h\in H(\frak{k}),$ let 
\begin{equation*}
\partial _{h}f(x)=\left. \frac{d}{dt}\right| _{0}f\left( e^{th}x\right) ,
\end{equation*}
provided the limit exists, where $\left( e^{th}x\right) (s)=e^{th(s)}x(s)$
for all $s\in \lbrack 0,1].$
\end{definition}

\begin{definition}[Hessians]
\label{d.8.3}Suppose that both $F:\mathcal{A}\rightarrow \mathbb{C}$ and $%
f:H(K)\rightarrow \mathbb{C}$ are twice continuously differentiable at $A\in 
\mathcal{A}$ and $x\in K,$ respectively. The Hessians of $F$ and $f$ at $A$
and $x,$ respectively, are the quadratic forms $D^{2}F(A)$ on $\mathcal{A}$
and $D^{2}f(x)$ on $H(K)$ defined by 
\begin{equation*}
D^{2}F(A)(B_{1},B_{2})=\left( \partial _{B_{1}}\partial _{B_{2}}F\right) (A)
\end{equation*}
and 
\begin{equation*}
D^{2}f(x)(h_{1},h_{2})=\left( \partial _{h_{1}}\partial _{h_{2}}f\right) (x).
\end{equation*}
Notice that $D^{2}f(x)(h_{1},h_{2})$ is \textbf{not} symmetric in $h_{1}$
and $h_{2},$ a reflection of the fact that we did not use the Levi-Civita
connection on $H(K)$ to define $D^{2}f.$ See Remark \ref{r.8.6} below.
\end{definition}

Using the above notation, it would be natural to define $\Delta _{\mathcal{A}%
}F(A)$ and $\Delta _{H(K)}f(x)$ by 
\begin{equation}
\Delta _{\mathcal{A}}F(A)=\text{tr}_{\mathcal{A}}D^{2}F(A)\quad \text{and
\quad }\Delta _{H(K)}f(x)=\text{tr}_{H(\frak{k})}D^{2}f(x),  \label{e.8.3}
\end{equation}
\textit{provided} that the quadratic forms $D^{2}F(A)$ and $D^{2}f(x)$ were
trace-class. The above definitions certainly would be suitable if $F$ and $f$
were smooth cylinder functions on $\mathcal{A}$ and $H(K),$ respectively.
However, the definition in (\ref{e.8.3}) is too restrictive for our
purposes. In particular, we are interested in \textit{non-cylinder}
functions on $\mathcal{A}$ of the form $F=f\circ \theta ,$ where $f$ is a
cylinder function on $H(K).$ For such a function $F,$ $D^{2}F(A)$ is
typically not trace-class and hence the $\Delta _{\mathcal{A}}F$ would not
be defined. Definition \ref{d.8.5} below overcomes this problem by using a
more inclusive notion of the trace of $D^{2}F(A).$

\begin{notation}
\label{n.8.4}Let $\beta =\{e_{i}\}_{i=1}^{d}$ be an orthonormal basis for $%
\frak{k},$ $\Gamma $ an orthonormal basis of $L^{2}\left( [0,1];\mathbb{R}%
\right) ,$ and $\gamma $ the orthonormal basis of $H(\mathbb{R)}$ given by 
\begin{equation*}
\gamma =\{v(\cdot )=\int_{0}^{\cdot }V(\tau )d\tau |V\in \Gamma \}.
\end{equation*}
Notice that $\Gamma \beta :=\{Ve_{i}\left| V\in \Gamma \text{ and }%
i=1,\ldots ,d\right. \}$ and $\gamma \beta :=\{ve_{i}\left| v\in \gamma 
\text{ and }i=1,\ldots ,d\right. \}$ are orthonormal bases for $\mathcal{A}$
and $H(\frak{k})$ respectively.
\end{notation}

\begin{definition}[Laplacians]
\label{d.8.5}Let $F:\mathcal{A}\rightarrow \mathbb{C},$ $f:H(K)\rightarrow 
\mathbb{R},$ $A\in \mathcal{A}$ and $x\in K.$ Then 
\begin{equation}
(\Delta _{\mathcal{A}}F)(A)=\sum_{V\in \Gamma }\left( \sum_{i=1}^{\dim \frak{%
k}}D^{2}F(A)(Ve_{i},Ve_{i})\right) =\sum_{V\in \Gamma }\left(
\sum_{i=1}^{\dim \frak{k}}(\partial _{Ve_{i}}^{2}F)(A)\right)   \label{e.8.4}
\end{equation}
and 
\begin{equation}
(\Delta _{H(K)}f)(x)=\sum_{v\in \gamma }\left( \sum_{i=1}^{\dim \frak{k}%
}D^{2}f(x)(ve_{i},ve_{i})\right) =\sum_{v\in \gamma }\left( \sum_{i=1}^{\dim 
\frak{k}}(\partial _{ve_{i}}^{2}f)(x)\right) ,  \label{e.8.5}
\end{equation}
provided the derivatives and the sums exist and are independent of the
choice of bases.
\end{definition}

\begin{remark}
\label{r.8.6}This two-step procedure for defining an infinite-dimensional
trace appears already in Freed \cite{F} and \cite{DL}. Moreover, it is shown
in \cite{DL} that $D^{2}f(x)(ve_{i},ve_{i})$ is the same as Hess$%
f(x)(ve_{i},ve_{i})$ where Hess$f$ denotes the Hessian of $f$ relative to
the Levi-Civita connection on $H(K).$ So despite the fact that Hess$f(x)$ is 
\textbf{not }trace-class (\cite[Remark 3.13]{DL}) when $f$ is a cylinder
function on $H(K),$ it is reasonable to interpret $\Delta _{H(K)}$ defined
in (\ref{e.8.5}) as the Levi-Civita Laplacian.
\end{remark}

\begin{definition}
\label{d.8.8}Given a partition $\mathcal{P=\{}0=\tau _{0}<\tau _{1}<\cdots
<\tau _{n}=1\}$ of $[0,1],$ let $K^{\mathcal{P}}=K^{n}$ and $x_{\mathcal{P}%
}=(x_{\tau _{1}},x_{\tau _{2}},\ldots ,x_{\tau _{n}})\in K^{\mathcal{P}}$
for all $x\in H(K).$ We also define a second-order elliptic operator $\Delta
_{\mathcal{P}}$ acting on $C^{\infty }(K^{\mathcal{P}})$ by 
\begin{equation*}
\left( \Delta _{\mathcal{P}}\phi \right) (x_{1},x_{2},\ldots
,x_{n})=\sum_{i,j=1}^{n}\sum_{m=1}^{\dim \frak{k}}\min (\tau _{i},\tau
_{j})\left( D_{e_{m}}^{(i)}D_{e_{m}}^{(j)}\phi \right) (x_{1},x_{2},\ldots
,x_{n})
\end{equation*}
where 
\begin{equation*}
\left( D_{A}^{(i)}\phi \right) (x_{1},x_{2},\ldots ,x_{n})=\left. \frac{d}{ds%
}\right| _{s=0}\phi (x_{1},x_{2},\ldots ,x_{i-1},e^{sA}x_{i},x_{i+1},\ldots
,x_{n})
\end{equation*}
for all $A\in \frak{k}$ and $i=1,\ldots ,n.$
\end{definition}

\begin{example}
\label{ex.8.9}Suppose that $\mathcal{P=\{}0=\tau _{0}<\tau _{1}<\cdots <\tau
_{n}=1\}$ is a partition of $[0,1]$ and $f:H(K)\rightarrow \mathbb{C}$ is a
smooth cylinder function of the form $f(x)=\phi (x_{\mathcal{P}})$ where $%
\phi :K^{n}\rightarrow \mathbb{C}$ is a smooth function.

\begin{enumerate}
\item  Then $\Delta _{H(K)}f$ exists and 
\begin{equation}
\left( \Delta _{H(K)}f\right) (x)=\left( \Delta _{\mathcal{P}}\phi \right)
(x_{\mathcal{P}})  \label{e.8.9}
\end{equation}
for all $x\in H(K).$ See the proof of Proposition 4.19 in \cite{DL} for
details.

\item  For $x\in H(K)$ let 
\begin{equation*}
x_{\mathcal{P}}^{\prime }=(x_{\tau _{1}},x_{\tau _{1}}^{-1}x_{\tau
_{2}},\ldots ,x_{\tau _{n-1}}^{-1}x_{\tau _{n}})
\end{equation*}
be the ``incremental coordinates'' of $x$ relative to the partition $%
\mathcal{P}.$ If $f:H(K)\rightarrow \mathbb{C}$ is a smooth cylinder
function of the form $f(x)=\psi (x_{\mathcal{P}}^{\prime })$ with $\psi
:K^{n}\rightarrow \mathbb{C}$ being a smooth function, then 
\begin{equation}
\Delta _{H(K)}f(x)=\sum_{i=1}^{n}(\tau _{i}-\tau _{i-1})\left( \Delta
_{K}^{(i)}\psi \right) (x_{\mathcal{P}}^{\prime }\mathbf{).}  \label{e.8.10}
\end{equation}
Here $\Delta _{K}^{(i)}\psi $ denotes $\Delta _{K}$ acting on the i$^{th}$
variable of $\psi $ while holding the remaining variables fixed. This could
be proved by a finite-dimensional calculation showing that 
\begin{equation*}
(\Delta _{\mathcal{P}}\phi )(x_{1},x_{2},\dots ,x_{n})=\sum_{i=1}^{n}(\tau
_{i}-\tau _{i-1})\left( \Delta _{K}^{(i)}\psi \right)
(x_{1},x_{1}^{-1}x_{2},\dots ,x_{n-1}^{-1}x_{n})
\end{equation*}
where $\phi (x_{1},x_{2},\dots ,x_{n})=\psi \left(
x_{1},x_{1}^{-1}x_{2},\dots ,x_{n-1}^{-1}x_{n}\right) $ or by a calculation
similar to the proof of Proposition 4.19 in \cite{DL}.
\end{enumerate}
\end{example}

Since $\theta :\mathcal{A}\rightarrow H(K)$ is an isometry by Theorem \ref
{t.8.1} and $\Delta _{\mathcal{A}}$ and $\Delta _{H(K)}$ deserve to be
thought of as Laplace-Beltrami operator on $\mathcal{A}$ and $H(K)$
respectively (see Remark \ref{r.8.6} above), we should expect that $\Delta _{%
\mathcal{A}}(f\circ \theta )=(\Delta _{H(K)}f)\circ \theta $ for all
``nice'' functions $f$ on $H(K).$ This would certainly be true in finite
dimensions. It remains true in this infinite-dimensional context when we use
the ``two step'' trace in Definition \ref{d.8.5} of the Laplacians.

\begin{theorem}
\label{t.8.11}Suppose that $f:H(K)\rightarrow \mathbb{C}$ is a smooth
cylinder function. Then $\Delta _{\mathcal{A}}(f\circ \theta )$ and $\Delta
_{H(K)}f$ exist and 
\begin{equation}
\Delta _{\mathcal{A}}(f\circ \theta )=(\Delta _{H(K)}f)\circ \theta .
\label{e.8.11}
\end{equation}
In particular if $\phi $ is a smooth function on $K,$ then 
\begin{equation}
\Delta _{\mathcal{A}}\left( \phi \circ h\right) =\left( \Delta _{K}\phi
\right) \circ h.
\end{equation}
\end{theorem}

Recall that $h\left( A\right) =\theta _{1}\left( A\right) .$ The following
lemma will be needed in the proof of this theorem.

\begin{lemma}
\label{l.8.12}Let $a,b\in \frak{k,}$ $x\in H(K)$ and $k:[0,1]^{2}\rightarrow 
\frak{k}$ be given by $k(r,\tau )=[Ad_{x(r)}a,Ad_{x(\tau )}b].$ Then, for $%
t\in \lbrack 0,1],$ 
\begin{equation}
\sum_{V\in \Gamma }\int_{[0,t]^{2}}1_{r\leq \tau }k(r,\tau )V(r)V(\tau
)\,drd\tau =\frac{1}{2}\int_{0}^{t}k(\tau ,\tau )\,d\tau .  \label{e.8.13}
\end{equation}
\end{lemma}

\textit{Proof.} Let $S_{t}$ be the left member of (\ref{e.8.13}) and for $%
V\in \Gamma $ let $v(\cdot )=\int_{0}^{\cdot }V(\tau )d\tau $. Also define 
\begin{equation*}
A_{V}(t)=\int_{[0,t]^{2}}1_{r\leq \tau }k(r,\tau )V(r)V(\tau )\,drd\tau ,
\end{equation*}
so that $S_{t}=\sum_{V\in \Gamma }A_{V}(t).$ Integration by parts shows: 
\begin{eqnarray*}
A_{V}(t) &=&\int_{0}^{t}k(\tau ,\tau )v(\tau )V(\tau )d\tau
-\int_{[0,t]^{2}}1_{r\leq \tau }k_{r}(r,\tau )v(r)V(\tau )\,drd\tau \\
&=&\frac{1}{2}\int_{0}^{t}k(\tau ,\tau )\frac{dv^{2}(\tau )}{d\tau }\,d\tau
+\int_{[0,t]^{2}}1_{r\leq \tau }k_{r,\tau }(r,\tau )v(r)v(\tau )\,drd\tau \\
&&\qquad -\int_{0}^{t}\left( k_{r}(r,t)v(r)v(t)\,-k_{r}(r,r)v(r)v(r)\right)
dr \\
&=&\frac{1}{2}k(t,t)v^{2}(t)-\frac{1}{2}\int_{0}^{t}\frac{dk(\tau ,\tau )}{%
d\tau }v^{2}(\tau )\,d\tau +\int_{[0,t]^{2}}1_{r\leq \tau }k_{r,\tau
}(r,\tau )v(r)v(\tau )\,drd\tau \\
&&\qquad -\int_{0}^{t}\left( k_{r}(r,t)v(r)v(t)\,-k_{r}(r,r)v(r)v(r)\right)
dr,
\end{eqnarray*}
where $k_{r}=\partial k/\partial r$ and $k_{r,\tau }=\partial ^{2}k/\partial
r\partial \tau .$ Summing this equation on $V\in \Gamma $ implies that 
\begin{eqnarray*}
S_{t} &=&\frac{1}{2}k(t,t)t-\frac{1}{2}\int_{0}^{t}\frac{dk(\tau ,\tau )}{%
d\tau }\tau \,d\tau +\int_{[0,t]^{2}}1_{r\leq \tau }k_{r,\tau }(r,\tau
)r\,drd\tau \\
&&\qquad -\int_{0}^{t}\left( k_{r}(r,t)r\,-k_{r}(r,r)r\right) dr,
\end{eqnarray*}
wherein we have used the identity $\sum_{V\in \Gamma }v(r)v(\tau )=\min
(r,\tau ).$ (This identity is a consequence of the reproducing kernel
property of $\min (r,\tau )$ and Bessel's equality, see for example the
proof of Lemma 3.8 in \cite{DL}.) An integration by parts on the second and
third terms above shows that $S_{t}=\frac{1}{2}\int_{0}^{t}k(\tau ,\tau
)\,d\tau .$ \qed

\textit{Proof of Theorem \ref{t.8.11}.}\ Let $A,B\in \mathcal{A},$ $x=\theta
(A),$ and $w(\cdot )=\int_{0}^{\cdot }Ad_{\theta _{\tau }(A)}B(\tau )d\tau .$
By (\ref{e.8.1}) and the chain rule, $\partial _{B}\left( f\circ \theta
\right) (A)=\left( \partial _{w}f\right) (\theta (A))$ and 
\begin{equation}
\partial _{B}^{2}\left( f\circ \theta \right) =(\partial
_{_{w}}^{2}f)(x)+\partial _{\int_{\lbrack 0,\cdot ]^{2}}1_{r\leq \tau
}[Ad_{x(r)}B(r),Ad_{x(\tau )}B(\tau )]\,drd\tau }f(x).  \label{e.8.14}
\end{equation}
Because of the second term on the right side of (\ref{e.8.14}) it may be
seen that $D^{2}(f\circ \theta )$ is not trace-class--see \cite[Remark 3.13]
{DL}. On the other hand the two-step trace does exist. To compute this
trace, let $B=Ve_{i}\in \Gamma \beta $ and sum (\ref{e.8.14}) on $i=1,\ldots
,d$ and then on $V\in \Gamma .$ Letting 
\begin{equation}
k(r,\tau )=\sum_{i=1}^{\dim \frak{k}}[Ad_{x(r)}e_{i},Ad_{x(\tau )}e_{i}]
\label{e.8.15}
\end{equation}
and noting that $k(\tau ,\tau )=0,$ we may apply Lemma \ref{l.8.12} to find 
\begin{equation}
\Delta _{\mathcal{A}}(f\circ \theta )(A)=\sum_{V\in \Gamma }\sum_{i=1}^{\dim 
\frak{k}}\left( \partial _{_{\int_{0}^{\cdot }Ad_{x(\tau )}V(\tau
)e_{i}d\tau }}^{2}f\right) (x).  \label{e.8.16}
\end{equation}
Since $Ad_{x_{\tau }}$ is an isometry on $\frak{k,}$ $\{\int_{0}^{\cdot
}Ad_{x_{\tau }}V(\tau )e_{i}d\tau :V\in \Gamma ,i=1,\ldots ,d\}$ is an
orthonormal basis for $H(\frak{k}).$ Therefore by Example \ref{ex.8.9}, the
sum appearing in (\ref{e.8.16}) is precisely $\Delta _{H(K)}f(x).$\qed

\end{document}